\documentclass[fleqn,3p,onecolumn,11pt,nofootinbib,showkeys,showpacs]{revtex4-1}

\newcommand{\versionnumber}{v1.00}

\renewcommand{\today}{01.11.2012 (\versionnumber)}

\usepackage{graphicx}
\usepackage{bm}
\usepackage{mathrsfs}
\usepackage[latin1]{inputenc}
\usepackage{stmaryrd}
\usepackage{array}
\usepackage{psfrag}
\usepackage{dsfont}
\usepackage{epsfig}
\usepackage{titletoc}
\usepackage{float}   
\usepackage{wrapfig} 
\usepackage{amsmath}
\usepackage[latin1]{inputenc}
\usepackage{amsmath}\usepackage{amssymb,stmaryrd}
\usepackage{mathrsfs}
\usepackage{array}
\usepackage{graphicx}
\usepackage{psfrag}
\usepackage{dsfont}
\usepackage{epsfig}
\usepackage{cancel}
\usepackage[dvips]{feynmp}
\usepackage{upgreek}
\usepackage{subfigure}

\usepackage{textfit} 

\usepackage{hyperref}

\begin{document}

\title{A simple model of pointlike spacetime defects \\ and implications for photon propagation}

\author{M. Schreck} \email{marco.schreck@kit.edu}
\author{F. Sorba} \email{fabrizio.sorba@kit.edu}
\author{S. Thambyahpillai} \email{shiyamala.thambyahpillai@kit.edu}
\affiliation{Institute for Theoretical Physics \\ Karlsruhe Institute of Technology \\
76128 Karlsruhe, Germany}

\begin{abstract}
A model in which pointlike defects are randomly embedded in Minkowski spacetime is considered. The distribution of spacetime defects is
constructed to be Lorentz-invariant. It does not introduce a preferred reference frame, because it is based on a sprinkling process.
A field-theoretic action for the photon and a fermion is set up, in which the photon is assumed not to couple to the defects directly,
but via a scalar field. We are interested in signs for Lorentz violation caused by the spacetime defects, which are expected to reveal
themselves in the photon sector. A modification of the photon dispersion relation may result as a quantum effect and we compute it at
leading order perturbation theory.

The outcome of the calculation is that the photon dispersion law remains conventional, if the defect distribution is dense, homogeneous,
and isotropic. This result sheds some new light on Lorentz violation in the framework of a small-scale structure of spacetime. It shows that
Lorentz invariance can be preserved even in the presence of a spacetime structure that is supposed to emerge at the Planck scale. This conclusion
has already been drawn on general grounds in other publications, where the current paper delivers a demonstration by a direct computation in a
simple model.
\end{abstract}
\keywords{Lorentz violation; Gauge field theories; Photon properties; Theory of quantized fields}
\pacs{11.30.Cp, 11.15.-q, 14.70.Bh, 03.70.+k}

\maketitle

\newpage
\setcounter{equation}{0}
\setcounter{section}{0}
\renewcommand{\theequation}{\arabic{section}.\arabic{equation}}

\begin{fmffile}{Feynman}
  \fmfcmd{
  style_def fermion expr p =
    cdraw p;
    shrink (0.75);
        cfill (arrow p);
    endshrink;
  enddef;}
\fmfcmd{
  style_def dashes_arrow expr p =
    draw_dashes p;
    shrink (0.75);
        cfill (arrow p);
    endshrink;
  enddef;}
\fmfcmd{
  style_def dots_arrow expr p =
    draw_dots p;
    shrink (0.75);
        cfill (arrow p);
    endshrink;
  enddef;}
\fmfcmd{
    vardef line (expr p, lenone, lentwo, ang) = ((-lenone/2,lentwo/2)--(lenone/2,lentwo/2))
        rotated (ang + angle
            direction 0.5length(p) of p)
       shifted point 0.5length(p) of p
        enddef;
    vardef middir(expr p,ang) = dir (angle (direction length(p) of p) + ang)
        enddef;
    style_def fermion_momentum expr p =
        shrink(0.7);
            cfill(arrow p shifted(4.0mm*middir(p,25.982)));  
        endshrink;
        draw_fermion p;
        ccutdraw line(p, 8mm, 3.5mm, 0);
    enddef;
    style_def antifermion_momentum expr p =
        shrink(0.7);
            cfill(arrow reverse p shifted(4.0mm*middir(p,154.018)));  
        endshrink;
        draw_fermion p;
        ccutdraw line(p, 8mm, 3.5mm, 0);
            enddef;
    style_def boson_momentum expr p =
        shrink(0.7);
            cfill(arrow p shifted(4.0mm*middir(p,25.982)));  
        endshrink;
        draw_boson p;
        ccutdraw line(p, 8mm, 3.5mm, 0);
enddef;}


\section{Introduction}
\label{sec:introduction}

Physics at the Planck scale is considered to be the holy grail in present fundamental research. Up to now it has not been rigorously shown which
physical phenomena occur at the Planck scale and how they can be described within a mathematical formalism. For this reason it makes sense to
construct simple models in order to mimic effects expected to occur at this scale. The property ``simple'' means using underlying concepts of
established theories such as general relativity and quantum theory, which are well understood and hold for energies much smaller than the Planck
energy.

One fundamental concept are Einstein's field equations in general relativity linking energy density to spacetime geometry. Another one is
Heisenberg's uncertainty principle in quantum theory saying that two complementary particle properties are always endowed with uncertainties. Assuming
that both still hold at the Planck scale, energy uncertainty will result in an uncertainty of spacetime geometry or even topology. Thus, under this
assumption, spacetime metric coefficients and spacetime curvature will begin to fluctuate at the Planck energy. Such fluctuations are often referred
to as spacetime foam or spacetime defects \cite{Wheeler:1957mu,Wheeler:1968,Hawking:1979zw,Hawking:1979pi,Friedman:1990xc,Visser:1996}.

There exist various models for spacetime defects with nontrivial topology. One way in which to obtain a spacetime defect is to cut an open set
out of Minkowski space and to impose certain conditions on the remaining boundary. This then leads to a spacetime $\mathcal{M}$ having a different
topology. For example, cutting out an open ball of $\mathbb{R}^3$ and identifying antipodal points on the boundary results in
$\mathcal{M}=\mathbb{R}\times (\mathbb{R}P^3-\{\mathrm{point}\})$, where $\mathbb{R}P^3$ is the three-dimensional orientable real projective space 
\cite{Schwarz:2010}.

However, the interest not only lies in the defects themselves, but in their influence upon the propagation properties of particles, for example photons.
In \cite{BernadotteKlinkhamer2007}, the modification of the photon dispersion law is investigated for certain classes of defects. The method used is
to consider the scattering of an electromagnetic wave at one single defect. Certain conditions for the physical fields are then set on the boundary
of the defect and Maxwell's equations are solved by introducing a correction field. Even for a single defect this is a difficult task and the
approach would be more difficult for two defects and impractical for a large number of defects nearby.

Since we are interested in the propagation of photons through a spacetime foam made up of many defects, we proceed with an alternative possibility
that was initiated in \cite{hep-th/0312032}. Here the \textit{CPT}-anomaly is derived for a non-Abelian gauge group \textit{SO}(10) on two spacetime
manifolds with nontrivial topology: a spacetime with a linear defect $\mathcal{M}=\mathbb{R}\times (\mathbb{R}\times (\mathbb{R}^2\setminus \{0\}))=
\mathbb{R}^4\setminus \mathbb{R}^2$ and a spacetime with two identical open balls removed from $\mathbb{R}^3$ and points on their boundary
properly identified (wormhole). It is shown that the \textit{CPT}-anomaly occurs for the Abelian subgroup $\mathit{U}(1)\subset \mathit{SO}(10)$, as
well. The anomaly gives a contribution to the effective action as an $F\widetilde{F}$-term with the field strength tensor $F$ and its dual $\widetilde{F}$.
Since it is tremendously difficult to obtain the anomaly for several defects, the introduction of a background field is mandatory. This field does
not describe any microscopic defect properties and, hence, serves as an effective approach for the case when the photon wave length is much larger
than the defect size.

We will follow this idea and describe a single defect as pointlike, where it is assumed to be time-dependent --- contrary to Ref.~\cite{hep-th/0312032}.
Such defects are distributed randomly in Minkowski space resulting in an effective ``random'' background field. Furthermore, the distribution of defects
is taken as being Lorentz-invariant. We study whether and how the dispersion relation of photons is affected by such a time-dependent and Lorentz-invariant
background.

The outline of this paper is as follows. In Sec.~\ref{sec:action-effective-theory} we introduce the action of the effective theory that forms the basis
of the article. This action describes the interaction between photons and the defects that is mediated via a real scalar field. Section~\ref{sec:statistical-treatment}
gives a description of how to treat photon propagation through a distribution of many pointlike defects that are put randomly at distinct points in Minkowski
spacetime. In Sec.~\ref{sec:solution-field-equation} the focus will be on the perturbative solution of the photon field equation. In the first part the solution
is obtained by inserting a perturbative \textit{Ansatz} into the field equation. In the second part we demonstrate how the same result follows from a diagrammatic
approach. As a next step we compute the leading-order solution of the photon field equation in Sec.~\ref{sec:caluclation-photon-field-perturbation}, where in this
context a renormalization procedure has to be performed. In the follow-up section, \ref{sec:caluclation-scalar-field-perturbation}, the scalar field equation is
solved to leading order in perturbation theory as well. Both results are combined to calculate the modified photon dispersion relation in
Sec.~\ref{sec:dispersion-relation-photon-modified}, and the physical meaning of the result is then discussed. Throughout the paper certain assumptions
will be made so that the calculation is feasible. In Sec.~\ref{sec:energy-transfer-to-defects} we make a couple of remarks on how the result may change if
certain assumptions are dropped. In the penultimate section, \ref{sec:pt-symmetric-extension}, we go on a brief excursion to \textit{PT}-symmetric quantum field theory
in the context of the special model proposed. The last section, \ref{sec:summary-and-conclusions}, gives a summary and a conclusion on the results. The most important
computational steps can be found in Apps. \ref{sec:table-dense-distribution}, \ref{sec:perturbative-feynman-rules}, \ref{sec:derivation-passarino-veltman}, and \ref{sec:computation-scalar-integrals}.

Throughout the paper natural units are used with $\hbar=c=1$. For some occurrences, $c$ and $\hbar$ will be reinstated for clarity.

\section{Action of the effective theory}
\label{sec:action-effective-theory}
\setcounter{equation}{0}

We wish to describe photon propagation through a Lorentz-invariant distribution of time-dependent, pointlike spacetime defects. The action upon which our
considerations are based is given as follows \cite{Klinkhamer:2012pr}:
\begin{align}
\label{eq:action-effective-theory}
S_{\mathrm{eff}}&=\int_{\mathbb{R}^4} \mathrm{d}^4x\,\bigg[-\frac{1}{4}F_{\mu\nu}(x)F^{\mu\nu}(x)-\frac{1}{2}\big(\partial_{\mu}A^{\mu}(x)\big)^2 \notag \\
&\phantom{{}={}\int_{\mathbb{R}^4} \mathrm{d}^4x\bigg[}+\frac{1}{2(b^{(0)})^2}\Big(\partial_{\mu}\phi(x)\partial^{\mu}\phi(x)-\frac{1}{(b^{(0)})^2}\phi(x)^2\Big) \notag \\
&\phantom{{}={}\int_{\mathbb{R}^4} \mathrm{d}^4x\bigg[}+\phi(x)\sum^{\infty}_{n=1} \varepsilon_n\delta^{(4)}(x-x_n)-\frac{\lambda^{(0)}}{4} f\big(\phi(x)\big)F_{\mu\nu}(x)\widetilde{F}^{\mu\nu}(x)\bigg]\,,
\end{align}
where $F_{\mu\nu}(x)\equiv \partial_{\mu}A_{\nu}(x)-\partial_{\nu}A_{\mu}(x)$ is the field strength tensor of the \textit{U}(1) gauge field $A_{\mu}(x)$.
The first term in the action is the standard kinetic term of the free photon field and the second fixes the gauge (we use Feynman gauge).
The third contribution contains the kinetic and mass terms of a free real scalar field $\phi(x)$, where $b^{(0)}$ is a parameter with mass dimension $-1$.
The scalar field $\phi(x)$ itself has mass dimension zero and its bare mass is given by $1/b^{(0)}$.

The fourth expression involves the pointlike spacetime defects sitting at distinct spacetime points $x_n$. The defects are effectively described as randomly
distributed single four-dimensional $\delta$-functions, each carrying some charge $\varepsilon\in \{-1,1\}$. Concretely, this means that each defect
appears at a single point in three-space for an infinitesimally short amount of time before disappearing again. This illustrative picture corresponds to what
a theorist has in mind when thinking on a simple spacetime foam.

Only the field $\phi(x)$ is assumed to couple directly to the defects via the charge $\varepsilon$. Finally the interaction of the photon field with the defects
is described by the last term, where the interaction is mediated by the scalar field $\phi$. The latter term involves the dual field strength tensor
\begin{equation}
\widetilde{F}^{\mu\nu}(x)\equiv \frac{1}{2}\varepsilon^{\mu\nu\varrho\sigma}F_{\varrho\sigma}(x)\,,
\end{equation}
with the four-dimensional Levi-Civita tensor $\varepsilon^{\mu\nu\varrho\sigma}$. The motivation behind such an interaction is that it appears in the effective
action in the context of the \textit{CPT}-anomaly \cite{hep-th/0312032}. The function $f(\phi(x))$ in the last term of Eq.~\eqref{eq:action-effective-theory}
can be arbitrary, in principle, but it is assumed to be sufficiently well-behaved. In order to keep the model simple we choose $f(\phi(x))=\phi(x)$. Note that
the fifth and sixth terms explicitly break gauge invariance.\footnote{They are not invariant under a gauge transformation of the field $\phi$, namely
$\phi(x)\mapsto -\phi(x)$. To be crystal clear, this gauge transformation has nothing to do with the \textit{U}(1) gauge transformation of the photon
field.} All fields are defined in Minkowski spacetime with metric $\eta_{\mu\nu}=\mathrm{diag}(1,-1,-1,-1)$. To summarize, the field content of the theory
is presented in Table \ref{tab:field-content}.
\begin{table}[t]
\centering
\begin{tabular}{c|c|m{1.6cm}|m{1.6cm}|m{2.0cm}m{0.1cm}}
\cline{1-5}
field/object & mass dimension & \multicolumn{3}{c}{coupling constant (charge) to} & \\
\cline{3-5}
 & & \centering $A_{\mu}$ & \centering $\phi$ & \centering defect & \\
\cline{1-5}
\cline{1-5}
photon $A_{\mu}$ & 1 & \multicolumn{1}{c|}{0} & \multicolumn{1}{c|}{$\lambda^{(0)}$} & \multicolumn{1}{c}{0} & \\
scalar $\phi$ & 0 & \multicolumn{1}{c|}{} & \multicolumn{1}{c|}{0} & \multicolumn{1}{c}{$\varepsilon$} & \\
defect & 4 & \multicolumn{1}{c|}{} & \multicolumn{1}{c|}{} & \multicolumn{1}{c}{0} & \\
\cline{1-5}
\end{tabular}
\caption{Fields and nondynamical objects (defects) appearing in the action of Eq.~\eqref{eq:action-effective-theory} with corresponding mass dimension,
bare coupling constant $\lambda^{(0)}\ll 1$ and $\varepsilon=\pm 1$.}
\label{tab:field-content}
\end{table}

If we wanted to couple photons to a conserved\footnote{By quantum corrections the explicit violation of gauge invariance in the
action $S_{\mathrm{eff}}$ of Eq.~\eqref{eq:action-effective-theory} may give rise to an anomalous nonconserved current. However, this effect is expected
to be suppressed by $(\lambda^{(0)})^2$.} fermionic current $j^{\mu}(x)=\overline{\psi}(x)\gamma^{\mu}\psi(x)$ with the Dirac field $\psi(x)$ and
standard Dirac matrices $\gamma^{\mu}$, the modified theory could be coupled to the Dirac theory of standard spin-1/2 fermions of charge $e$ and mass $M$:
\begin{equation}
\label{eq:standDirac-action}
S_{\mathrm{Dirac}}=\int_{\mathbb{R}^4} \mathrm{d}^4 x \, \overline\psi(x) \Big\{
\gamma^\mu \big[\mathrm{i}\,\partial_\mu -e A_\mu(x) \big] -M\Big\} \psi(x)\,.
\end{equation}
This would make the complete action of the theory
\begin{equation}
\label{eq:action-modified-theory-complete}
S=S_{\mathrm{eff}}+S_{\mathrm{Dirac}}\,,
\end{equation}
with $S_{\mathrm{eff}}$ given by Eq.~\eqref{eq:action-effective-theory} and $S_{\mathrm{Dirac}}$ by Eq.~\eqref{eq:standDirac-action}. The
description of the spacetime foam model by the action given is the fundamental assumption of this paper. It will be referred to
as Assumption~(1).\footnote{In what follows, several further assumptions will be taken. In such a context the word ``assumption'' is
abbreviated as ``Ass.'' in combination with a number and optional small Latin letters.}

\section{Statistical treatment of a large number of spacetime defects}
\label{sec:statistical-treatment}
\setcounter{equation}{0}

\subsection{Distribution of defects in Minkowski spacetime (sprinkling)}
\label{sec:sprinkling}

\begin{figure}[t]
\centering
\subfigure[$\,\,\,\beta=0$]{\includegraphics[scale=0.49]{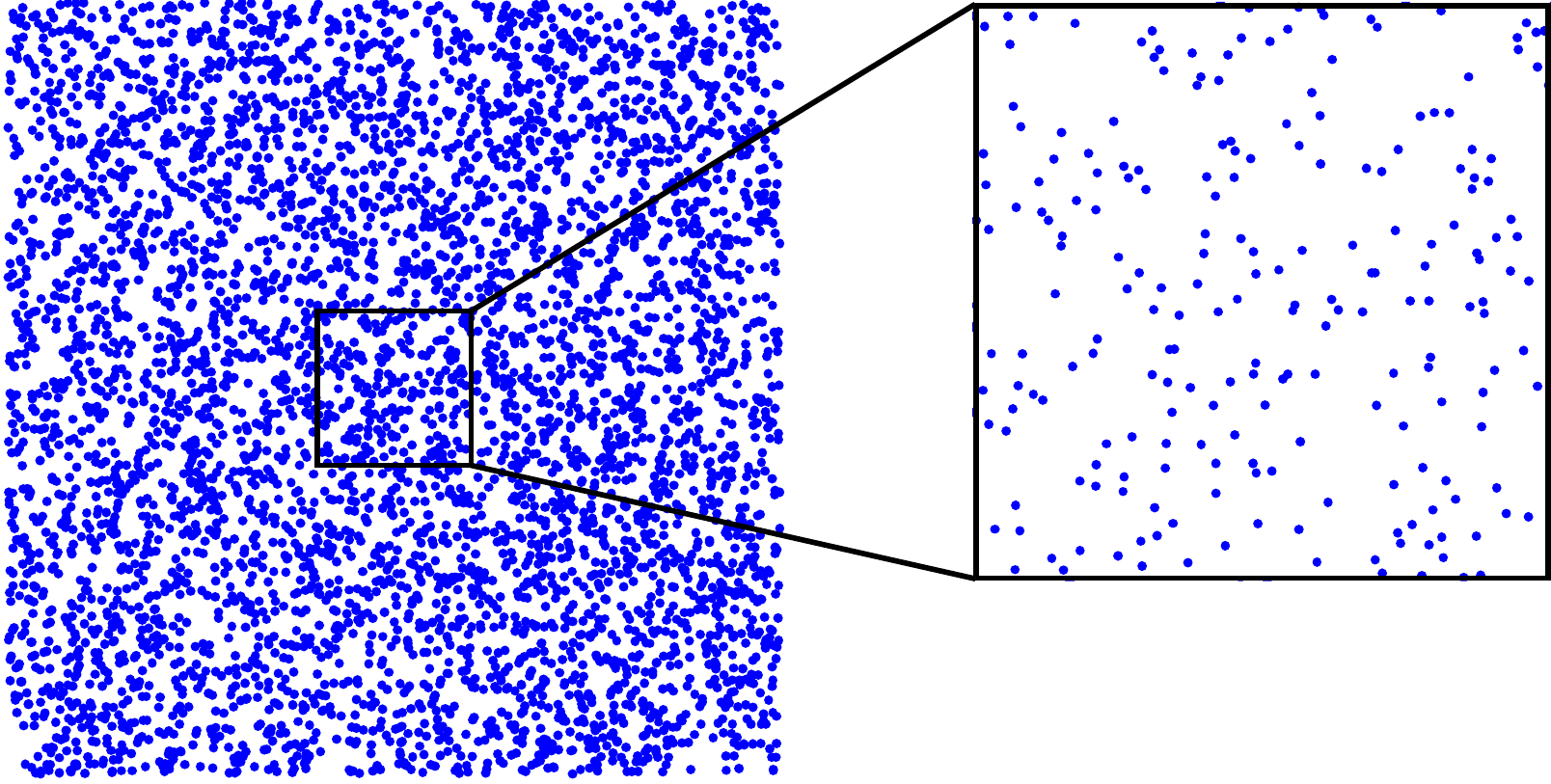}}
\subfigure[$\,\,\,\beta=0.7$]{\includegraphics[scale=0.49]{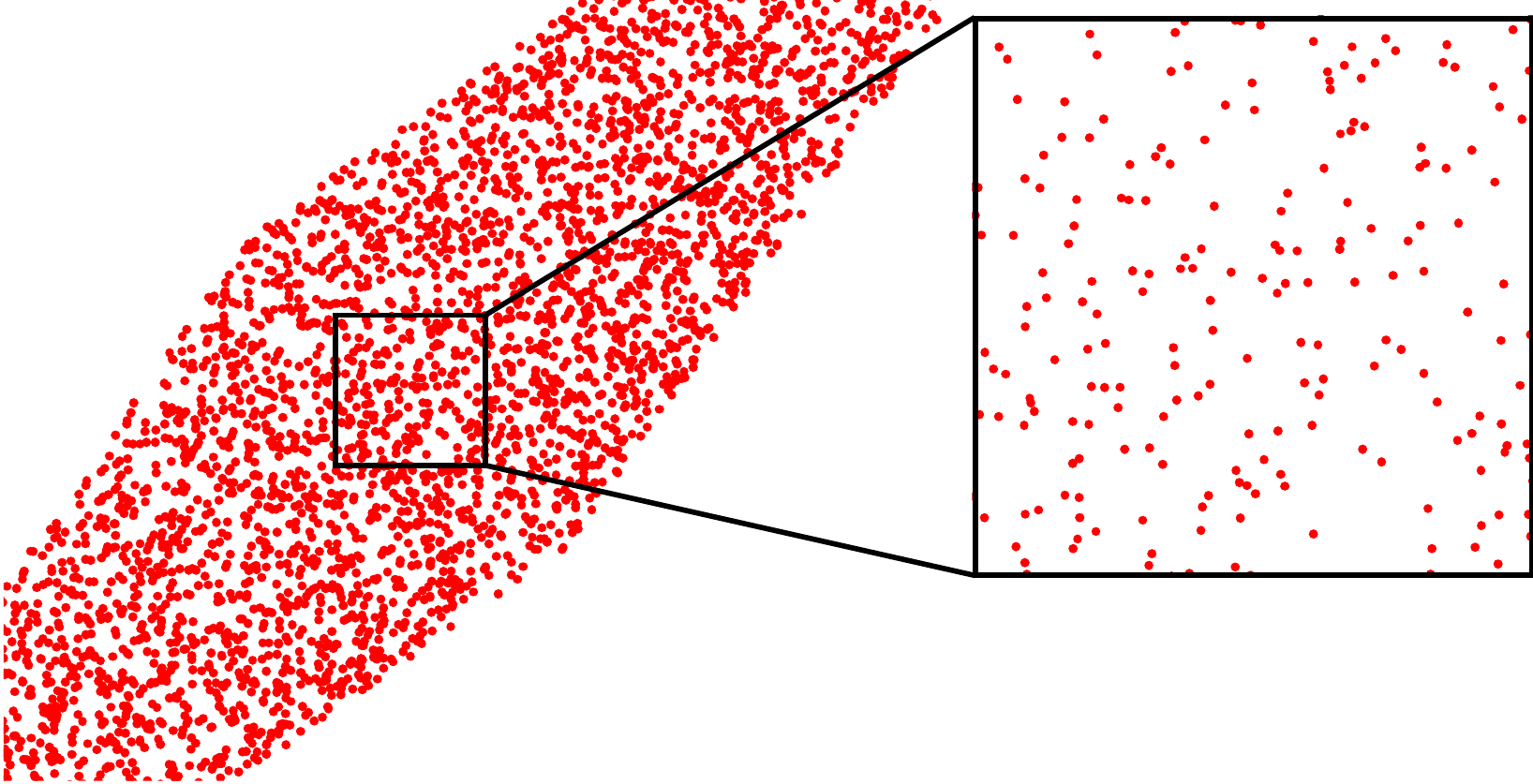}}
\caption{Example of sprinkling in a finite region of a two-dimensional spacetime (1D-space $\times$ time) as it looks in two different inertial
frames. The second frame is boosted along the positive spatial axis with a Lorentz boost factor $\beta=0.7$ with respect to the first frame. This
boost changes the shape of the spacetime region. However, the mean density $\langle\rho_{\mathrm{obs}}\rangle$ of defects is the same in both frames.
As an illustration compare the two enlarged regions of both distributions. Note that in the boosted distribution not all defects are shown.
}
\label{fig:defect-sprinkling}
\end{figure}

We intend to distribute spacetime defects in four-dimensional Minkowski spacetime $\mathcal{M}$ in a Lorentz-invariant manner. This will be
possible if defects are distributed according to a ``Poisson process'' (i.e. a sprinkling). The result of the Poisson process is a Poisson
distribution of defects throughout the spacetime. This means that the probability of observing $n$ defects in a rectangular spacetime region
with side length $\mathscr{R}$ and volume
\begin{equation}
\label{eq:minkowski-space-subset-volume}
\mathcal{V}=\int\limits_{\mathrm{region}} \mathrm{d}^4x\,\sqrt{-\det(\eta_{\mu\nu})}=\int\limits_{\substack{|x^{\mu}|\leq \mathscr{R}/2 \\ \mu\in \{0,1,2,3\}}} \mathrm{d}^4x=\mathscr{R}^4\,,
\end{equation}
is given by:
\begin{equation}
\label{eq:poisson-distribution}
P_{n}(\mathcal{V})=\frac{(\varrho \mathcal{V})^{n}\exp(-\varrho \mathcal{V})}{n!}\,.
\end{equation}
Herein, $\varrho$ is at first a parameter that characterizes the distribution. Note that Eq.~\eqref{eq:poisson-distribution} is a valid
description for a probability distribution, which is both isotropic and homogeneous. The most natural choice for a distribution of pointlike
--- i.e. zero-dimensional --- spacetime defects is an isotropic one if we do not take into account any mechanism producing defects that make space
anisotropic (for example, defects similar to cosmic strings \cite{Kibble:1976sj}). When the volume of the region approaches zero
($\mathcal{V}\mapsto \delta \mathcal{V}$ with an infinitesimal value $\delta \mathcal{V}$), the probability of finding a single defect in that
region is proportional to the volume
\begin{equation}
P_{n=1}(\delta \mathcal{V})=\varrho \delta \mathcal{V}+\mathcal{O}(\delta \mathcal{V}^2)\,.
\end{equation}
On the other hand, the probability of finding more than one defect is negligible:
\begin{equation}
P_{n>1}(\delta \mathcal{V})=\mathcal{O}(\delta \mathcal{V}^n)\,.
\end{equation}
An explicit realization of the Poisson process is then described by the following steps \cite{Dowker:2003hb}:
\begin{itemize}

\item[1)] Divide $\mathcal{V}$ into small boxes with volume $V$.
\item[2)] Then place a defect into each box with probability $P=\varrho V$.
\item[3)] The Poisson process is obtained in the limit $V \mapsto 0$.

\end{itemize}
The Poisson process or ``sprinkling'' is invariant under any volume-preserving linear transformation and in particular it is invariant under
Lorentz transformations (this happens because the process only depends on the spacetime volume $V$). Moreover, it has been shown in
\cite{Bombelli:2006nm} that the realizations of the Poisson process are Lorentz-invariant individually as well. Lorentz invariance in this
context has the following meaning
\begin{center}
``\emph{The discrete set of sprinkled points must not, in and of itself, serve to pick out a preferred reference frame.}'' \cite{Dowker:2003hb}
\end{center}
That is, the statistical properties of the distribution of defects (e.g. the mean density of defects) do not depend on which reference frame
we choose to measure them in (see Fig.~\ref{fig:defect-sprinkling}).

We emphasize that the number of defects contained in different regions of equal volume $V$ is not constant but fluctuates from region to
region
\begin{equation}
N(V)=\langle N(V)\rangle\pm\delta N(V)\,,
\end{equation}
where the mean number of defects and the standard deviation are respectively
\begin{equation}
\langle N(V)\rangle=\sum_{n=0}^{\infty} nP_{n}(V)=\varrho V\,, \quad \delta N(V)=\sqrt{\sum_{n=0}^{\infty} \big(n-\langle n\rangle\big)^2P_n(V)}=\sqrt{\varrho V}\,.
\end{equation}
These results can be explicitly obtained by using Eq.~\eqref{eq:poisson-distribution}. Therefore we can identify the parameter $\varrho$ with the
mean density of defects: $\langle \rho_{\mathrm{obs}}\rangle=\varrho$. Moreover fluctuations in the number of defects also imply fluctuations in
the density $\rho_{\mathrm{obs}}$ from a region to another
\begin{equation}
\rho_{\mathrm{obs}}=\langle\rho_{\mathrm{obs}}\rangle\pm\delta\rho=\frac{\langle N(V)\rangle}{V}\pm\frac{\delta N(V)}{V}=\varrho\pm\sqrt{\frac{\varrho}{V}}\,.
\end{equation}
Nevertheless, these fluctuations become negligible when the mean volume occupied by the defects, namely $V_d=1/\varrho$, is much smaller than the
volume $V$ of the region considered
\begin{equation}
V_d=\frac{1}{\varrho}\ll V\Longrightarrow\frac{\delta\rho}{\varrho}=\frac{1}{\sqrt{\varrho V}}\sim 0\,.
\end{equation}
Thus we can regard the density as being constant so long as we consider scales that are much larger than the mean separation between defects:
$N(V)=\varrho V=\mathrm{const.}$ when $V\gg V_d$. Being able to perform computations with a globally constant volume means that the distribution
is homogeneous as well. Otherwise the density $\varrho$ would only be defined locally. We will refer to the isotropic and homogeneous distribution
of spacetime defects as Ass.~(2a) and Ass.~(2b), respectively.

\subsection{Derivation of auxiliary functions}
\label{sec:auxiliary-functions}

In this section we intend to derive a set of functions that will be extensively needed for the solution of the scalar and photon field equations. First
of all, we investigate the problem of photon propagation through a spacetime with $\mathcal{N}$ defects in the finite subset with volume $\mathcal{V}$
given by Eq.~\eqref{eq:minkowski-space-subset-volume}. The corresponding functions and fields are denoted with indices `$\mathcal{N}$' or `$\mathscr{R}$'.
We now consider the ``free'' field equation for $\phi$, i.e. neglecting the coupling to the photon but not to the defects. In order to solve the field equation,
the truncated Fourier transform of $\phi(x)$ when restricted to the box is necessary
\begin{equation}
\widetilde{g}_{\scriptscriptstyle{\mathcal{N}}}(k)\equiv \int\limits_{\substack{|x^{\mu}|\leq \mathscr{R}/2 \\ \mu\in \{0,1,2,3\}}} \mathrm{d}^4x\,\exp(\mathrm{i}k\cdot x)\phi(x)\,.
\end{equation}
Analogously we define the inverse Fourier transform by
\begin{equation}
\label{eq:truncated-fourier-transform-inverse}
\phi(x)=\int\limits_{\substack{|k^{\mu}|\geq 1/\mathscr{R} \\ \mu\in \{0,1,2,3\}}} \frac{\mathrm{d}^4k}{(2\pi)^4}\,\exp(\mathrm{i}k\cdot x)\widetilde{g}_{\scriptscriptstyle{\mathcal{N}}}(k)\,.
\end{equation}
In configuration space the coordinates are restricted by the side length of the box considered. This corresponds to a minimum value in momentum
space, which is manifest in the integration limits. In principle, for a finite volume the Fourier transform would correspond to a Fourier series
with discrete coordinates or momenta. However for simplicity we will assume a continuous spectrum. This does not play any role as we will eventually
generalize the results to the whole of Minkowski spacetime anyway. We use Eq.~\eqref{eq:truncated-fourier-transform-inverse} as an \textit{Ansatz} to obtain
the following solution of the field equation of $\phi(x)$ in momentum space
\begin{subequations}
\begin{equation}
\label{eq:definition-gn-function}
\widetilde{g}_{\scriptscriptstyle{\mathcal{N}}}(k)=\sqrt{\mathcal{N}}\widetilde{H}(k)\widetilde{G}_{\mathcal{N}}(k)\,,
\end{equation}
\begin{equation}
\label{eq:definition-gn-function-2}
\widetilde{H}(k)\equiv \frac{-(b^{(0)})^2}{k^2-1/(b^{(0)})^2+\mathrm{i}\epsilon}\,,\quad \widetilde{G}_{\mathcal{N}}(k)\equiv \frac{1}{\sqrt{\mathcal{N}}}\sum^{\mathcal{N}}_{i=1} \varepsilon_i\exp(\mathrm{i}k\cdot x_i)\,.
\end{equation}
\end{subequations}
The solution depends on the bare mass of the scalar field and the distribution of spacetime defects. Furthermore it consists of two contributions.
The first, $\widetilde{G}_{\mathcal{N}}(k)$, solely describes the defects and the second, $\widetilde{H}(k)$, serves as a mediator between the
defects and the photons. In principle Eq.~\eqref{eq:definition-gn-function} can be regarded as being a solution of the classical field equations
restricted to a finite box. Later on we will need to consider the product of two such solutions evaluated at different momenta --- especially for the
perturbative photon field. The function $\widetilde{g}_{\scriptscriptstyle{\mathcal{N}}}(k)$ then serves as an effective background that describes
the influence of the spacetime defects on the photons. Therefore we will refer to it as an effective background field in what follows.

Since Eq.~\eqref{eq:definition-gn-function} gives the background field with respect to the number of defects, we define the corresponding
background field with regard to the volume:
\begin{equation}
\label{eq:definition-gr-function}
\widetilde{g}_{\scriptscriptstyle{\mathscr{R}}}(k)\equiv \sqrt{\mathcal{V}}\widetilde{H}(k)\widetilde{G}_{\mathscr{R}}(k)\,,\quad \widetilde{G}_{\mathscr{R}}(k)\equiv \frac{1}{\sqrt{\mathcal{V}}}\sum^{\mathcal{N}}_{i=1} \varepsilon_i\exp(\mathrm{i}k\cdot x_i)\,.
\end{equation}
We keep in mind that $\phi$ couples to the photon via the last term of Eq.~\eqref{eq:action-effective-theory}. After establishing all intermediate
results we compute the limits $\mathcal{N}\mapsto \infty$ and $\mathscr{R}\mapsto \infty$, respectively.

In the subsequent paragraphs we are going to perform a statistical treatment of photon propagation through the background field. In light of this we
will encounter products of functions defined by Eqs.~\eqref{eq:definition-gn-function}, \eqref{eq:definition-gn-function-2}
(or Eq.~\eqref{eq:definition-gr-function}), where each depends on a different four-momentum. Such products will have to be summed over the total number
$\mathcal{N}$ of defects distributed in the whole of Minkowski space.

According to Eq.~\eqref{eq:poisson-distribution} the defect density does not change with respect to Lorentz transformations. However this does not mean
that the defect positions are Lorentz-invariant. Therefore, besides summing over the number of defects we are also going to integrate over all Lorentz
transformations from zero to an upper limit $\Lambda'$. We introduce this cutoff to avoid difficulties from integrating over the noncompact Lorentz
group. We are then ready to compute the sum of complex exponential functions (random phases), each evaluated at the spacetime point of a defect.
The corresponding result will be needed later:
\begin{align}
\label{eq:sum-of-random-phases}
\int\limits^{\Lambda'} \mathrm{d}\Lambda\,\sum^{\mathcal{N}}_{n=1} \exp[\mathrm{i}k\cdot (\Lambda x_n)]&\simeq\int\limits^{\Lambda'} \mathrm{d}\Lambda\,\varrho\int\mathrm{d}^4x\, \exp[\mathrm{i}k\cdot (\Lambda x)] \notag \displaybreak[0]\\
&=\int\limits^{\Lambda'} \mathrm{d}\Lambda\,\varrho\int\mathrm{d}^4x'\,|\det(\Lambda^{-1})|\exp[\mathrm{i}k\cdot x'] \notag \displaybreak[0]\\
&=\Xi\,(2\pi)^4\varrho\,\delta^{(4)}(k)\,,\quad \Xi\equiv \int\limits^{\Lambda'} \mathrm{d}\Lambda\,.
\end{align}
In the second step we have approximated the sum over $n$ as an integral over $x$:
\begin{equation}
\sum^{\mathcal{N}}_{n=1} \mapsto \int \mathrm{d}n=\varrho\int \mathrm{d}^4x\,.
\end{equation}
This is possible since the sprinkling procedure ensures the proportionality between the number of defects and the volume: $\mathrm{d}n=\varrho\,\mathrm{d}^4x$.
Furthermore the defect distribution is assumed to be \textit{dense}. The latter is an important issue not only for calculational but for physical reasons as
well. In Ref.~\cite{BernadotteKlinkhamer2007} a classical spacetime foam with topologically nontrivial defects having a particular size $\overline{b}$ and a mean
separation $\overline{l}$ is considered. Bounds obtained from the absence of vacuum Cherenkov radiation lead to the constraint $\overline{b}/\overline{l}
\lesssim 10^{-7}$ within the spacetime foam model considered in this reference. Hence, for spacetime defects of the size $10^2\times L_{\mathrm{Pl}}$ with
the Planck length $L_{\mathrm{Pl}}\equiv \sqrt{G\hbar/c^3}\approx 1.62\cdot 10^{-35}\,\mathrm{m}$, where a classical approach is supposed to be valid, the
defects would be separated by at least $10^9\times L_{\mathrm{Pl}}\approx 1.62\cdot 10^{-26}\,\mathrm{m}$. Even if their separation is larger by several
orders of magnitude, the approximation of a dense distribution still makes sense. We assume that this conclusion can be applied to the spacetime foam model
with pointlike defects investigated here. This is fortified by Table \ref{tab:separation-between-defects} in App.~\ref{sec:table-dense-distribution}. The 
dense distribution of defects will be referred to as Ass.~(2c).

The last assumption in Eq.~\eqref{eq:sum-of-random-phases} when computing the remaining integral is an infinite spacetime. Current cosmological data implies that
we live in a flat universe of finite age. For the curvature radius of the universe, which is related to its size, only upper bounds can be given within the
$\Lambda\mathrm{CDM}$ model \cite{Komatsu:2008hk}. Therefore it has not as yet been clarified whether the universe has a finite or infinite volume. However
considering photon propagation on time scales that are much smaller than cosmological time scales the Friedmann--Robertson--Walker metric for a flat universe
corresponds to the Minkowski metric to a reasonable approximation. In order to keep the model as simple as possible, we do not describe cosmological
effects and assume the spacetime volume to be infinite. In what follows we will refer to the infinite spacetime volume as Ass.~(2d).

After writing the sum as an infinite spatial integral, we change the variables to $x\mapsto x'=\Lambda x$ and employ the fact that
$|\det(\Lambda^{-1})|=1$. The integral leads to a four-dimensional $\delta$-function. The result does not depend on the total number $\mathcal{N}$ of defects
any more but only on the density $\varrho$. The latter acts as a constant of proportionality which is independent of the reference frame. Because of this the
final result is Lorentz-invariant. The Lorentz transformation of the defect positions does not play a role and what remains is the integral over all
Lorentz transformations. It corresponds to the ``volume'' $\Xi$ of the Lorentz group when restricted to an upper cutoff $\Lambda'$ and it is a mere number.

We can now compute the product of two functions $\widetilde{G}_{\mathcal{N}}$ --- followed by an integration over all Lorentz transformations up to the
cutoff $\Lambda'$ --- in the limit of large~$\mathcal{N}$:
\begin{align}
\label{eq:product-two-gn}
\int\limits^{\Lambda'} \mathrm{d}\Lambda\lim_{\mathcal{N}\mapsto \infty} \!\widetilde{G}_{\mathcal{N}}(k)\widetilde{G}_{\mathcal{N}}(p)=\int\limits^{\Lambda'} \mathrm{d}\Lambda\lim_{\mathcal{N}\mapsto \infty} \frac{1}{\mathcal{N}}\Bigg[\sum^{\mathcal{N}}_{i=1} \exp[\mathrm{i}(k+p)(\Lambda x_i)]+\sum_{m\neq n} P_{mn}\Bigg]\,.
\end{align}
The quantity $P_{mn}$ involves the product of charges of different defects:
\begin{equation}
P_{mn}=\varepsilon_m\varepsilon_n\exp[\mathrm{i}k\cdot (\Lambda x_m)]\exp[\mathrm{i}p\cdot (\Lambda x_n)]\,.
\end{equation}
The sum over $P_{mn}$ has to be evaluated for a large number of defects. This will be done in the following few lines. We transform coordinates to Euclidian
space via a Wick rotation, where these coordinates will be marked by an index `$E$'. Now consider a small hypercube $H_{a_n}(x_{E,n})$ around a defect at $x_{E,n}$
with side length
\begin{equation}
a_n \equiv \left(\frac{\mathcal{V}}{\mathcal{N}}\right)^{1/4}=\left(\frac{1}{\varrho}\right)^{1/4}\,,
\quad n\in \{1,\dots,\mathcal{N}\}\,.
\end{equation}
On average every defect lies within such a hypercube. For $\mathcal{N}\gg 1$, given a defect at $x_n$ with charge $\varepsilon=\pm 1$, a
partner with a charge of opposite sign $\varepsilon=\mp 1$ can be found at a distance $\delta x_{E,n}$ that is of the order of the side
length $a_n$ of a small hypercube (see Table \ref{tab:neighboring-defects-of-opposite-sign}).
\begin{table}[t]
\centering
\begin{tabular}{c|c|c}
\hline
defect index & sign of charge & defect position \\
\hline
\hline
$i $ & $\pm 1$ & $x_{E,i}$ \\
$j \neq i$ & $\mp 1$ & $x_{E,j}=x_{E,i}+\delta x_{E,i}$ \\
\hline
\end{tabular}
\caption{For a large number of defects we consider a single defect at $x_{E,i}$ with a specific charge. Then the probability of finding a
neighboring defect at $x_{E,j}$ with a charge of the opposite sign,
where $x_{E,j}$ has a small Euclidian distance to $x_{E,i}$, is close to
1.}
\label{tab:neighboring-defects-of-opposite-sign}
\end{table}
We can therefore perform a Taylor expansion for $P_{mn}$ in the small parameters $\delta x_{E,n}$. For the sum over $P_{mn}$ for $m\neq n$ we
consider defects at $x_{E,m}$ and $x_{E,n}$ with charges $\varepsilon=1$ and their neighbors at $\delta x_{E,m}$ and $\delta x_{E,n}$ with
charges $\varepsilon=-1$. Substituting $x_{E,i}'\equiv \Lambda x_{E,i}$ we obtain
\begin{align}
\sum_{m\neq n} P_{mn}&=\sum_{m\neq n} \varepsilon_m\varepsilon_n\exp(\mathrm{i}k\cdot x_{E,m}')\exp(\mathrm{i}p\cdot x_{E,n}') \notag\displaybreak[0] \\
&=\sum_{m\neq n} \left\{(+1)^2\exp(\mathrm{i}k\cdot x_{E,m}')\exp(\mathrm{i}p\cdot x_{E,n}')\right. \notag\displaybreak[0] \\
&\phantom{{}={}\sum\Big\{}\,+(-1)^2\exp[\mathrm{i}k\cdot (x_{E,m}'+\delta x_{E,m}')]\exp[\mathrm{i}p\cdot (x_{E,n}'+\delta x_{E,n}')] \notag\displaybreak[0] \\
&\phantom{{}={}\sum\Big\{}+(+1)\cdot (-1)\exp(\mathrm{i}k\cdot x_{E,m}')\exp[\mathrm{i}p\cdot (x_{E,n}'+\delta x_{E,n}')] \notag\displaybreak[0] \\
&\phantom{{}={}\sum\Big\{}+\left.(-1)\cdot (+1)\exp[\mathrm{i}k\cdot (x_{E,m}'+\delta x_{E,m}')]\exp(\mathrm{i}p\cdot x_{E,n}')\right\} \notag\displaybreak[0] \\
&=\mathcal{O}(\delta x_{E,m}'\cdot \delta x_{E,n}')\,.
\end{align}
As a result the linear term vanishes and all further contributions are suppressed by small distances. Thus for a large number of defects
in the sum over all $P_{mn}$, contributions from neighboring defects with opposite charges compensate each other. The second term of
Eq.~\eqref{eq:product-two-gn} averages out
\begin{equation}
\lim_{\mathcal{N}\mapsto \infty} \frac{1}{\mathcal{N}}\sum_{m\neq n} P_{mn}=0\,.
\end{equation}
We obtain the following result for the product of two functions $\widetilde{G}_{\mathcal{N}}$, each evaluated at a different momentum
\begin{equation}
\label{eq:momentum-conservation-defect-vertex-1}
\int\limits^{\Lambda'} \mathrm{d}\Lambda\,\lim_{\mathcal{N}\mapsto \infty} \widetilde{G}_{\mathcal{N}}(k)\widetilde{G}_{\mathcal{N}}(p)
=\Xi \,\lim_{\mathcal{N}\mapsto \infty} \frac{1}{\mathcal{N}}(2\pi)^4\varrho\delta^{(4)}(k+p)\,.
\end{equation}
The integral over all Lorentz transformations up to a boost limit $\Lambda'$ can be separated out again.

Furthermore, the functions $\widetilde{G}_{\mathscr{R}}(k)$ and $\widetilde{g}_{\scriptscriptstyle{\mathscr{R}}}(k)$ defined by
Eq.~\eqref{eq:definition-gr-function}
obey a similar relation
\begin{equation}
\label{eq:momentum-conservation-defect-vertex-2}
\int\limits^{\Lambda'} \mathrm{d}\Lambda\,\lim_{\mathscr{R}\mapsto \infty} \widetilde{G}_{\mathscr{R}}(k)\widetilde{G}_{\mathscr{R}}(p)\equiv \Xi \lim_{\mathcal{V}\mapsto \infty} \frac{1}{\mathcal{V}}(2\pi)^4\varrho\delta^{(4)}(k+p)\,.
\end{equation}
The physical interpretation of the result obtained is as follows. If the scalar field $\phi$ scatters at a defect it can either transfer
momentum to the defect or absorb momentum from the defect. Averaging over many defects --- implying the limits $\mathcal{N}\mapsto \infty$,
$\mathcal{V}\mapsto \infty$ or $\mathscr{R}\mapsto \infty$ --- leads to zero average momentum transfer at each defect. This means
momentum conservation for the $\phi$-field. Averaging over infinitely many randomly distributed defects results in a translation invariant
theory that clearly obeys the property of momentum conservation.

The results obtained depend only on the density of defects and (for some) $\mathcal{N}$ or $\mathcal{V}$. Hence these can be generalized to
the whole of Minkowski spacetime. As mentioned this corresponds to blowing up the spacetime region of Eq.~\eqref{eq:minkowski-space-subset-volume},
namely to the limit $\mathscr{R}\mapsto \infty$.

\section{Perturbative solution of the field equations}
\label{sec:solution-field-equation}
\setcounter{equation}{0}

The dispersion relations of both the scalar field and the photon follow from the appropriate field equations that are modified by the presence
of the spacetime defects. We would like to set up the modified field equations at a perturbative level. In the first part of the current
section we will follow the lines of Ref.~\cite{hep-th/0312032}. In the second part we will show that the results obtained can be reproduced
with the help of the perturbative Feynman rules. These are given in App.~\ref{sec:perturbative-feynman-rules}, where
Eqs.~\eqref{eq:Feynman-rule-1} -- \eqref{eq:Feynman-rule-4a} can be directly derived from the action ~\eqref{eq:action-effective-theory}. The
Feynman rule \eqref{eq:Feynman-rule-4b} follows from the deliberations of the previous section.

\subsection{Perturbative \textit{Ansatz} for the photon field}

In order to obtain the modification of the photon field originating in the interaction with spacetime defects via the scalar field $\phi$ 
we have to solve the modified photon field equation resulting from the action \eqref{eq:action-effective-theory}. In momentum space it can
be written as an integral equation
\begin{equation}
\label{eq:interacting-field-equation-exact}
k^2\widetilde{A}_{\mathscr{R}}^{\nu}(k)=-\frac{\lambda^{(0)}}{(2\pi)^4}\int\limits_{\substack{|q^{\mu}|\geq 1/\mathscr{R} \\ \mu\in \{0,1,2,3\}}}
\mathrm{d}^4q\,\widetilde{g}_{\scriptscriptstyle{\mathscr{R}}}(q)\varepsilon^{\mu\nu\varrho\sigma}q_{\mu}(k-q)_{\varrho}\widetilde{A}_{\mathscr{R},\sigma}(k-q)\,.
\end{equation}
The index `$\mathscr{R}$' denotes that the system is at first considered in a finite rectangular region of side length $\mathscr{R}$. The
exact solution to the latter equation is out of reach. Therefore we make the following perturbative \textit{Ansatz} for the full solution
$\widetilde{A}_{\mathscr{R}}^{\nu}$ in powers of the bare coupling constant $\lambda^{(0)}$. This is reasonable if we expect a modified
photon dispersion law since current experimental bounds on Lorentz symmetry violation --- and thus a modified dispersion relation for the
photon --- are very tight (see \cite{Kostelecky:2008ts}).
\begin{equation}
\label{eq:solution-power-expansion}
\widetilde{A}_{\mathscr{R}}^{\nu}=\widetilde{A}^{(0)\,\nu}+\lambda^{(0)}\widetilde{A}_{\mathscr{R}}^{(1)\,\nu}+(\lambda^{(0)})^2\widetilde{A}_{\mathscr{R}}^{(2)\,\nu}+\hdots\,.
\end{equation}
Herein $\widetilde{A}^{(0)\,\nu}$ is a solution of the free-field equation
\begin{equation}
k^2\widetilde{A}^{(0)\,\nu}(k)=0\,.
\end{equation}
By successively inserting the power expansion of Eq.~\eqref{eq:solution-power-expansion} in Eq.~\eqref{eq:interacting-field-equation-exact}, we
obtain a perturbative expansion of the exact solution. Now the first step is to insert $\widetilde{A}^{(0)}$. We identify each perturbative order
remembering that $k^2\widetilde{A}^{(0)\,\nu}=0$. Using the definition
\begin{equation}
\widetilde{\Delta}\equiv \frac{1}{k^2+\mathrm{i}\epsilon}\,,
\end{equation}
with an infinitesimal real parameter $\epsilon$ to avoid the pole at $k^2=0$, we find the following first order perturbative solution, where
we now take the limit $\mathscr{R}\mapsto \infty$
\begin{equation}
\label{eq:modified-photon-field-first-order}
\lambda^{(0)}\widetilde{A}^{(1)\,\nu}(k)=\lim_{\mathscr{R}\mapsto \infty} -\frac{\lambda^{(0)}}{(2\pi)^4}\widetilde{\Delta}(k)\int \mathrm{d}^4q\,\widetilde{g}_{\scriptscriptstyle{\mathscr{R}}}(q)\varepsilon^{\mu\nu\varrho\sigma}q_{\mu}(k-q)_{\varrho}\widetilde{A}^{(0)}_{\sigma}(q)\,.
\end{equation}
The first order photon field correction vanishes in the limit considered. This is clear from Eq.~\eqref{eq:momentum-conservation-defect-vertex-2}
since $\widetilde{g}_{\scriptscriptstyle{\mathscr{R}}\mapsto \infty}(q)$ does not come together with a second background field, thus only producing a
contribution for $q=0$.

The second order solution of the photon field equation reads as follows
\begin{align}
\label{eq:photon-field-equation-second-order}
(\lambda^{(0)})^2\widetilde{A}^{(2)\,\nu}(k)&=\lim_{\mathscr{R}\mapsto \infty} -\frac{\lambda^{(0)}}{(2\pi)^4}\widetilde{\Delta}(k)\int \mathrm{d}^4q\,\widetilde{g}_{\scriptscriptstyle{\mathscr{R}}}(q)\varepsilon^{\mu\nu\varrho\sigma}q_{\mu}(k-q)_{\varrho}\big(\lambda^{(0)}\widetilde{A}^{(1)}_{\mathscr{R},\sigma}(q)\big) \notag\displaybreak[0] \\
&=\lim_{\mathscr{R}\mapsto \infty} \frac{(\lambda^{(0)})^2}{(2\pi)^8}\widetilde{\Delta}(k)\int \mathrm{d}^4q\,\widetilde{g}_{\scriptscriptstyle{\mathscr{R}}}(q)\varepsilon^{\mu\nu\varrho\sigma}q_{\mu}(k-q)_{\varrho}\widetilde{\Delta}(k-q) \notag\displaybreak[0] \\
&\phantom{{}={}\lim_{\mathscr{R}\mapsto \infty}}\times\int \mathrm{d}^4p\,\widetilde{g}_{\scriptscriptstyle{\mathscr{R}}}(k-p-q)\varepsilon_{\alpha\beta\gamma\sigma}(k-p-q)^{\alpha}p^{\beta}\widetilde{A}^{(0)\,\gamma}(p)\,.
\end{align}
We contract the Levi-Civita tensors, perform the limit $\mathscr{R}\mapsto \infty$, and use Eq.~\eqref{eq:momentum-conservation-defect-vertex-2}
to simplify the latter result which finally leads to
\begin{subequations}
\begin{equation}
\label{eq:definition-curly-c}
(\lambda^{(0)})^2\widetilde{A}^{(2)\,\nu}(k)=
-\mathcal{C}^{(0)}\widetilde{\Delta}(k)B^{\nu}_{\phantom{\nu}\gamma}(k)\widetilde{A}^{(0)\,\gamma}(k)\,,\quad \mathcal{C}^{(0)}\equiv \Xi (b^{(0)})^4(\lambda^{(0)})^2\varrho\,,
\end{equation}
\begin{equation}
\label{eq:definition-bnugamma}
B^{\nu}_{\phantom{\nu}\gamma}(k)=\int \frac{\mathrm{d}^4q}{(2\pi)^4}\,\left[\frac{1}{(b^{(0)})^2}\widetilde{H}(q)\right]^2\frac{1}{(k-q)^2+\mathrm{i}\epsilon}K^{\nu}_{\phantom{\nu}\gamma}\,,
\end{equation}
with
\begin{equation}
\label{eq:definition-knugamma}
K^{\nu}_{\phantom{\nu}\gamma}=\delta^{\phantom{[}\mu}_{[\alpha}\delta^{\nu}_{\beta}\delta^{\varrho}_{\gamma]}q_{\mu}(k-q)_{\varrho}q^{\alpha}k^{\beta}\,.
\end{equation}
\end{subequations}
In Eq.~\eqref{eq:definition-knugamma}, $[\alpha,\beta,\gamma]$ denotes a totally antisymmetric permutation of the indices $\alpha$, $\beta$,
and $\gamma$. The tensor $K^{\nu}_{\phantom{\nu}\gamma}$ also appears in the modified photon field that is obtained in the context of the
effective background field model in \cite{hep-th/0312032}. The quantities $\Xi$, $(b^{(0)})^4\varrho$, and $\lambda^{(0)}$ have no mass
dimension and hence $\mathcal{C}^{(0)}$ is also a dimensionless parameter.

\subsection{Establishing the perturbation from the Feynman rules of the modified theory}
\label{subsec:perturbation-from-feynman-rules}

Analogously, the perturbative series of the full solution of the modified photon field equation can be obtained in terms of Feynman diagrams.
The corresponding Feynman rules follow from the action of Eq.~\eqref{eq:action-effective-theory} and are given in App.~\ref{sec:perturbative-feynman-rules}.

We couple the photon field to a conserved fermion current $j_{\nu}$ (with $k^{\nu}j_{\nu}(k)=0$) that is represented by a plain line with an arrow,
where the scalar field is denoted by a plain line without any arrow. The photon field is drawn as a single wiggly line and a double wiggly line
stands for the full field. Ordinary vertices are represented by dots and the scattering at a defect (``defect vertex'') is shown as a cross:
\begin{equation}
\label{eq:photon-field-equation-second-order-diagrammatical}
\begin{array}{c}
\begin{fmfgraph*}(37.5,75)
\fmfpen{thin}
\fmfleft{i1,i2}
\fmfright{o1}
\fmf{dbl_wiggly,label=$\overrightarrow{\phantom{i}k\phantom{i}}$}{v1,o1}
\fmf{fermion,label=$j_{\nu}$,label.side=left}{i2,v1}
\fmf{fermion}{v1,i1}
\fmfforce{0.0w,1.0h}{i1}
\fmfforce{0.0w,0.0h}{i2}
\fmfforce{0.0w,0.465h}{v1}
\fmfforce{1.0w,0.465h}{o1}
\fmfv{label=$\nu$,label.angle=45}{v1}
\fmfv{label=$\widetilde{A}^{\nu}(k)$,label.angle=90}{o1}
\fmfdot{v1}
\end{fmfgraph*}
\end{array}\,\,=\,\,\begin{array}{c}
\begin{fmfgraph*}(37.5,75)
\fmfpen{thin}
\fmfleft{i1,i2}
\fmfright{o1}
\fmf{photon,label=$\overrightarrow{\phantom{i}k\phantom{i}}$}{v1,o1}
\fmf{fermion,label=$j_{\nu}$,label.side=left}{i2,v1}
\fmf{fermion}{v1,i1}
\fmfforce{0.0w,1.0h}{i1}
\fmfforce{0.0w,0.0h}{i2}
\fmfforce{0.0w,0.465h}{v1}
\fmfforce{1.0w,0.465h}{o1}
\fmfv{label=$\nu$,label.angle=45}{v1}
\fmfdot{v1}
\end{fmfgraph*}
\end{array}\,\,+\,\,
\begin{array}{c}
\begin{fmfgraph*}(75,75)
\fmfpen{thin}
\fmfleft{i1,i2}
\fmfright{o1}
\fmf{photon,label=$\overrightarrow{\phantom{i}k\phantom{i}}$}{v1,v2}
\fmf{photon,label=$\overrightarrow{\phantom{i}k\phantom{i}}$}{v2,o1}
\fmf{fermion,label=$j_{\nu}$,label.side=left}{i2,v1}
\fmf{fermion}{v1,i1}
\fmf{plain,label=$q\uparrow$}{v2,v3}
\fmfforce{0.0w,1.0h}{i1}
\fmfforce{0.0w,0.0h}{i2}
\fmfforce{0.0w,0.465h}{v1}
\fmfforce{0.5w,0.465h}{v2}
\fmfforce{0.5w,1.0h}{v3}
\fmfforce{1.0w,0.465h}{o1}
\fmfv{label=$\nu$,label.angle=45}{v1}
\fmfv{decoration.shape=cross,decoration.filled=full,decoration.size=15,label.angle=90,label.dist=0.25cm}{v3}
\fmfdot{v1,v2}
\end{fmfgraph*}
\end{array}\,\,+\,\,
\begin{array}{c}
\begin{fmfgraph*}(125,75)
\fmfpen{thin}
\fmfleft{i1,i2}
\fmfright{o1}
\fmfv{label=$\widetilde{A}^{\gamma\,,(0)}(k)$,label.angle=90}{o1}
\fmf{photon,label=$\overrightarrow{\phantom{i}k\phantom{i}}$}{v1,v2}
\fmf{photon,label=$\overrightarrow{\phantom{i}k-q\phantom{i}}$}{v2,v3}
\fmf{photon,label=$\overrightarrow{\phantom{i}k\phantom{i}}$}{v3,o1}
\fmf{plain,label=$q\nearrow$,left=0.45}{v2,v4}
\fmfv{decoration.shape=cross,decoration.filled=full,decoration.size=15,label.angle=90,label.dist=0.25cm}{v4}
\fmf{plain,label=$\searrow{}q$,left=0.45}{v4,v3}
\fmfforce{0.0w,0.465h}{v1}
\fmfforce{1.0w,0.465h}{o1}
\fmfforce{0.333w,0.465h}{v2}
\fmfforce{0.667w,0.465h}{v3}
\fmfdot{v1,v2,v3}
\fmfv{label=$\gamma$,label.angle=45}{v3}
\fmfforce{0.0w,1.0h}{i1}
\fmfforce{0.0w,0.0h}{i2}
\fmfforce{0.5w,0.75h}{v4}
\fmf{fermion,label=$j_{\nu}$,label.side=left}{i2,v1}
\fmf{fermion}{v1,i1}
\fmfv{label=$\nu$,label.angle=45}{v1}
\end{fmfgraph*}
\end{array}+\hdots
\end{equation}
Contributions proportional to odd powers of $\lambda^{(0)}$ (containing a defect vertex connected to only a single $\phi$ field) vanish
because of four-momentum conservation. For example, this is the case for the second diagram on the right-hand side of the diagrammatical
equation above, cf. the discussion below Eq.~\eqref{eq:modified-photon-field-first-order} in the previous section. The second order
perturbative solution $\widetilde{A}^{(2)\,\nu}(k)$ --- corresponding to the third diagram on the right-hand side of
Eq.~\eqref{eq:photon-field-equation-second-order-diagrammatical} --- leads to a nonvanishing correction of the photon field.

A resummation of all one-particle reducible diagrams at one-loop order leads to a resummed photon field
\begin{equation}
\widehat{A}^{\mu}(k)\equiv\begin{array}{c}
\begin{fmfgraph*}(50,25)
\fmfpen{thin}
\fmfleft{i1}
\fmfright{o1}
\fmf{photon}{i1,o1}
\fmfforce{0.0w,0.405h}{i1}
\fmfforce{1.0w,0.405h}{o1}
\fmfdot{i1}
\end{fmfgraph*}
\end{array}+\begin{array}{c}
\begin{fmfgraph*}(50,25)
\fmfpen{thin}
\fmfleft{i1}
\fmfright{o1}
\fmf{photon}{i1,o1}
\fmf{plain,left=1}{v1,v2}
\fmfforce{0.0w,0.405h}{i1}
\fmfforce{0.25w,0.405h}{v1}
\fmfforce{0.75w,0.405h}{v2}
\fmfforce{0.5w,0.905h}{v3}
\fmfforce{1.0w,0.405h}{o1}
\fmfv{decoration.shape=cross,decoration.filled=full,decoration.size=10,label.angle=90,label.dist=0.25cm}{v3}
\fmfdot{i1,v1,v2}
\end{fmfgraph*}
\end{array}+\begin{array}{c}
\begin{fmfgraph*}(100,25)
\fmfpen{thin}
\fmfleft{i1}
\fmfright{o1}
\fmf{photon}{i1,o1}
\fmf{plain,left=1}{v1,v2}
\fmf{plain,left=1}{v3,v4}
\fmfforce{0.0w,0.405h}{i1}
\fmfforce{0.125w,0.405h}{v1}
\fmfforce{0.375w,0.405h}{v2}
\fmfforce{0.625w,0.405h}{v3}
\fmfforce{0.875w,0.405h}{v4}
\fmfforce{0.25w,0.905h}{v5}
\fmfforce{0.75w,0.905h}{v6}
\fmfforce{1.0w,0.405h}{o1}
\fmfv{decoration.shape=cross,decoration.filled=full,decoration.size=10,label.angle=90,label.dist=0.25cm}{v5}
\fmfv{decoration.shape=cross,decoration.filled=full,decoration.size=10,label.angle=90,label.dist=0.25cm}{v6}
\fmfdot{i1,v1,v2,v3,v4}
\end{fmfgraph*}
\end{array}+\hdots\,.
\end{equation}
Multiplying the inverse standard photon propagator $\widetilde{\Delta}^{-1}(k)=k^2$ with $\widehat{A}^{\mu}(k)$ nullifies the zeroth order
contribution corresponding to the free photon field $\widetilde{A}^{\mu\,(0)}(k)$. Furthermore, it cancels a propagator in each further
term and therefore an overall contribution can be factored out:
\begin{align}
\label{eq:field-equation-diagrammatical}
&\widetilde{\Delta}^{-1}(k)\left(\begin{array}{c}
\begin{fmfgraph*}(50,25)
\fmfpen{thin}
\fmfleft{i1}
\fmfright{o1}
\fmf{photon}{i1,o1}
\fmfdot{i1}
\fmfforce{0.0w,0.405h}{i1}
\fmfforce{1.0w,0.405h}{o1}
\end{fmfgraph*}
\end{array}+\begin{array}{c}
\begin{fmfgraph*}(50,25)
\fmfpen{thin}
\fmfleft{i1}
\fmfright{o1}
\fmf{photon}{i1,o1}
\fmf{plain,left=1}{v1,v2}
\fmfforce{0.0w,0.405h}{i1}
\fmfforce{0.25w,0.405h}{v1}
\fmfforce{0.75w,0.405h}{v2}
\fmfforce{0.5w,0.905h}{v3}
\fmfforce{1.0w,0.405h}{o1}
\fmfv{decoration.shape=cross,decoration.filled=full,decoration.size=10,label.angle=90,label.dist=0.25cm}{v3}
\fmfdot{i1,v1,v2}
\end{fmfgraph*}
\end{array}+\begin{array}{c}
\begin{fmfgraph*}(100,25)
\fmfpen{thin}
\fmfleft{i1}
\fmfright{o1}
\fmf{photon}{i1,o1}
\fmf{plain,left=1}{v1,v2}
\fmf{plain,left=1}{v3,v4}
\fmfforce{0.0w,0.405h}{i1}
\fmfforce{0.125w,0.405h}{v1}
\fmfforce{0.375w,0.405h}{v2}
\fmfforce{0.625w,0.405h}{v3}
\fmfforce{0.875w,0.405h}{v4}
\fmfforce{0.25w,0.905h}{v5}
\fmfforce{0.75w,0.905h}{v6}
\fmfforce{1.0w,0.405h}{o1}
\fmfv{decoration.shape=cross,decoration.filled=full,decoration.size=10,label.angle=90,label.dist=0.25cm}{v5}
\fmfv{decoration.shape=cross,decoration.filled=full,decoration.size=10,label.angle=90,label.dist=0.25cm}{v6}
\fmfdot{i1,v1,v2,v3,v4}
\end{fmfgraph*}
\end{array}+\hdots\right) \notag \\
&\qquad=\begin{array}{c}
\begin{fmfgraph*}(30,25)
\fmfpen{thin}
\fmfleft{i1}
\fmfright{o1}
\fmf{photon}{i1,o1}
\fmf{plain,left=1}{i1,o1}
\fmfforce{0.0w,0.405h}{i1}
\fmfforce{0.075w,0.405h}{i1}
\fmfforce{0.925w,0.405h}{o1}
\fmfforce{0.5w,0.955h}{v2}
\fmfforce{1.0w,0.405h}{o1}
\fmfv{decoration.shape=cross,decoration.filled=full,decoration.size=10,label.angle=90,label.dist=0.25cm}{v2}
\fmfdot{i1,o1}
\end{fmfgraph*}
\end{array}\left(\begin{array}{c}
\begin{fmfgraph*}(50,25)
\fmfpen{thin}
\fmfleft{i1}
\fmfright{o1}
\fmf{photon}{i1,o1}
\fmfdot{i1}
\fmfforce{0.0w,0.405h}{i1}
\fmfforce{1.0w,0.405h}{o1}
\end{fmfgraph*}
\end{array}+\begin{array}{c}
\begin{fmfgraph*}(50,25)
\fmfpen{thin}
\fmfleft{i1}
\fmfright{o1}
\fmf{photon}{i1,o1}
\fmf{plain,left=1}{v1,v2}
\fmfforce{0.0w,0.405h}{i1}
\fmfforce{0.25w,0.405h}{v1}
\fmfforce{0.75w,0.405h}{v2}
\fmfforce{0.5w,0.905h}{v3}
\fmfforce{1.0w,0.405h}{o1}
\fmfv{decoration.shape=cross,decoration.filled=full,decoration.size=10,label.angle=90,label.dist=0.25cm}{v3}
\fmfdot{i1,v1,v2}
\end{fmfgraph*}
\end{array}+\begin{array}{c}
\begin{fmfgraph*}(100,25)
\fmfpen{thin}
\fmfleft{i1}
\fmfright{o1}
\fmf{photon}{i1,o1}
\fmf{plain,left=1}{v1,v2}
\fmf{plain,left=1}{v3,v4}
\fmfforce{0.0w,0.405h}{i1}
\fmfforce{0.125w,0.405h}{v1}
\fmfforce{0.375w,0.405h}{v2}
\fmfforce{0.625w,0.405h}{v3}
\fmfforce{0.875w,0.405h}{v4}
\fmfforce{0.25w,0.905h}{v5}
\fmfforce{0.75w,0.905h}{v6}
\fmfforce{1.0w,0.405h}{o1}
\fmfv{decoration.shape=cross,decoration.filled=full,decoration.size=10,label.angle=90,label.dist=0.25cm}{v5}
\fmfv{decoration.shape=cross,decoration.filled=full,decoration.size=10,label.angle=90,label.dist=0.25cm}{v6}
\fmfdot{i1,v1,v2,v3,v4}
\end{fmfgraph*}
\end{array}+\hdots\right)\,.
\end{align}
This prefactor is the one-loop diagram in front of the round brackets on the right-hand side of Eq.~\eqref{eq:field-equation-diagrammatical}.
It is given by the contribution $-\mathcal{C}^{(0)}B^{\nu}_{\phantom{\nu}\gamma}(k)$ with $B^{\nu}_{\phantom{\nu}\gamma}(k)$ of
Eq.~\eqref{eq:definition-bnugamma}. Hence, the modified field equation that results from the resummation of all one-loop photon field
corrections is
\begin{equation}
k^2\widehat{A}^{\nu}(k)=-\mathcal{C}^{(0)}B^{\nu}_{\phantom{\nu}\gamma}(k)\widehat{A}^{\gamma}(k)\,.
\end{equation}
A modified photon dispersion law following from this will be derived later. Note that one-particle irreducible higher-order corrections,
for example
\begin{center}
\begin{fmfgraph*}(100,45)
\fmfpen{thin}
\fmfleft{i1}
\fmfright{o1}
\fmf{photon}{i1,o1}
\fmf{plain,left=1}{v1,v3}
\fmf{plain,right=1}{v2,v4}
\fmfforce{0.0w,0.5h}{i1}
\fmfforce{0.2w,0.5h}{v1}
\fmfforce{0.4w,0.5h}{v2}
\fmfforce{0.6w,0.5h}{v3}
\fmfforce{0.8w,0.5h}{v4}
\fmfforce{1.0w,0.5h}{o1}
\fmfforce{0.4w,0.945h}{v5}
\fmfforce{0.6w,0.055h}{v6}
\fmfv{decoration.shape=cross,decoration.filled=full,decoration.size=10,label.angle=90,label.dist=0.25cm}{v5}
\fmfv{decoration.shape=cross,decoration.filled=full,decoration.size=10,label.angle=90,label.dist=0.25cm}{v6}
\fmfdot{i1,v1,v2,v3,v4}
\end{fmfgraph*}
\end{center}
are not covered by the resummed photon field $\widehat{A}^{\mu}$. They are assumed to give a contribution to the modified dispersion relation
of the photons as well. However since these are suppressed by at least one further factor $(\lambda^{(0)})^2$ we will neglect them in our
calculations.

\section{Leading-order perturbation of the photon field}
\label{sec:caluclation-photon-field-perturbation}
\setcounter{equation}{0}

\subsection{Dimensional regularization}

We now want to compute the one-loop contribution $B^{\nu}_{\phantom{\nu}\gamma}(k)$ to the photon field that was set up in Eq.~\eqref{eq:definition-bnugamma}.
We begin by contracting the indices in Eq.~\eqref{eq:definition-knugamma}
\begin{align}
K^{\nu}_{\phantom{\nu}\gamma}\widetilde{A}^{(0)\,\gamma}(k)&=\delta^{\phantom{[}\mu}_{[\alpha}\delta^{\nu}_{\beta}\delta^{\kappa}_{\gamma]}q_{\mu}(k-q)_{\kappa}q^{\alpha}k^{\beta}\widetilde{A}^{(0)\,\gamma} \notag \\
&=k^{\nu}\big(k\cdot \widetilde{A}^{(0)}\big)q^2-k^{\nu}k^{\varrho}\widetilde{A}^{(0)\,\sigma}q_{\varrho}q_{\sigma}-\widetilde{A}^{(0)\,\nu}k^2q^2 \notag\displaybreak[0] \\
&\quad\,+\widetilde{A}^{(0)\,\nu}k^{\varrho}k^{\sigma}q_{\varrho}q_{\sigma}-q^{\nu}q_{\varrho}k^{\varrho}\big(k\cdot \widetilde{A}^{(0)}\big)+q^{\nu}q_{\varrho}\widetilde{A}^{(0)\,\varrho}k^2.
\end{align}
The second step is to perform the four-dimensional momentum integral over $q$:
\begin{align}
(2\pi)^4\,B^{\nu}_{\phantom{\nu}\gamma}\widetilde{A}^{(0)\,\gamma}(k)&= \int \mathrm{d}^4q\,\frac{1}{(q^2-1/b^2+\mathrm{i}\epsilon)^2[(k-q)^2+\mathrm{i}\epsilon]}K^{\nu}_{\phantom{\nu}\gamma}\widetilde{A}^{(0)\,\gamma}(k) \notag \\
&=\widehat{I}_{\varrho\sigma}\Big\{k^{\varrho}k^{\sigma}\widetilde{A}^{(0)\,\nu}-k^{\nu}k^{\varrho}\widetilde{A}^{(0)\,\sigma}+\eta^{\nu\sigma}\Big[k^2\widetilde{A}^{(0)\,\varrho}-k^{\varrho}\big(k\cdot\widetilde{A}^{(0)}\big)\Big]\Big\} \notag\displaybreak[0] \\
&\quad\,+\widehat{I}_0\Big[k^{\nu}\big(k\cdot \widetilde{A}^{(0)}\big)-k^2\widetilde{A}^{(0)\,\nu}\Big]\,,
\end{align}
where $\widehat{I}_{\varrho\sigma}$ is a tensor one-loop and $\widehat{I}_0$ a scalar one-loop integral
\begin{subequations}
\begin{equation}
\label{eq:tensor-one-loop-integral}
\widehat{I}_{\varrho\sigma}\equiv \int \mathrm{d}^4q\,\frac{q_{\varrho}q_{\sigma}}{(q^2-1/b^2+\mathrm{i}\epsilon)^2[(k-q)^2+\mathrm{i}\epsilon]}\,,
\end{equation}
\begin{equation}\widehat{I}_0\equiv \widehat{I}_{\varrho}^{\phantom{\varrho}\varrho}=\int \mathrm{d}^4q\,\frac{q^2}{(q^2-1/b^2+\mathrm{i}\epsilon)^2[(k-q)^2+\mathrm{i}\epsilon]}\,.
\end{equation}
\end{subequations}
By power-counting we see that the integrals $\widehat{I}_{\varrho\sigma}$ and $\widehat{I}_0$ are ultraviolet-divergent. Therefore they have
to be regularized and we decide to use dimensional regularization. The basic principle is to analytically continue the integrals to
$d$ spacetime dimensions, where $d\neq 4$ is a real number. If we use the convention $d=4-2\widehat{\varepsilon}$, with four spacetime
dimensions to be recovered in the limit $\widehat{\varepsilon}\mapsto 0$, the divergences become manifest as poles in $\widehat{\varepsilon}$.
Via
\begin{equation}
\int \mathrm{d}^4q=(2\pi)^4\int \frac{\mathrm{d}^4q}{(2\pi)^4} \mapsto (2\pi)^4\mu^{4-d}\int \frac{\mathrm{d}^dq}{(2\pi)^d}=(2\pi\mu)^{4-d} \int \mathrm{d}^dq\,,
\end{equation}
the renormalization scale $\mu$, of mass dimension 1, is introduced to conserve the dimension of the integral.

\subsection{Passarino--Veltman decomposition}
\label{subsec:passarino-veltman-decomposition}

Equation \eqref{eq:tensor-one-loop-integral} gives a tensor integral that can be reduced to scalar integrals $\widehat{I}_1$ and $\widehat{I}_2$
with the following \textit{Ansatz}
\begin{equation}
\widehat{I}_{\varrho\sigma}=\eta_{\varrho\sigma}\widehat{I}_1+k_{\varrho}k_{\sigma}\widehat{I}_2\,.
\end{equation}
The integrals $\widehat{I}_1$, $\widehat{I}_2$ follow from the contractions $K_1\equiv k^{\varrho}k^{\sigma}\widehat{I}_{\varrho\sigma}$, $K_2\equiv \eta^{\varrho\sigma}\widehat{I}_{\varrho\sigma}$
\begin{equation}
\label{eq:scalar-integrals-expressed-by-contractions}
\widehat{I}_1=\frac{K_1-k^2K_2}{(1-d)k^2}\,,\quad \widehat{I}_2=\frac{-dK_1+k^2K_2}{(1-d)k^4}\,.
\end{equation}
What remains is the reduction of the contractions $K_1$ and $K_2$ to scalar master integrals via a Passarino--Veltman decomposition. This leads to
the following result (see App.~\ref{sec:derivation-passarino-veltman} for a detailed calculation)
\begin{subequations}
\begin{align}
\label{eq:result-contraction-1}
K_1&=\frac{1}{4}\left\{-A_0\left(\frac{1}{(b^{(0)})^2}\right)\right. \notag \\
&\phantom{{}={}\frac{1}{4}\left\{\right.}\hspace{-0.05cm}+\left(k^2+\frac{1}{(b^{(0)})^2}\right)\left[2B_0\left(-k,\frac{1}{(b^{(0)})^2},0\right)-B_0\left(0,\frac{1}{(b^{(0)})^2},\frac{1}{(b^{(0)})^2}\right)\right] \notag \displaybreak[0] \\
&\phantom{{}={}\frac{1}{4}\left\{\right.}\hspace{-0.05cm}+\left.\left[k^4+\frac{2k^2}{(b^{(0)})^2}+\frac{1}{(b^{(0)})^4}\right]C_0\left(-k,0,\frac{1}{(b^{(0)})^2},0,\frac{1}{(b^{(0)})^2}\right)\right\}\,,
\end{align}
\begin{align}
\label{eq:result-contraction-2}
K_2=\widehat{I}_0=B_0\left(-k,\frac{1}{(b^{(0)})^2},0\right)+\frac{1}{(b^{(0)})^2}C_0\left(-k,0,\frac{1}{(b^{(0)})^2},0,\frac{1}{(b^{(0)})^2}\right)\,.
\end{align}
\end{subequations}
Here the contractions $K_1$, $K_2$ are expressed solely in terms of master integrals given by
\begin{subequations}
\label{eq:master-integrals}
\begin{align}
A_0\left(\frac{1}{(b^{(0)})^2}\right) &= (2\pi\mu)^{4-d}\int \mathrm{d}^dq\,\frac{1}{q^2-1/(b^{(0)})^2+\mathrm{i}\epsilon}\,, \\
B_0\left(0,\frac{1}{(b^{(0)})^2},\frac{1}{(b^{(0)})^2}\right) &= (2\pi\mu)^{4-d}\int \mathrm{d}^dq\,\frac{1}{(q^2-1/(b^{(0)})^2+\mathrm{i}\epsilon)^2}\,, \\
B_0\left(-k,\frac{1}{(b^{(0)})^2},0\right) &= (2\pi\mu)^{4-d}\int \mathrm{d}^dq\,\frac{1}{(q^2-1/(b^{(0)})^2+\mathrm{i}\epsilon)[(k-q)^2+\mathrm{i}\epsilon]}\,, \\
C_0\left(-k,0,\frac{1}{(b^{(0)})^2},0,\frac{1}{(b^{(0)})^2}\right) &= \int \mathrm{d}^4q\,\frac{1}{(q^2-1/(b^{(0)})^2+\mathrm{i}\epsilon)^2[(k-q)^2+\mathrm{i}\epsilon]}\,.
\end{align}
\end{subequations}
Note that the $C_0$-integral has neither infrared nor ultraviolet
divergences so there is no need to regularize it. The integrals can be
computed by standard methods such as Feynman parametrization (see for example \cite{Steinhauser:2003}). The results are as follows
\begin{subequations}
\begin{eqnarray}
\frac{1}{\mathrm{i}\pi^2}A_0\left(\frac{1}{(b^{(0)})^2}\right) &=& \frac{1}{(b^{(0)})^2}\left[\frac{1}{\varepsilon}-\ln\left(\frac{1}{(b^{(0)})^2\mu^2}\right)+1\right]+\mathcal{O}(\varepsilon)\,, \\
\frac{1}{\mathrm{i}\pi^2}B_0\left(0,\frac{1}{(b^{(0)})^2},\frac{1}{(b^{(0)})^2}\right) &=& \frac{1}{\varepsilon}-\ln\left(\frac{1}{(b^{(0)})^2\mu^2}\right)+\mathcal{O}(\varepsilon)\,, \\
\frac{1}{\mathrm{i}\pi^2}B_0\left(-k,\frac{1}{(b^{(0)})^2},0\right) &=& \frac{1}{\varepsilon}-\int_0^1\mathrm{d}x\,\ln\left[\frac{k^2x^2-\left(k^2+1/(b^{(0)})^2\right)x+1/(b^{(0)})^2-\mathrm{i}\epsilon}{\mu^2}\right]+\mathcal{O}(\varepsilon) \notag \\
 &=& \frac{1}{\varepsilon}-\ln\left(\frac{1}{(b^{(0)})^2\mu^2}\right)+2 \notag \\
 &\phantom{{}={}}&\phantom{\frac{1}{\varepsilon}}-\frac{k^2-1/(b^{(0)})^2}{k^2}\ln\left[1-(b^{(0)})^2k^2-\mathrm{i}\epsilon\right]+\mathcal{O}(\varepsilon)\,, \\
\frac{1}{\mathrm{i}\pi^2}C_0\left(-k,0,\frac{1}{(b^{(0)})^2},0,\frac{1}{(b^{(0)})^2}\right) &=& -\int_0^1 \mathrm{d}x\,\frac{1-x}{k^2x^2-\left(k^2+1/(b^{(0)})^2\right)x+1/(b^{(0)})^2-\mathrm{i}\epsilon} \notag \\
 &=&\frac{1}{k^2}\ln\Big[1-(b^{(0)})^2k^2-\mathrm{i}\epsilon\Big]\,.
\end{eqnarray}
\end{subequations}
We have used
\begin{equation}
\frac{1}{\varepsilon}\equiv \frac{1}{\widehat{\varepsilon}}-\gamma_{\scriptscriptstyle{E}}+\ln(4\pi)\,,
\end{equation}
with the Euler-Mascheroni constant $\gamma_{\scriptscriptstyle{E}}\approx 0.577216$, which is a reasonable redefinition of the regularization
parameter. Terms of $\mathcal{O}(\varepsilon)$ have been discarded since they are not needed. An elaborate computation of the
scalar integrals is presented in App.~\ref{sec:computation-scalar-integrals}.

To summarize, we obtain the following photon field correction at second order in perturbation theory
\begin{align}
\label{eq:final-result-photon-field-correction}
(2\pi)^4B^{\nu\gamma}(k)\widetilde{A}^{(0)}_{\gamma}(k)&=\big(\eta_{\varrho\sigma}\widehat{I}_1+k_{\varrho}k_{\sigma}\widehat{I}_2\big) \notag \\
&\phantom{{}={}}\times\Big\{k^{\varrho}k^{\sigma}\widetilde{A}^{(0)\,\nu}-k^{\nu}k^{\varrho}\widetilde{A}^{(0)\,\sigma}+\eta^{\nu\sigma}\Big[k^2\widetilde{A}^{(0)\,\varrho}-k^{\varrho}\big(k\cdot\widetilde{A}^{(0)}\big)\Big]\Big\} \notag\displaybreak[0] \\
&\phantom{{}={}}+\widehat{I}_0\Big[k^{\nu}\big(k\cdot \widetilde{A}^{(0)}\big)-k^2\widetilde{A}^{(0)\,\nu}\Big] \notag\displaybreak[0] \\
&=\big(k^{\nu}k^{\gamma}-\eta^{\nu\gamma}k^2\big)\Big[\widehat{I}_0-2\widehat{I}_1-k^2\widehat{I}_2\Big]\widetilde{A}^{(0)}_{\gamma}(k)\,.
\end{align}
The scalar integrals $\widehat{I}_1$, $\widehat{I}_2$ result from the contractions $K_1$, $K_2$ via Eq.~\eqref{eq:scalar-integrals-expressed-by-contractions},
where these are given by Eqs. \eqref{eq:result-contraction-1}, \eqref{eq:result-contraction-2}. The bare correction to the photon field is transverse
and contains $1/\varepsilon$ poles. In order to obtain a physically meaningful result Eq.~\eqref{eq:final-result-photon-field-correction} has to
be renormalized.

\subsection{Renormalization procedure}

The second order solution of the photon field equation can now be written as follows
\begin{subequations}
\begin{eqnarray}
(\lambda^{(0)})^2\widetilde{A}^{(2)\,\nu}(k) &=& -\mathcal{C}^{(0)}\widetilde{\Delta}(k)B^{\nu\gamma}(k)\widetilde{A}_{\gamma}^{(0)}(k)=-\mathcal{C}^{(0)}\widetilde{\Delta}(k)\mathrm{i}\Pi^{\nu\gamma}(k)\widetilde{A}_{\gamma}^{(0)}(k)\,, \\
\mathrm{i}\Pi^{\nu\gamma}(k) &=& \mathrm{i}(k^{\nu}k^{\gamma}-\eta^{\nu\gamma}k^2)\Pi(k^2)\,, \\
\Pi(k^2) &=& -\mathrm{i}(\widehat{I}_0-2\widehat{I}_1-k^2\widehat{I}_2)\,,
\end{eqnarray}
with the explicit result
\begin{align}
16\pi^2\,\Pi(k^2)&=\frac{1}{2\varepsilon}-\frac{1}{2}\ln\left(\frac{1}{(b^{(0)}\mu)^2}\right)+\frac{1}{2}\left(1-\frac{1}{(b^{(0)}k)^2}\right) \notag \\
&\phantom{{}={}\frac{1}{2\varepsilon}}-\frac{1}{2}\left(1-\frac{2}{(b^{(0)}k)^2}+\frac{1}{(b^{(0)}k)^4}\right)\ln\Big[1-(b^{(0)}k)^2-\mathrm{i}\epsilon\Big]\,.
\end{align}
\end{subequations}
Since the one-loop diagram computed resembles the vacuum polarization contribution of standard quantum electrodynamics (QED), we perform a renormalization
of the coupling constant $\lambda^{(0)}$. We construct the modified photon propagator from the modified photon field by successively
inserting one-loop corrections. This gives an infinite resummation of one-particle reducible diagrams at one-loop order leading to the full
propagator. In this procedure we neglect higher order perturbative corrections $\mathcal{O}((\lambda^{(0)})^4)$ that are one-particle irreducible.
Finally the photon propagator is coupled to a conserved current. This procedure will not be performed explicitly, but it is needed in order to drop all terms
proportional to the four-momentum $k^{\mu}$. In a diagrammatical notation our approach appears as follows
\begin{equation}
\begin{array}{c}
\begin{fmfgraph*}(50,35)
\fmfpen{thin}
\fmfleft{i1}
\fmfright{o1}
\fmf{dbl_wiggly}{i1,v1,v2,o1}
\fmf{fermion}{v3,i1,v4}
\fmfforce{0.0w,0.425h}{i1}
\fmfforce{1.0w,0.425h}{o1}
\fmfforce{0.25w,0.425h}{v1}
\fmfforce{0.75w,0.425h}{v2}
\fmfforce{0.0w,0.0h}{v3}
\fmfforce{0.0w,1.0h}{v4}
\fmfdot{i1,o1}
\end{fmfgraph*}
\end{array}=
\begin{array}{c}
\begin{fmfgraph*}(50,35)
\fmfpen{thin}
\fmfleft{i1}
\fmfright{o1}
\fmf{photon}{i1,v1,v2,o1}
\fmf{fermion}{v3,i1,v4}
\fmfforce{0.0w,0.425h}{i1}
\fmfforce{1.0w,0.425h}{o1}
\fmfforce{0.25w,0.425h}{v1}
\fmfforce{0.75w,0.425h}{v2}
\fmfforce{0.0w,0.0h}{v3}
\fmfforce{0.0w,1.0h}{v4}
\fmfdot{i1,o1}
\end{fmfgraph*}
\end{array}+
\begin{array}{c}
\begin{fmfgraph*}(50,35)
\fmfpen{thin}
\fmfleft{i1}
\fmfright{o1}
\fmf{photon}{i1,v1,v2,o1}
\fmf{plain,left=1}{v1,v2}
\fmf{fermion}{v3,i1,v4}
\fmfforce{0.25w,0.425h}{v1}
\fmfforce{0.75w,0.425h}{v2}
\fmfforce{0.0w,0.425h}{i1}
\fmfforce{1.0w,0.425h}{o1}
\fmfforce{0.0w,0.0h}{v3}
\fmfforce{0.0w,1.0h}{v4}
\fmfv{decoration.shape=cross,decoration.filled=full,decoration.size=10,label.angle=90,label.dist=0.25cm}{v5}
\fmfforce{0.5w,0.775h}{v5}
\fmfdot{i1,v1,v2,o1}
\end{fmfgraph*}
\end{array}+
\begin{array}{c}
\begin{fmfgraph*}(100,35)
\fmfpen{thin}
\fmfleft{i1}
\fmfright{o1}
\fmf{photon}{i1,v1,v2,v3,v4,o1}
\fmf{plain,left=1}{v1,v2}
\fmf{plain,left=1}{v3,v4}
\fmf{fermion}{v5,i1,v6}
\fmfforce{0.125w,0.425h}{v1}
\fmfforce{0.375w,0.425h}{v2}
\fmfforce{0.625w,0.425h}{v3}
\fmfforce{0.875w,0.425h}{v4}
\fmfforce{0.0w,0.0h}{v5}
\fmfforce{0.0w,1.0h}{v6}
\fmfforce{0.0w,0.425h}{i1}
\fmfforce{1.0w,0.425h}{o1}
\fmfv{decoration.shape=cross,decoration.filled=full,decoration.size=10,label.angle=90,label.dist=0.25cm}{v7}
\fmfforce{0.25w,0.775h}{v7}
\fmfv{decoration.shape=cross,decoration.filled=full,decoration.size=10,label.angle=90,label.dist=0.25cm}{v8}
\fmfforce{0.75w,0.775h}{v8}
\fmfdot{i1,v1,v2,v3,v4,o1}
\end{fmfgraph*}
\end{array}+\hdots
\end{equation}
In terms of equations this corresponds to the vacuum expectation value of the time-ordered product of field operators
$\langle T\widehat{A}^{\mu}(k)\widehat{A}^{\nu}(k)\rangle$, which is the resummed Feynman propagator\footnote{The coupling to the conserved
current is suppressed.}
\begin{align}
\big\langle T\widehat{A}^{\mu}(k)\widehat{A}^{\nu}(k)\big\rangle&=\frac{-\mathrm{i}\eta^{\mu\nu}}{k^2+\mathrm{i}\epsilon}+\frac{-\mathrm{i}\eta^{\mu\varrho}}{k^2+\mathrm{i}\epsilon}\mathrm{i}\Pi_{\varrho\sigma}(k)\frac{-\mathrm{i}\eta^{\sigma\nu}}{k^2+\mathrm{i}\epsilon} \notag\displaybreak[0] \\
&\phantom{{}={}\frac{-\mathrm{i}\eta^{\mu\varrho}}{k^2+\mathrm{i}\epsilon}}+\frac{-\mathrm{i}\eta^{\mu\varrho}}{k^2+\mathrm{i}\epsilon}\mathrm{i}\Pi_{\varrho\sigma}(k)\frac{-\mathrm{i}\eta^{\sigma\alpha}}{k^2+\mathrm{i}\epsilon}\mathrm{i}\Pi_{\alpha\beta}(k)\frac{\mathrm{-i}\eta^{\beta\nu}}{k^2+\mathrm{i}\epsilon}+\hdots \notag\displaybreak[0] \\
&=\frac{-\mathrm{i}\eta^{\mu\nu}}{k^2+\mathrm{i}\epsilon}-\frac{\mathrm{i}\Pi^{\mu\nu}(k)}{k^4+\mathrm{i}\epsilon}-\frac{\mathrm{i}\Pi^{\mu}_{\phantom{\mu}\sigma}(k)\Pi^{\sigma\nu}(k)}{k^6+\mathrm{i}\epsilon}+\hdots\,.
\end{align}
The contraction of two transverse structures $\Pi^{\mu\nu}(k)$ results in
\begin{equation}
\Pi^{\nu\alpha}(k)\Pi_{\alpha}^{\phantom{\alpha}\varrho}(k)=-k^2\Pi^{\nu\varrho}\,.
\end{equation}
So the resummation of all one-particle reducible diagrams at one-loop order corresponds to a geometric series and leads to the following
result:\footnote{In the calculations performed within this section, $(\lambda^{(0)})^2$ is understood to be extracted from the dimensionless constant
$\mathcal{C}^{(0)}$ defined in Eq.~\eqref{eq:definition-curly-c}. Consequently, it appears together with the one-loop correction $\Pi(k^2)$ in order
to keep track of all powers of $\lambda^{(0)}$.}
\begin{align}
\label{eq:full-propagator-from-one-particle-reducible}
\big\langle T\widehat{A}^{\mu}(k)\widehat{A}^{\nu}(k)\big\rangle&=\frac{-\mathrm{i}\eta^{\mu\nu}}{k^2+\mathrm{i}\epsilon}\left\{1-(\lambda^{(0)})^2\Pi(k^2)+\big[(\lambda^{(0)})^2\Pi(k^2)\big]^2\mp\hdots\right\} \notag \\
&=\frac{-\mathrm{i}\eta^{\mu\nu}}{k^2+\mathrm{i}\epsilon}\frac{1}{1+(\lambda^{(0)})^2\Pi(k^2)} \notag\displaybreak[0] \\
&=\big\langle T\widehat{A}^{\mathrm{(0)}\,\mu}(k)\widehat{A}^{\mathrm{(0)}\,\nu}(k)\big\rangle \frac{1}{1+(\lambda^{(0)})^2\Pi(k^2)}\,.
\end{align}
The full propagator can be expressed via the bare propagator multiplied by a prefactor that contains the bare one-loop correction
$\Pi(k^2)$ to the photon field. In order to perform the renormalization we consider a physical process that contains a photon propagator,
e.g. the scattering of $\phi$ at a photon and its subsequent emission. We use the full propagator for setting up the amplitude of the
process with the propagator momentum squared $k^2$ corresponding to the squared center of mass energy $\sqrt{s}$, namely $k^2=s$.
\begin{align}
\begin{array}{c}
\begin{fmfgraph*}(50,50)
\fmfpen{thin}
\fmfleft{i1,i2}
\fmfright{o1,o2}
\fmf{photon}{i1,v1}
\fmf{dbl_wiggly}{v1,v2}
\fmf{photon}{v2,o1}
\fmf{plain}{i2,v1}
\fmf{plain}{v2,o2}
\fmfforce{0.0w,0.0h}{i1}
\fmfforce{0.0w,1.0h}{i2}
\fmfforce{0.25w,0.45h}{v1}
\fmfforce{0.75w,0.45h}{v2}
\fmfforce{1.0w,0.0h}{o1}
\fmfforce{1.0w,1.0h}{o2}
\fmfdot{v1,v2}
\end{fmfgraph*}
\end{array}
=(\lambda^{(0)})^2\frac{-\mathrm{i}\eta^{\mu\nu}}{s+\mathrm{i}\epsilon}\frac{1}{1+(\lambda^{(0)})^2\Pi(s)}S_{\mu\nu} \equiv \lambda^2\frac{-\mathrm{i}\eta^{\mu\nu}}{s+\mathrm{i}\epsilon}S_{\mu\nu}\,.
\end{align}
Here $S_{\mu\nu}$ contains the remainder of the amplitude, which is not important for current considerations. In the course of renormalization
the bare coupling $\lambda^{(0)}$ is replaced by the renormalized coupling $\lambda$, such that the renormalized amplitude is finite
\begin{equation}
\lambda^2\equiv\frac{(\lambda^{(0)})^2}{1+(\lambda^{(0)})^2\Pi(s)}\,,\quad
(\lambda^{(0)})^2=\frac{\lambda^2}{1-\lambda^2\Pi(s)}=\lambda^2+\mathcal{O}(\lambda^4)\,.
\end{equation}
If the propagator momentum squared $k^2$ differs from the scale $s$ we obtain an expression similar to the bare
amplitude at order $(\lambda^{(0)})^2$
\begin{align}
\label{eq:renormalization-procedure}
(\lambda^{(0)})^2\frac{-\mathrm{i}\eta^{\mu\nu}}{k^2+\mathrm{i}\epsilon}\frac{1}{1+(\lambda^{(0)})^2\Pi(k^2)}S_{\mu\nu} &=\lambda^2\frac{-\mathrm{i}\eta^{\mu\nu}}{k^2+\mathrm{i}\epsilon}\frac{1}{\big[1-\lambda^2\Pi(s)\big]\big[1+\lambda^2\Pi(k^2)\big]}S_{\mu\nu}+\mathcal{O}(\lambda^4) \notag\displaybreak[0] \\
&=\lambda^2\frac{-\mathrm{i}\eta^{\mu\nu}}{k^2+\mathrm{i}\epsilon}\frac{1}{1+\lambda^2\Pi_{\mathrm{ren}}(k^2)}S_{\mu\nu}+\mathcal{O}(\lambda^4)\,,
\end{align}
but with $\lambda^{(0)}$ replaced by $\lambda$ and $\Pi(k^2)$ replaced by the renormalized one-loop correction
\begin{equation}
\label{eq:one-loop-correction-renormalized}
\Pi_{\mathrm{ren}}(k^2)=\Pi(k^2)-\Pi(s)\,.
\end{equation}
Equation \eqref{eq:full-propagator-from-one-particle-reducible} shows that the imaginary unit in front of the one-loop correction $\Pi(k^2)$ is
put into the propagator. What in fact matters for the photon dispersion relation is the real quantity $\Pi(k^2)$. From the bare one-loop
correction we obtain the renormalized correction according to Eq.~\eqref{eq:one-loop-correction-renormalized}:
\begin{align}
\label{eq:one-loop-correction-renormalized-explicit}
16\pi^2\,\Pi_{\mathrm{ren}}(k^2)&=\frac{1}{2(b^{(0)})^2}\left(\frac{1}{s}-\frac{1}{k^2}\right) \notag \\
&\phantom{{}={}}+\frac{1}{2}\left\{-\left[1-\frac{2}{(b^{(0)}k)^2}+\frac{1}{(b^{(0)}k)^4}\right]\ln\Big[1-(b^{(0)}k)^2-\mathrm{i}\epsilon\Big]\right. \notag \\
&\phantom{{}={}+\frac{1}{2}\Big\{}\left.+\left[1-\frac{2}{(b^{(0)})^2s}+\frac{1}{[(b^{(0)})^2s]^2}\right]\ln\Big[1-(b^{(0)})^2s-\mathrm{i}\epsilon\Big]
\right\}\,,
\end{align}
where the renormalization scale $\mu$ obviously cancels. The physically important quantity $\Pi_{\mathrm{ren}}(k^2)$ is the difference between
the bare self-energy correction evaluated at $k^2$ and the same quantity evaluated at an arbitrary scale $s$. In principle $s$ can be
chosen such that the last expression of Eq.~\eqref{eq:one-loop-correction-renormalized-explicit} in rectangular brackets vanishes. This
holds for $s=1/(b^{(0)})^2$ which then leads to
\begin{align}
16\pi^2\,\widehat{\Pi}_{\mathrm{ren}}(k^2)&\equiv 16\pi^2\left.\Pi_{\mathrm{ren}}(k^2)\right|_{s=1/(b^{(0)})^2} \notag \\
&=\frac{1}{2}\left(1-\frac{1}{(b^{(0)}k)^2}\right)-\frac{\big[1-(b^{(0)}k)^2\big]^2}{2(b^{(0)}k)^4}\ln\Big[1-(b^{(0)}k)^2-\mathrm{i}\epsilon\Big]\,.
\end{align}
The choice $s=1/(b^{(0)})^2$ is not unreasonable since $1/(b^{(0)})^2$ is the only parameter of the action \eqref{eq:action-effective-theory}
that has the same mass dimension as $s$.

\section{Leading order perturbation of the scalar field}
\label{sec:caluclation-scalar-field-perturbation}
\setcounter{equation}{0}

Having obtained the modified photon field we now consider the modification of the scalar field $\phi$ resulting from the
action~\eqref{eq:action-effective-theory}. There are two corrections to the bare scalar field. First of all there is a quantum correction
involving the photon (see Fig.~\ref{fig:scalar-field-correction1}) and, secondly, the scattering of the scalar field with the defects has to be
taken into account (see Fig.~\ref{fig:scalar-field-correction2}). We will now compute the self-energy correction given by the diagram in
Fig.~\ref{fig:scalar-field-correction1}.
\begin{subequations}
\begin{align}
\label{eq:self-energy-scalar-field-bare}
-\mathrm{i}\Sigma(k^2,m^2)&=\begin{array}{c}
\begin{fmfgraph*}(100,50)
\fmfpen{thin}
\fmfleft{i1}
\fmfright{o1}
\fmf{plain,label=$\overrightarrow{\phantom{i}k\phantom{i}}$,label.side=left}{i1,v1}
\fmf{photon,left=1,label=$\overleftarrow{\phantom{i}q\phantom{i}}$,label.side=right}{v1,v2}
\fmf{photon,left=1}{v2,v1}
\fmf{plain,label=$\overrightarrow{\phantom{i}k\phantom{i}}$,label.side=left}{v2,o1}
\fmfv{label=$\mu$,label.dist=0.2cm,label.angle=135}{v1}
\fmfv{label=$\varrho$,label.dist=0.2cm,label.angle=-135}{v3}
\fmfv{label=$\nu$,label.dist=0.2cm,label.angle=45}{v2}
\fmfv{label=$\sigma$,label.dist=0.2cm,label.angle=-45}{v4}
\fmfforce{0.0w,0.45h}{i1}
\fmfforce{0.333w,0.45h}{v1}
\fmfforce{0.667w,0.45h}{v2}
\fmfforce{0.333w,0.45h}{v3}
\fmfforce{0.667w,0.45h}{v4}
\fmfforce{1.0w,0.45h}{o1}
\fmfdot{v1,v2}
\end{fmfgraph*}
\end{array} \notag \\
&=(\mathrm{i}\lambda)^2(2\pi\mu)^{4-d}\int \mathrm{d}^dq\,\varepsilon^{\alpha\mu\beta\varrho}q_{\alpha}(-k-q)_{\beta}\varepsilon^{\gamma\nu\delta\sigma}(-q)_{\gamma}(k+q)_{\delta} \notag \\
&\hspace{4cm}\,\times \frac{-\mathrm{i}\eta_{\mu\nu}}{q^2+m^2+\mathrm{i}\epsilon}\frac{-\mathrm{i}\eta_{\varrho\sigma}}{(k+q)^2+m^2+\mathrm{i}\epsilon} \notag \\
&=2\lambda^2[k^2\widehat{I}_3(m)-\widehat{I}_4(m,k)]\,,
\end{align}
with
\begin{equation}
\widehat{I}_3(k^2,m^2)=(2\pi\mu)^{4-d}\int\mathrm{d}^dq\,\frac{q^2}{(q^2-m^2+\mathrm{i}\epsilon)[(k+q)^2-m^2+\mathrm{i}\epsilon]}\,,
\end{equation}
\begin{equation}
\widehat{I}_4(k^2,m^2)=(2\pi\mu)^{4-d}\int \mathrm{d}^dq\,\frac{(k\cdot q)^2}{(q^2-m^2+\mathrm{i}\epsilon)[(k+q)^2-m^2+\mathrm{i}\epsilon]}\,.
\end{equation}
\end{subequations}
The ultraviolet divergences are again regularized by dimensional regularization with renormalization scale $\mu$. Furthermore a photon mass
$m$ has been introduced to regularize possible infrared divergences. The results of the integrals $\widehat{I}_3$, $\widehat{I}_4$ in the
limit $m\mapsto 0$ are given by
\begin{figure}[b]
\centering
\subfigure[\,\,\,One-loop photon self-energy contribution to the scalar field.]{\label{fig:scalar-field-correction1}
\begin{fmfgraph*}(100,50)
\fmfpen{thin}
\fmfleft{i1}
\fmfright{o1}
\fmf{plain,label=$\overrightarrow{\phantom{i}k\phantom{i}}$}{i1,v1}
\fmf{photon,left=1}{v1,v2}
\fmf{photon,left=1}{v2,v1}
\fmf{plain,label=$\overrightarrow{\phantom{i}k\phantom{i}}$}{v2,o1}
\fmfforce{0.0w,0.5h}{i1}
\fmfforce{0.333w,0.5h}{v1}
\fmfforce{0.667w,0.5h}{v2}
\fmfforce{1.0w,0.5h}{o1}
\fmfdot{v1,v2}
\end{fmfgraph*}
}
\hspace{0.5cm}
\subfigure[\,\,\,Correction to the scalar field originating from the scattering of $\phi$ at the defects.]{\label{fig:scalar-field-correction2}
\begin{fmfgraph*}(100,50)
\fmfpen{thin}
\fmfleft{i1}
\fmfright{o1}
\fmf{plain,label=$\overrightarrow{\phantom{i}k\phantom{i}}$}{i1,v1}
\fmf{plain,label=$\overrightarrow{\phantom{i}k\phantom{i}}$}{v1,o1}
\fmfv{decoration.shape=cross,decoration.filled=full,decoration.size=15,label.angle=90,label.dist=0.25cm}{v1}
\fmfforce{0.0w,0.5h}{i1}
\fmfforce{0.5w,0.5h}{v1}
\fmfforce{1.0w,0.5h}{o1}
\end{fmfgraph*}
}
\caption{Possible corrections to the bare scalar field $\phi(k)$.}
\label{fig:scalar-field-corrections}
\end{figure}
\begin{equation}
\widehat{I}_3(k^2,0)=0\,,\quad \widehat{I}_4(k^2,0)=\mathrm{i}\pi^2\frac{k^4}{4}\left\{\frac{1}{\varepsilon}+\left[2-\ln\left(-\frac{k^2}{\mu^2}-\mathrm{i}\epsilon\right)\right]\right\}\,.
\end{equation}
Fortunately the integrals are infrared-finite. Hence we consider $\Sigma(k^2,0)\equiv \Sigma(k^2)$ from now on. A resummation of all one-loop
photon self-energy corrections leads to a modification of the scalar field propagator
\begin{align}
\frac{\mathrm{i}b^2}{k^2-1/(b^{(0)})^2-(b^{(0)})^2\Sigma(k^2,0)+\mathrm{i}\epsilon}&=\begin{array}{c}
\begin{fmfgraph*}(50,25)
\fmfpen{thin}
\fmfleft{i1}
\fmfright{o1}
\fmf{plain}{i1,o1}
\fmfforce{0.0w,0.405h}{i1}
\fmfforce{1.0w,0.405h}{o1}
\fmfdot{i1,o1}
\end{fmfgraph*}
\end{array}+\begin{array}{c}
\begin{fmfgraph*}(75,25)
\fmfpen{thin}
\fmfleft{i1}
\fmfright{o1}
\fmf{plain}{i1,v1}
\fmf{photon,left=1}{v1,v2}
\fmf{photon,left=1}{v2,v1}
\fmf{plain}{v2,o1}
\fmfforce{0.0w,0.405h}{i1}
\fmfforce{0.333w,0.405h}{v1}
\fmfforce{0.667w,0.405h}{v2}
\fmfforce{1.0w,0.405h}{o1}
\fmfdot{i1,v1,v2,o1}
\end{fmfgraph*}
\end{array} \notag \\
&\phantom{{}={}} +\begin{array}{c}
\begin{fmfgraph*}(125,25)
\fmfpen{thin}
\fmfleft{i1}
\fmfright{o1}
\fmf{plain}{i1,v1}
\fmf{photon,left=1}{v1,v2}
\fmf{photon,left=1}{v2,v1}
\fmf{plain}{v2,v3}
\fmf{photon,left=1}{v3,v4}
\fmf{photon,left=1}{v4,v3}
\fmf{plain}{v4,o1}
\fmfforce{0.0w,0.405h}{i1}
\fmfforce{0.2w,0.405h}{v1}
\fmfforce{0.4w,0.405h}{v2}
\fmfforce{0.6w,0.405h}{v3}
\fmfforce{0.8w,0.405h}{v4}
\fmfforce{1.0w,0.405h}{o1}
\fmfdot{i1,v1,v2,v3,v4,o1}
\end{fmfgraph*}
\end{array}+\hdots\,.
\end{align}
Contrary to Sec.~\ref{sec:caluclation-photon-field-perturbation} the renormalization of the correction to the scalar field will be performed with the help
of counterterms, which is a more convenient procedure here. We will follow the lines of \cite{PeskinSchroeder1995}. Both the photon mass and the field are 
renormalized according to
\begin{equation}
\phi=\sqrt{Z_2}\phi_{\mathrm{ren}}\,,\quad \delta_{1/b}=\frac{Z_2}{(b^{(0)})^2}-\frac{1}{b^2}\,,\quad \delta_{Z_2}=Z_2-1\,,
\end{equation}
where $\phi$ is the bare scalar field, $\phi_{\mathrm{ren}}$ the renormalized field, $Z_2$ the field renormalization constant, $\delta_{Z_2}$
the field renormalization counterterm, and $\delta_{1/b}$ the mass counterterm. Furthermore, we use the renormalization conditions
\begin{equation}
\Sigma(k^2)|_{k^2=1/b^2}=0\,,\quad \left.\frac{\mathrm{d}}{\mathrm{d}k^2}\Sigma(k^2)\right|_{k^2=1/b^2}=0\,,
\end{equation}
From these we obtain the counterterm which can be used for both the mass and the field renormalization
\begin{subequations}
\begin{equation}
\begin{array}{c}
\begin{fmfgraph*}(50,25)
\fmfpen{thin}
\fmfleft{i1}
\fmfright{o1}
\fmf{plain}{i1,v1}
\fmf{plain}{v1,o1}
\fmfv{decoration.shape=circle,decoration.filled=empty,decoration.size=15}{v1}
\fmfv{decoration.shape=cross,decoration.filled=full,decoration.size=15}{v2}
\fmfforce{0.0w,0.405h}{i1}
\fmfforce{0.5w,0.405h}{v1}
\fmfforce{0.5w,0.405h}{v2}
\fmfforce{1.0w,0.405h}{o1}
\end{fmfgraph*}
\end{array}=\mathrm{i}\left[\delta_{1/b^2}\left(\frac{k^4}{4}-\frac{1}{b^4}\right)+\frac{\delta_{Z_2}}{b^4}\right]\,,
\end{equation}
with
\begin{equation}
\delta_{1/b}=\pi^2\left\{\frac{2\lambda^2}{\varepsilon}+\lambda^2\left[3-2\ln\left(-\frac{1}{b^2\mu^2}-\mathrm{i}\epsilon\right)\right]\right\}\,,
\end{equation}
\begin{equation}
\delta_{Z_2}=\pi^2\left\{\frac{2\lambda^2}{\varepsilon}+\frac{\lambda^2}{4}\left[13-8\ln\left(-\frac{1}{b^2\mu^2}-\mathrm{i}\epsilon\right)\right]\right\}\,.
\end{equation}
\end{subequations}
The renormalized photon one-loop self-energy contribution to the scalar field is then
\begin{align}
\label{eq:self-energy-scalar-field-renormalized}
-\mathrm{i}\Sigma_{\mathrm{ren}}(k^2,b^2)&=-\mathrm{i}\Sigma(k^2)+\mathrm{i}\left[\delta_{1/b^2}\left(\frac{k^4}{4}-\frac{1}{b^4}\right)+\frac{\delta_{Z_2}}{b^4}\right] \notag \\
&=\frac{\mathrm{i}\pi^2\lambda^2}{4b^4}\left\{1-(bk)^4\left[1+2\ln\left(\frac{1}{(bk)^2}+\mathrm{i}\epsilon\right)\right]\right\}\,.
\end{align}
This correction leads to a renormalization of the $\phi$-field mass. The correction is of order $\lambda^2$ and depends on
both $k^2$ and the renormalized mass $1/b^2$
\begin{equation}
\frac{1}{(b^{(0)})^2}\mapsto \frac{1}{b^2}+b^2\Sigma_{\mathrm{ren}}(k^2,b^2)\,.
\end{equation}
As in the case of the photon field we now define the resummation of all one-particle reducible one-loop corrections to the scalar field
\begin{align}
\widehat{\phi}(k)&\equiv \begin{array}{c}
\begin{fmfgraph*}(37.5,25)
\fmfpen{thin}
\fmfleft{i1}
\fmfright{o1}
\fmf{plain}{i1,o1}
\fmfforce{0.0w,0.405h}{i1}
\fmfforce{1.0w,0.405h}{o1}
\fmfdot{i1}
\end{fmfgraph*}
\end{array}+\begin{array}{c}
\begin{fmfgraph*}(75,25)
\fmfpen{thin}
\fmfleft{i1}
\fmfright{o1}
\fmf{plain}{i1,v1}
\fmf{photon,left=1}{v1,v2}
\fmf{photon,left=1}{v2,v1}
\fmf{plain}{v2,o1}
\fmfforce{0.0w,0.405h}{i1}
\fmfforce{0.333w,0.405h}{v1}
\fmfforce{0.667w,0.405h}{v2}
\fmfforce{1.0w,0.405h}{o1}
\fmfdot{i1,v1,v2}
\end{fmfgraph*}
\end{array}+\begin{array}{c}
\begin{fmfgraph*}(125,25)
\fmfpen{thin}
\fmfleft{i1}
\fmfright{o1}
\fmf{plain}{i1,v1}
\fmf{photon,left=1}{v1,v2}
\fmf{photon,left=1}{v2,v1}
\fmf{plain}{v2,v3}
\fmf{photon,left=1}{v3,v4}
\fmf{photon,left=1}{v4,v3}
\fmf{plain}{v4,o1}
\fmfforce{0.0w,0.405h}{i1}
\fmfforce{0.2w,0.405h}{v1}
\fmfforce{0.4w,0.405h}{v2}
\fmfforce{0.6w,0.405h}{v3}
\fmfforce{0.8w,0.405h}{v4}
\fmfforce{1.0w,0.405h}{o1}
\fmfdot{i1,v1,v2,v3,v4}
\end{fmfgraph*}
\end{array} \notag \\
&\phantom{{}={}}+\begin{array}{c}
\begin{fmfgraph*}(175,25)
\fmfpen{thin}
\fmfleft{i1}
\fmfright{o1}
\fmf{plain}{i1,v1}
\fmf{photon,left=1}{v1,v2}
\fmf{photon,left=1}{v2,v1}
\fmf{plain}{v2,v3}
\fmf{photon,left=1}{v3,v4}
\fmf{photon,left=1}{v4,v3}
\fmf{plain}{v4,v5}
\fmf{photon,left=1}{v5,v6}
\fmf{photon,left=1}{v6,v5}
\fmf{plain}{v6,o1}
\fmfforce{0.0w,0.405h}{i1}
\fmfforce{0.143w,0.405h}{v1}
\fmfforce{0.286w,0.405h}{v2}
\fmfforce{0.429w,0.405h}{v3}
\fmfforce{0.571w,0.405h}{v4}
\fmfforce{0.714w,0.405h}{v5}
\fmfforce{0.857w,0.405h}{v6}
\fmfforce{1.0w,0.405h}{o1}
\fmfdot{i1,v1,v2,v3,v4,v5,v6}
\end{fmfgraph*}
\end{array}+\dots \notag \\
&\phantom{{}={}}+\text{counterterms}+\hdots\,.
\end{align}
This resummation fulfills a modified field equation with renormalized mass $1/b$ of the $\phi$-field
\begin{equation}
\label{eq:field-equation-phi-diagrammatical}
\frac{1}{b^2}\left(k^2-\frac{1}{b^2}\right)\widehat{\phi}(k)=\begin{array}{c}
\begin{fmfgraph*}(25,25)
\fmfpen{thin}
\fmfleft{i1}
\fmfright{o1}
\fmf{photon,left=1}{i1,o1}
\fmf{photon,left=1}{o1,i1}
\fmfforce{0.0w,0.405h}{i1}
\fmfforce{1.0w,0.405h}{o1}
\fmfdot{i1,o1}
\end{fmfgraph*}
\end{array}\Big(\widehat{\phi}(k)\Big)\,.
\end{equation}
Inserting this mass correction into the photon field equation leads to $\mathcal{O}(\lambda^4)$-corrections. Note that the imaginary part
resulting from both Eq.~\eqref{eq:self-energy-scalar-field-bare} and Eq.~\eqref{eq:self-energy-scalar-field-renormalized} is given by
\begin{equation}
\mathrm{Im}(\Sigma)=\frac{\pi^3}{2}\lambda^2k^4\,.
\end{equation}
This result does not depend on the renormalization program since by the optical theorem it is linked to the total cross section for the
decay of an excitation of the $\phi$-field (a scalar particle) into two photons.

Finally, we take into account the second contribution in Fig.~\ref{fig:scalar-field-correction2}, namely the scattering of $\phi$ at a defect.
In so doing we define the full solution $\widehat{\phi}_{\varrho}(k)$ of the corresponding field equation as the resummation of all one-loop
corrections with defect vertex insertions
\begin{align}
\begin{array}{c}
\begin{fmfgraph*}(37.5,25)
\fmfpen{thin}
\fmfleft{i1}
\fmfright{o1}
\fmf{dbl_plain}{i1,o1}
\fmfforce{0.0w,0.405h}{i1}
\fmfforce{1.0w,0.405h}{o1}
\fmfdot{i1}
\end{fmfgraph*}
\end{array}&\equiv \begin{array}{c}
\begin{fmfgraph*}(37.5,25)
\fmfpen{thin}
\fmfleft{i1}
\fmfright{o1}
\fmf{plain}{i1,o1}
\fmfforce{0.0w,0.405h}{i1}
\fmfforce{1.0w,0.405h}{o1}
\fmfdot{i1}
\end{fmfgraph*}
\end{array}+\begin{array}{c}
\begin{fmfgraph*}(50,25)
\fmfpen{thin}
\fmfleft{i1}
\fmfright{o1}
\fmf{plain}{i1,v1,o1}
\fmfv{decoration.shape=cross,decoration.filled=full,decoration.size=15,label.angle=90,label.dist=0.25cm}{v1}
\fmfforce{0.0w,0.405h}{i1}
\fmfforce{0.5w,0.405h}{v1}
\fmfforce{1.0w,0.405h}{o1}
\fmfdot{i1}
\end{fmfgraph*}
\end{array}+\begin{array}{c}
\begin{fmfgraph*}(100,25)
\fmfpen{thin}
\fmfleft{i1}
\fmfright{o1}
\fmf{plain}{i1,v1}
\fmf{plain}{v1,v2}
\fmf{photon,left=1}{v2,v3}
\fmf{photon,left=1}{v3,v2}
\fmf{plain}{v3,o1}
\fmfv{decoration.shape=cross,decoration.filled=full,decoration.size=15,label.angle=90,label.dist=0.25cm}{v1}
\fmfforce{0.0w,0.405h}{i1}
\fmfforce{0.25w,0.405h}{v1}
\fmfforce{0.50w,0.405h}{v2}
\fmfforce{0.75w,0.405h}{v3}
\fmfforce{1.0w,0.405h}{o1}
\fmfdot{i1,v2,v3}
\end{fmfgraph*}
\end{array}+\begin{array}{c}
\begin{fmfgraph*}(100,25)
\fmfpen{thin}
\fmfleft{i1}
\fmfright{o1}
\fmf{plain}{i1,v1}
\fmf{photon,left=1}{v1,v2}
\fmf{photon,left=1}{v2,v1}
\fmf{plain}{v2,o1}
\fmfv{decoration.shape=cross,decoration.filled=full,decoration.size=15,label.angle=90,label.dist=0.25cm}{v3}
\fmfforce{0.0w,0.405h}{i1}
\fmfforce{0.25w,0.405h}{v1}
\fmfforce{0.50w,0.405h}{v2}
\fmfforce{0.75w,0.405h}{v3}
\fmfforce{1.0w,0.405h}{o1}
\fmfdot{i1,v1,v2}
\end{fmfgraph*}
\end{array} \notag \\
&\phantom{{}={}}+\begin{array}{c}
\begin{fmfgraph*}(150,25)
\fmfpen{thin}
\fmfleft{i1}
\fmfright{o1}
\fmf{plain}{i1,v1,v2}
\fmf{photon,left=1}{v2,v3}
\fmf{photon,left=1}{v3,v2}
\fmf{plain}{v3,v4}
\fmf{photon,left=1}{v4,v5}
\fmf{photon,left=1}{v5,v4}
\fmf{plain}{v5,o1}
\fmfv{decoration.shape=cross,decoration.filled=full,decoration.size=15,label.angle=90,label.dist=0.25cm}{v1}
\fmfforce{0.0w,0.405h}{i1}
\fmfforce{0.167w,0.405h}{v1}
\fmfforce{0.333w,0.405h}{v2}
\fmfforce{0.5w,0.405h}{v3}
\fmfforce{0.667w,0.405h}{v4}
\fmfforce{0.833w,0.405h}{v5}
\fmfforce{1.0w,0.405h}{o1}
\fmfdot{i1,v2,v3,v4,v5}
\end{fmfgraph*}
\end{array}+\begin{array}{c}
\begin{fmfgraph*}(150,25)
\fmfpen{thin}
\fmfleft{i1}
\fmfright{o1}
\fmf{plain}{i1,v1}
\fmf{photon,left=1}{v1,v2}
\fmf{photon,left=1}{v2,v1}
\fmf{plain}{v2,v3,v4}
\fmf{photon,left=1}{v4,v5}
\fmf{photon,left=1}{v5,v4}
\fmf{plain}{v5,o1}
\fmfv{decoration.shape=cross,decoration.filled=full,decoration.size=15,label.angle=90,label.dist=0.25cm}{v3}
\fmfforce{0.0w,0.405h}{i1}
\fmfforce{0.167w,0.405h}{v1}
\fmfforce{0.333w,0.405h}{v2}
\fmfforce{0.5w,0.405h}{v3}
\fmfforce{0.667w,0.405h}{v4}
\fmfforce{0.833w,0.405h}{v5}
\fmfforce{1.0w,0.405h}{o1}
\fmfdot{i1,v1,v2,v4,v5}
\end{fmfgraph*}
\end{array}+\dots \notag \\
&\phantom{{}={}}+\begin{array}{c}
\begin{fmfgraph*}(225,25)
\fmfpen{thin}
\fmfleft{i1}
\fmfright{o1}
\fmf{plain}{i1,v1}
\fmf{photon,left=1}{v1,v2}
\fmf{photon,left=1}{v2,v1}
\fmf{plain}{v2,v3,v4}
\fmf{photon,left=1}{v4,v5}
\fmf{photon,left=1}{v5,v4}
\fmf{plain}{v5,v6,v7}
\fmf{photon,left=1}{v7,v8}
\fmf{photon,left=1}{v8,v7}
\fmf{plain}{v8,o1}
\fmfv{decoration.shape=cross,decoration.filled=full,decoration.size=15,label.angle=90,label.dist=0.25cm}{v3}
\fmfv{decoration.shape=cross,decoration.filled=full,decoration.size=15,label.angle=90,label.dist=0.25cm}{v6}
\fmfforce{0.0w,0.405h}{i1}
\fmfforce{0.111w,0.405h}{v1}
\fmfforce{0.222w,0.405h}{v2}
\fmfforce{0.333w,0.405h}{v3}
\fmfforce{0.444w,0.405h}{v4}
\fmfforce{0.556w,0.405h}{v5}
\fmfforce{0.667w,0.405h}{v6}
\fmfforce{0.778w,0.405h}{v7}
\fmfforce{0.889w,0.405h}{v8}
\fmfforce{1.0w,0.405h}{o1}
\fmfdot{i1,v1,v2,v4,v5,v7,v8}
\end{fmfgraph*}
\end{array}
+\hdots \notag \\
&\phantom{{}={}}+\text{counterterms}+\hdots\,.
\end{align}
\begin{subequations}
This leads to the following field equation for the scalar field
\begin{equation}
\label{eq:field-equation-phi-diagrammatical-defects}
\frac{1}{b^2}\left(k^2-\frac{1}{b^2}\right)\left(\begin{array}{c}
\begin{fmfgraph*}(37.5,25)
\fmfpen{thin}
\fmfleft{i1}
\fmfright{o1}
\fmf{dbl_plain}{i1,v1,o1}
\fmfforce{0.0w,0.405h}{i1}
\fmfforce{1.0w,0.405h}{o1}
\fmfdot{i1}
\end{fmfgraph*}
\end{array}\right)=\begin{array}{c}
\begin{fmfgraph*}(25,25)
\fmfpen{thin}
\fmfleft{i1}
\fmfright{o1}
\fmf{plain}{i1,v1}
\fmf{plain}{v1,o1}
\fmfv{decoration.shape=cross,decoration.filled=full,decoration.size=15,label.angle=90,label.dist=0.25cm}{v1}
\fmfforce{0.0w,0.405h}{i1}
\fmfforce{0.5w,0.405h}{v1}
\fmfforce{1.0w,0.405h}{o1}
\end{fmfgraph*}
\end{array}\left(
\begin{array}{c}
\begin{fmfgraph*}(37.5,25)
\fmfpen{thin}
\fmfleft{i1}
\fmfright{o1}
\fmf{dbl_plain}{i1,v1,o1}
\fmfforce{0.0w,0.405h}{i1}
\fmfforce{1.0w,0.405h}{o1}
\fmfdot{i1}
\end{fmfgraph*}
\end{array}\right)\,,
\end{equation}
or
\begin{equation}
\frac{1}{b^2}\left(k^2-\frac{1}{b^2}\right)\widehat{\phi}_{\varrho}(k)=\varrho\widehat{\phi}_{\varrho}(k)\,.
\end{equation}
\end{subequations}
Physically, the interaction of the $\phi$-field with the defects results
in a shift of the mass of $\phi$ that corresponds to the density of
defects as long as momentum transfer to the defects is neglected.
\begin{equation}
\label{eq:phi-field-modified-mass}
\frac{1}{b^2}\mapsto \frac{1}{b^2}+\varrho b^2\equiv \frac{1}{b_{\varrho}^2}\,.
\end{equation}

\section{Modified photon dispersion relation}
\label{sec:dispersion-relation-photon-modified}
\setcounter{equation}{0}

\subsection{Final result for the modified theory}

To summarize, the second order perturbative solution of the photon field equation is
\begin{subequations}
\label{eq:result-one-loop}
\begin{eqnarray}
\lambda^2\widetilde{A}^{(2)\,\nu}(k) &=& -\mathcal{C}\widetilde{\Delta}(k)B^{\nu\gamma}(k)\widetilde{A}_{\gamma}^{(0)}(k)=-\mathcal{C}\widetilde{\Delta}(k)\mathrm{i}\Pi^{\nu\gamma}(k)\widetilde{A}_{\gamma}^{(0)}(k)\,, \\
\label{eq:result-one-loop-eq2}
\mathrm{i}\Pi^{\nu\gamma}(k) &=& \mathrm{i}(k^{\nu}k^{\gamma}-\eta^{\nu\gamma}k^2)\widehat{\Pi}_{\mathrm{ren}}(k^2)\,, \\
16\pi^2\,\widehat{\Pi}_{\mathrm{ren}}(k^2) &=& \frac{1}{2}\left(1-\frac{1}{(b_{\varrho}k)^2}\right)-\frac{\big[1-(b_{\varrho}k)^2\big]^2}{2(b_{\varrho}k)^4}\ln\Big[1-(b_{\varrho}k)^2-\mathrm{i}\epsilon\Big]\,,
\end{eqnarray}
\end{subequations}
where $\mathcal{C}=\Xi b_{\varrho}^4\lambda^2\varrho$ and $\epsilon=0^+$. In Sec.~\ref{sec:caluclation-photon-field-perturbation}
we integrated out the scalar field at one-loop level. Using the results from Sec.~\ref{sec:caluclation-scalar-field-perturbation}
this leads to an effective vertex whose Feynman rule is given by Eq.~\eqref{eq:Feynman-rule-5} in
App.~\ref{sec:perturbative-feynman-rules}.

In order to obtain the modified photon field equation, we define the one-loop resummed photon field $\widehat{A}_{\varrho}^{\nu}(k)$ by employing
the full scalar field solution $\widehat{\phi}_{\varrho}(k)$ from the previous section.
\begin{equation}
\begin{array}{c}
\begin{fmfgraph*}(50,37.5)
\fmfpen{thin}
\fmfleft{i1}
\fmfright{o1}
\fmf{dbl_wiggly,label=$\overrightarrow{\phantom{i}k\phantom{i}}$,label.side=right}{i1,o1}
\fmfdot{i1}
\fmfv{label=$\widehat{A}_{\varrho}^{\nu}(k)$,label.angle=90}{o1}
\fmfforce{0.0w,0.425h}{i1}
\fmfforce{1.0w,0.425h}{o1}
\end{fmfgraph*}
\end{array}=\begin{array}{c}
\begin{fmfgraph*}(50,37.5)
\fmfpen{thin}
\fmfleft{i1}
\fmfright{o1}
\fmf{photon,label=$\overrightarrow{\phantom{i}k\phantom{i}}$,label.side=right}{i1,o1}
\fmfdot{i1}
\fmfforce{0.0w,0.425h}{i1}
\fmfforce{1.0w,0.425h}{o1}
\end{fmfgraph*}
\end{array}+\begin{array}{c}
\begin{fmfgraph*}(50,37.5)
\fmfpen{thin}
\fmfleft{i1}
\fmfright{o1}
\fmf{photon,label=$\overrightarrow{\phantom{i}k\phantom{i}}$,label.side=right}{i1,v1}
\fmf{photon}{v1,v2}
\fmf{photon,label=$\overrightarrow{\phantom{i}k\phantom{i}}$,label.side=right}{v2,o1}
\fmf{dbl_plain,left=1}{v1,v2}
\fmfforce{0.0w,0.425h}{i1}
\fmfforce{0.25w,0.425h}{v1}
\fmfforce{0.75w,0.425h}{v2}
\fmfforce{0.5w,0.755h}{v3}
\fmfforce{1.0w,0.425h}{o1}
\fmfv{decoration.shape=cross,decoration.filled=full,decoration.size=10,label.angle=90,label.dist=0.25cm}{v3}
\fmfdot{i1,v1,v2}
\end{fmfgraph*}
\end{array}+\begin{array}{c}
\begin{fmfgraph*}(100,37.5)
\fmfpen{thin}
\fmfleft{i1}
\fmfright{o1}
\fmf{photon,label=$\overrightarrow{\phantom{i}k\phantom{i}}$,label.side=right}{i1,v1}
\fmf{photon}{v1,v2}
\fmf{photon,label=$\overrightarrow{\phantom{i}k\phantom{i}}$,label.side=right}{v2,v3}
\fmf{photon}{v3,v4}
\fmf{photon,label=$\overrightarrow{\phantom{i}k\phantom{i}}$,label.side=right}{v4,o1}
\fmf{dbl_plain,left=1}{v1,v2}
\fmf{dbl_plain,left=1}{v3,v4}
\fmfforce{0.0w,0.425h}{i1}
\fmfforce{0.125w,0.425h}{v1}
\fmfforce{0.375w,0.425h}{v2}
\fmfforce{0.625w,0.425h}{v3}
\fmfforce{0.875w,0.425h}{v4}
\fmfforce{0.25w,0.755h}{v5}
\fmfforce{0.75w,0.755h}{v6}
\fmfforce{1.0w,0.425h}{o1}
\fmfv{label=$\hspace{0.6cm}\widetilde{A}^{(0),,\nu}(k)$,label.angle=90}{o1}
\fmfv{decoration.shape=cross,decoration.filled=full,decoration.size=10,label.angle=90,label.dist=0.25cm}{v5}
\fmfv{decoration.shape=cross,decoration.filled=full,decoration.size=10,label.angle=90,label.dist=0.25cm}{v6}
\fmfdot{i1,v1,v2,v3,v4}
\end{fmfgraph*}
\end{array}\qquad+\hdots
\end{equation}
The one-loop contribution to the photon field is transverse, and after renormalizing the coupling constant using
Eq.~\eqref{eq:renormalization-procedure} it is finite. Furthermore, it becomes imaginary for $k^2=1/b_{\varrho}^2$. This indicates that an
electromagnetic wave is damped when $k^2$ approaches the mass of the scalar field $\phi$ leading to a resonance behavior.

Pursuing the discussions at the end of Sec.~\ref{subsec:perturbation-from-feynman-rules} and below
Eq.~\eqref{eq:one-loop-correction-renormalized} we now consider the modified field equation
\begin{subequations}
\begin{equation}
\label{eq:field-equation-photon-diagrammatical}
k^2\left(\begin{array}{c}
\begin{fmfgraph*}(50,25)
\fmfpen{thin}
\fmfleft{i1}
\fmfright{o1}
\fmf{dbl_wiggly}{i1,o1}
\fmfdot{i1}
\fmfforce{0.0w,0.405h}{i1}
\fmfforce{1.0w,0.405h}{o1}
\end{fmfgraph*}
\end{array}\right)=\begin{array}{c}
\begin{fmfgraph*}(30,25)
\fmfpen{thin}
\fmfleft{i1}
\fmfright{o1}
\fmf{photon}{i1,o1}
\fmf{dbl_plain,left=1}{i1,o1}
\fmfforce{0.0w,0.405h}{i1}
\fmfforce{0.075w,0.405h}{i1}
\fmfforce{0.925w,0.405h}{o1}
\fmfforce{0.5w,0.955h}{v2}
\fmfforce{1.0w,0.405h}{o1}
\fmfv{decoration.shape=cross,decoration.filled=full,decoration.size=10,label.angle=90,label.dist=0.25cm}{v2}
\fmfdot{i1,o1}
\end{fmfgraph*}
\end{array}\left(\begin{array}{c}
\begin{fmfgraph*}(50,25)
\fmfpen{thin}
\fmfleft{i1}
\fmfright{o1}
\fmf{dbl_wiggly}{i1,o1}
\fmfdot{i1}
\fmfforce{0.0w,0.405h}{i1}
\fmfforce{1.0w,0.405h}{o1}
\end{fmfgraph*}
\end{array}\right)\,,
\end{equation}
\begin{align}
\label{eq:field-equation-photon}
k^2\widehat{A}_{\varrho}^{\nu}(k)&=-\mathcal{C} \big[-\mathrm{i}B^{\nu}_{\phantom{\nu}\gamma}(k)\big]\widehat{A}_{\varrho}^{\gamma}(k)=-\mathcal{C}\,(k^{\nu}k_{\gamma}-\delta^{\nu}_{\phantom{\nu}\gamma}k^2)\widehat{\Pi}_{\mathrm{ren}}(k^2)\widehat{A}_{\varrho}^{\gamma}(k) \notag \\
&=\mathcal{C}\,k^2\widehat{\Pi}_{\mathrm{ren}}(k^2)\widehat{A}_{\varrho}^{\nu}(k)\,.
\end{align}
\end{subequations}
Two distinct photon dispersion relations follow from Eq.~\eqref{eq:field-equation-photon}. First of all we obtain the standard dispersion
law $k^2=0$. Secondly, considering $k^2\neq 0$ Eq.~\eqref{eq:field-equation-photon} results in
\begin{equation}
\label{eq:modified-field-equation-result}
\widehat{A}_{\varrho}^{\nu}(k)=-\mathcal{C}\widetilde{\Delta}(k)\big[-\mathrm{i}B^{\nu}_{\phantom{\nu}\gamma}(k)\big]\widehat{A}_{\varrho}^{\gamma}(k)=\mathcal{C}\frac{k^2}{k^2+\mathrm{i}\epsilon}\widehat{\Pi}_{\mathrm{ren}}(k^2)\widehat{A}_{\varrho}^{\nu}(k)\,.
\end{equation}
Nontrivial solutions for the photon field will exist if the following transcendental equation holds
\begin{equation}
\label{eq:dispersion-relation-of-shell}
1-\mathcal{C}\,\widehat{\Pi}_{\mathrm{ren}}(k^2)=0\,.
\end{equation}
Unfortunately it cannot be solved analytically. 
Since we expect the solution to be a minuscule correction to the standard photon dispersion relation $k^2=0$, we assume $(b_{\varrho}k)^2\ll 1$. This 
leads to an approximate equation that can be investigated analytically:
\begin{equation}
\label{eq:dispersion-relation-of-shell-perturbative}
1+\frac{\mathcal{C}}{64\pi^2}\left[1-\frac{2b_{\varrho}^2}{3}\big(k_0^2-|\mathbf{k}|^2\big)\right]=0\,.
\end{equation}
Its solution is
\begin{equation}
\label{eq:dispersion-law-photon-modified}
k_0=\sqrt{|\mathbf{k}|^2+m_{\upgamma}^2}\,,\quad m_{\upgamma}^2=\frac{3}{2}\left(1+\frac{64\pi^2}{\mathcal{C}}\right)\frac{1}{b_{\varrho}^2}\,,
\end{equation}
with the three-momentum $\mathbf{k}=(k_1,k_2,k_3)$. Thus the photon acquires a mass $m_{\upgamma}$. Note that this mass depends inversely on
the parameters $b_{\varrho}$ and $\mathcal{C}$, which are supposed to be small.

\subsection{Physical meaning of the modified photon dispersion relation}

In this section we wish to discuss what the result obtained in the previous section means physically.
For our argument we choose exemplary values for the parameters of the theory. First of all, a reasonable value for the upper boost limit $\Lambda'$
has to be chosen. This is a dimensionless number, thus e.g. the ratio of two characteristic masses. As the first we choose the Planck mass
$M_{\mathrm{Pl}}=\sqrt{\hbar c/G} \approx 1.22\cdot 10^{19}\,\mathrm{GeV}/c^2$ and as the second the proton mass $M_{\mathrm{P}}\approx 0.94\,\mathrm{GeV}/c^2$
that is a characteristic scale for standard model physics. Since the Lorentz group acts on four-dimensional Minkowski space it can be shown that
$\Xi$ grows with the fourth power of the dimensionless number $M_{\mathrm{Pl}}/M_{\mathrm{P}}$ that has been previously mentioned.

Furthermore, since the scalar field is the mediator between the photon and the spacetime defects, which are Planck-scale effects, the scalar field
mass $1/b_{\varrho}$ is assumed to be large, perhaps some fraction of the Planck mass. Then $b_{\varrho}$ would be of the order of the Planck length.
Assuming the spacetime defects to have an average separation of $10^{10}\times L_{\mathrm{Pl}}$ (see the discussion below Eq.~\eqref{eq:sum-of-random-phases})
and setting $\lambda\approx 10^{-19}$ (the coupling constant $\lambda$ is taken to be a very small number) we obtain:
\begin{equation}
\mathcal{C}=\Xi \varrho b_{\varrho}^4\lambda^2\sim \frac{1}{100}\left(\frac{\Xi}{M_{\mathrm{Pl}}/M_P}\right) \left(\frac{\varrho}{1/(10^{10}\times L_{\mathrm{Pl}})^4}\right)\left( \frac{b_{\varrho}^4}{1/M_{\mathrm{Pl}}^4}\right)\left(\frac{\lambda^2}{(10^{-19})^2}\right)\,.
\end{equation}
Hence, the upper choice of the values is consistent with our procedure of performing a perturbative expansion with respect to $\lambda$, which in
principle corresponds to an expansion in $\mathcal{C}$.

Now it becomes evident that the behavior of Eq.~\eqref{eq:dispersion-law-photon-modified} as a photon mass is unusual for a perturbative approach.
The deviation from the standard photon dispersion law is expected to be a correction that vanishes for $b_{\varrho}\mapsto 0$ or $\mathcal{C}\mapsto 0$.
Physically, for $b_{\varrho}\mapsto 0$ the scalar mediator between the photon and the defects decouples, whereas for $\mathcal{C}\mapsto 0$ either
the density $\varrho$ of defects or the coupling $\lambda$ between the photon and the scalar vanishes (or again $b_{\varrho}\mapsto 0$). For both scenarios 
the modified photon mass is expected to vanish, since it is a perturbative result. However, as becomes evident from Eq.~\eqref{eq:dispersion-law-photon-modified}, 
the contrary happens.

Let us see how this behavior is connected to the limit $k^2\mapsto 0$ of the quantity $\widehat{\Pi}_{\mathrm{ren}}$ of
Eq.~\eqref{eq:modified-field-equation-result}, which is the mathematical basis of the modified photon dispersion law. Two cases have to be considered
here: $(b_{\varrho}k)^2\ll 1$ and $(b_{\varrho}k)^2\gg 1$.
\begin{subequations}
\begin{equation}
\label{eq:gauge-invariance-modified-theory}
\lim_{k^2\mapsto 0} 16\pi^2\,k^2\widehat{\Pi}_{\mathrm{ren}}(k^2)\Big|_{\substack{(b_{\varrho}k)^2\ll 1 \\ \text{fixed}}}=0\,,
\end{equation}
\begin{align}
\label{eq:gauge-invariance-violation-modified-theory}
\lim_{k^2\mapsto 0} 16\pi^2\,k^2\widehat{\Pi}_{\mathrm{ren}}(k^2)\Big|_{\substack{(b_{\varrho}k)^2\gg 1 \\ \text{fixed}}}&=\frac{1}{2b_{\varrho}^2}\left(\frac{1}{1-\mathrm{i}\epsilon}-1\right) \notag \\
&\phantom{{}={}}+\lim_{k^2\mapsto 0}\frac{1}{b_{\varrho}^2}\left(1-\frac{1}{2b_{\varrho}^2k^2}\right)\ln(1-\mathrm{i}\epsilon)\,.
\end{align}
\end{subequations}
Compare these results to the renormalized self-energy correction of the photon (vacuum polarization) $\Pi_{\mathrm{ren}}(k^2)|^{\mathrm{QED}}$
in ordinary QED, for which
\begin{equation}
\label{eq:gauge-invariance-standard-qed}
\lim_{k^2\mapsto 0} k^2 \Pi_{\mathrm{ren}}(k^2)\Big|^{\mathrm{QED}}=0\,.
\end{equation}
Equation \eqref{eq:gauge-invariance-standard-qed} means that gauge invariance is maintained for quantum corrections in QED and the dispersion law
of the photon remains as $k^2=0$. An analogous argument holds for the modified theory defined by the action \eqref{eq:action-modified-theory-complete}.
As long as $k^2\ll 1/b_{\varrho}^2$, for which the scalar mass is large and the modification of the photon dispersion law is a small deviation
from $k^2=0$, gauge invariance is maintained. Then the photon mass given by Eq.~\eqref{eq:dispersion-law-photon-modified} cannot appear in this case.
The tensor structure $(k^{\nu}k^{\gamma}-\eta^{\nu\gamma}k^2)$ of $B^{\nu\gamma}(k)$ in Eq.~\eqref{eq:result-one-loop} is crucial for this result, 
namely the conservation of gauge invariance. As long as this structure is conserved by the interaction with the defects, the photon dispersion relation 
stays $k^2=0$ (for $b_{\varrho}^2k^2\ll 1$).

However from Eq.~\eqref{eq:gauge-invariance-violation-modified-theory} it follows that gauge invariance is violated for large $b_{\varrho}$ (and small
scalar mass $1/b_{\varrho}$). The technical reason is that $\epsilon=0^+$ cannot be discarded in this case, since the argument of the logarithm
in Eq.~\eqref{eq:result-one-loop} may be negative. The complex logarithm has a branch cut on the negative real axis and, therefore, a small
imaginary part $\epsilon=0^+$ has to be added to its argument. Thus the limit $k^2\mapsto 0$ does not exist here and gauge invariance is violated
resulting in a photon mass. Because of the infinitesimal imaginary part in the logarithm, $\widehat{\Pi}_{\mathrm{ren}}(k^2)$ also has an imaginary
part. Physically this corresponds to the damping of electromagnetic waves when the modified photon momentum square approaches the mass square of the
scalar field.

\subsection{General remarks about the previous results}
\label{sec:general-remarks-about-results}

A graphical investigation of $\widehat{\Pi}_{\mathrm{ren}}(k^2)$ shows that Eq.~\eqref{eq:dispersion-relation-of-shell} does indeed not have a
solution. This means that the perturbative solution \eqref{eq:dispersion-law-photon-modified} only appears as a solution to the expanded equation
\eqref{eq:dispersion-relation-of-shell-perturbative}, but it is not a solution of Eq.~\eqref{eq:dispersion-relation-of-shell}, which is exact. Why did 
we not do a graphical analysis at first? The answer is that in this section we would like to argue on a very fundamental basis why the photon dispersion 
relation remains the conventional one within the simple spacetime foam model proposed. For this reason we will investigate a generic function $h$  
coming from a quantum correction that involves the interaction with the defects. 

However first of all we recall the following three general modifications of the photon dispersion law that originate from a violation of Lorentz or 
gauge invariance. They are denoted as cases (1), (2), and (3).

\begin{itemize}

\item[1)] Scaleless Lorentz-violating modification (cf. modified Maxwell theory
\cite{ChadhaNielsen1983,ColladayKostelecky1998,KosteleckyMewes2002,BaileyKostelecky2004}):

\noindent In the first case Lorentz invariance is violated by preferred spacetime directions. These are given by four-vectors $\alpha_1$,
$\alpha_2$ etc. with dimensionless components $\kappa_1$, $\kappa_2$ etc. The modified photon dispersion relation is obtained from a zero of a
function $f$ whose arguments are scalars. Therefore, they can only consist of scalar products with the momentum four-vector and scalar products
among themselves: $f(k^2,\alpha_1\cdot k,\alpha_2\cdot k,\alpha_1\cdot \alpha_2,\dots)=0$. The solution of such an equation is
\begin{equation}
\label{eq:modified-photon-dispersion-scaleless}
k_0=g(k_1,k_2,k_3,\kappa_1,\kappa_2,\dots)\approx |\mathbf{k}|+g_1(k_1,k_2,k_3)\kappa_1+g_2(k_1,k_2,k_3)\kappa_2+\hdots\,,
\end{equation}
with functions $g_1$, $g_2$ etc. that have mass dimension 1 and only depend on the three-momentum components. Higher-order terms in the scaleless
deformation parameters have been neglected.
\item[2)] Scale-dependent Lorentz-violating modification (cf. Maxwell-Chern-Simons theory
\cite{Carroll-etal1990,ColladayKostelecky1998,KosteleckyMewes2002,BaileyKostelecky2004}):

\noindent Preferred spacetime directions $a_1$, $a_2$, etc. appear in this case as well. They come together with physical scales $m_1$, $m_2$ etc.,
hence have dimensionful components. We will denote them with Latin letters in the following. The modified dispersion relation is a zero of a function
$\widetilde{f}$ containing only scalar products as arguments: $\widetilde{f}(k^2,a_1\cdot k,a_2\cdot k,a_1\cdot a_2,\dots)=0$.
We perform an expansion in the scales $m_1$, $m_2$ of the deformation by considering them to be much smaller than the remaining physical scales.
This leads to the following approximate solution of the latter equation:
\begin{equation}
\label{eq:modified-photon-dispersion-scale-dependent}
k_0=\widetilde{g}(k_1,k_2,k_3,m_1,m_2,\dots)\approx |\mathbf{k}|+\widetilde{g}_1(k_1,k_2,k_3)m_1+\widetilde{g}_2(k_1,k_2,k_3)m_2+\hdots\,,
\end{equation}
where the functions $\widetilde{g}_1$, $\widetilde{g}_2$ etc. are dimensionless and depend on the three-momentum components.
Thus the existence of preferred directions leads to a different behavior for the four-momentum components in the dispersion relation since
the scalar products project out certain momentum components.

\item[3)] Emergence of a photon mass:

The emergence of a photon mass $m_{\upgamma}$ by a violation of gauge invariance would result in the following dispersion relation
\begin{equation}
\label{eq:emergence-photon-mass}
k_0=\sqrt{\mathbf{k}^2+m^2_{\upgamma}}=|\mathbf{k}|+\frac{m^2_{\upgamma}}{2|\mathbf{k}|}+\hdots\,.
\end{equation}
Note that in Eq.~\eqref{eq:emergence-photon-mass} no term that is proportional to $m_{\upgamma}$ appears --- contrary to
Eq.~\eqref{eq:modified-photon-dispersion-scale-dependent} where linear terms in the dimensionful parameters $m_1$, $m_2$ etc. can
be found.

\end{itemize}
A modification of the photon dispersion relation in the context of spacetime foam is expected to emerge according to one of the
previous cases. For example, within the spacetime foam model considered in \cite{hep-th/0312032} the photon dispersion law is
modified according to (1). The reason is that the method of distributing spacetime defects in the latter reference is not
Lorentz-invariant, leading to the spatial dimensions behaving differently to the time dimension. This corresponds to
a preferred direction in spacetime and results in a modified photon dispersion law of the form of Eq.~\eqref{eq:modified-photon-dispersion-scaleless},
since in the photon model there is no quantity having a mass dimension.

On the contrary, cases (1) and (2) cannot play a role within the spacetime foam considered in the current article. In the action
\eqref{eq:action-modified-theory-complete} of the effective theory no preferred spacetime directions appear. Furthermore, no such
directions emerge, since the defects are distributed in a Lorentz-invariant way. As a result of this time and spatial dimensions
behave the same. Thus only the dimensionless quantity $b_{\varrho}k=b_{\varrho}(k_0^2-\mathbf{k}^2)^{1/2}$ appears in the equation
leading to the modified photon dispersion law. So it must be of the form $h(b_{\varrho}k)=0$, where $h$ is contains the expression
of the quantum correction involving a scattering with the defects. In our case $h$ corresponds to the left-hand side of 
Eq.~\eqref{eq:dispersion-relation-of-shell}. Assuming $b_{\varrho}^2k^2\ll 1$, which is reasonable within perturbation theory, 
$h(b_{\varrho}k)$ can be expanded with respect to its argument $b_{\varrho}k$ to quadratic order:
\begin{equation}
\label{eq:emergence-photon-mass2}
h(0)+b_{\varrho}kh'(0)+\frac{1}{2}b_{\varrho}^2k^2h''(0) \simeq 0 \Rightarrow k_0^2\simeq \mathbf{k}^2-\frac{2}{b_{\varrho}^2}\frac{h(0)}{h''(0)}\,,
\end{equation}
where we have used that $h'(0)=0$.\footnote{Avoiding branch cuts by infinitesimal imaginary parts at the appropriate places, perturbative
corrections are analytic functions in $k^2$, and so do not depend on $k$.} This would lead to a nonvanishing photon mass indicating a
violation of gauge invariance. Hence, the only modification may be according to (3). However the photon mass in Eq.~\eqref{eq:emergence-photon-mass2}
is proportional to the scalar mass $1/b_{\varrho}$, which is not a small perturbation when $1/b_{\varrho}$ is large.
Considering the left-hand side of Eq.~\eqref{eq:dispersion-relation-of-shell} as the function $h$ we, in fact, obtain the result of
Eq.~\eqref{eq:dispersion-law-photon-modified}. Hence even without a graphical analysis, which is not possible for a general function $h$, the
argument given is still valid. We conjecture that in such a case, where the photon interacts with the defects via a quantum correction 
with a scalar particle, its dispersion relation remains conventional.

What is the physical reason ruling out a modification (3)? Because of $k^2\widehat{\Pi}_{\mathrm{ren}}(k^2)=0$ for $k^2\mapsto 0$ (see
Eq.~\eqref{eq:gauge-invariance-modified-theory} gauge invariance is indeed conserved, although the action~\eqref{eq:action-effective-theory} seemed
to violate gauge invariance (with respect to the gauge transformation $\phi\mapsto -\phi$) explicitly. However, the sign change of the penultimate
term of \eqref{eq:action-effective-theory} can be absorbed into the defect charges $\varepsilon$. For an infinite number of defects this does not
change anything since there are equally many defects with $\varepsilon=1$ and $\varepsilon=-1$. Also for the last term in the action the sign can
be absorbed into the coupling constant $\lambda$. Since the correction computed is proportional to $\lambda^2$ this sign change has no physical
implications.

\section{Momentum transfer from particles to defects}
\label{sec:energy-transfer-to-defects}
\setcounter{equation}{0}

The final result of Sec.~\ref{sec:dispersion-relation-photon-modified} was obtained after making a series of assumptions in the course of the
calculation. Let us recap these assumptions:

\begin{itemize}

\item Ass.~(1): The interaction of the photon with the defects is mediated by a scalar field according to the effective theory of
Eq.~\eqref{eq:action-modified-theory-complete}. This forms the basis of the simple model proposed in this article and is, therefore, the most
important assumption.
\item Ass.~(2a) and (2b) + (2c) + (2d): Isotropic and homogenous (``random'') defect distribution (Sec.~\ref{sec:sprinkling}) +
dense defect distribution (Sec.~\ref{sec:auxiliary-functions}) + infinite spacetime volume (Sec.~\ref{sec:auxiliary-functions}). These have been
introduced not only in order to keep the calculation feasible, but also for physical reasons. In principle one or several of them can be
dropped leading to a more difficult calculation.
\item Ass.~(3): The modified photon momentum squared is supposed to be much smaller than the mass of the scalar field squared, i.e.
$k^2\ll 1/b_{\varrho}^2$. This assumption has been made only for physical reasons. The mass of the postulated scalar field $1/b_{\varrho}$ is
expected to be large and the deviation of the photon momentum squared from the standard result $k^2=0$ should be a small perturbation.

\end{itemize}
So far the momentum transfer from the photon to the defects and vice versa has been neglected. The reason for this is Ass.~(2) and
Eq.~\eqref{eq:momentum-conservation-defect-vertex-2} that followed from it.

Now we would like to investigate how the results obtained change when momentum is transferred from and to the defects. This means that we drop
Ass.~(2a), (2b), (2c) or (2d). The technical problem is that the free photon field in
Eq.~\eqref{eq:photon-field-equation-second-order} then depends on the integration momentum $p$, which renders the evaluation of the corresponding
integral impossible. However we will stick with Ass.~(1) and treat the propagation of photons through a spacetime foam via
the effective theory defined by Eq.~\eqref{eq:action-effective-theory}. Thus the external photon momentum is assumed to be much smaller than the
Planck scale.

On the one hand, the $\phi$-field with its renormalized mass $1/b$ directly interacts with the spacetime defects. We assume that $\phi$ will only
probe the defects if $1/b$ is of the order of the Planck scale. For $1/b$ much smaller than the Planck scale a momentum transfer to the defects
will be low at most, i.e. suppressed by that scale. As a result, the main contribution of the $p$-integral will come from the region $p\simeq k$.
The difference between $p$ and $k$ must be suppressed by a small dimensionless number that can be written as a ratio of two mass scales. The scale
where the influence of any spacetime foam may become especially important is the Planck mass $M_{\mathrm{Pl}}$. Since the $\phi$-field is assumed to
interact with the defects, the only other scale is the mass $1/b$.

On the other hand, the interaction of the photon with the spacetime defects is mediated via a quantum correction involving the virtual $\phi$-field
with the new mass $1/b_{\varrho}$ originating from the interaction of $\phi$ with the defects. As a result, the photon acquires a size from this
quantum correction that is inversely proportional to $1/b_{\varrho}$. Hence we assume the suppression of a momentum transfer from the photon to the
defects to be similar as for the scalar field, but with the mass $1/b$ replaced by $1/b_{\varrho}$. This behavior will be summarized in the following
paragraph.

\subsection{Assumption (4): Momentum transfer suppressed by the Planck scale}

In order to be able to compute the integral over $p$ in Eq.~\eqref{eq:photon-field-equation-second-order}, we introduce a fourth assumption. We
assume that the ratios
\begin{equation}
\label{eq:suppression-momentum-transfer-phi}
(k-p)^2\sim \frac{1}{b^3M_{\mathrm{Pl}}}\,,\quad (k-p)\cdot x\sim \frac{1}{bM_{\mathrm{Pl}}}k\cdot x\,,
\end{equation}
approximately hold for the momentum transfer $k-p$ of a low-energy $\phi$-field with initial momentum $k$ and final momentum $p$ to a spacetime
defect. Analogous ratios hold for the momentum transfer of a photon to a defect with the difference that $1/b$ must be replaced by $1/b_{\varrho}$
\begin{equation}
\label{eq:suppression-momentum-transfer-photon}
(k-p)^2\sim \frac{1}{b_{\varrho}^3M_{\mathrm{Pl}}}\,,\quad (k-p)\cdot x\sim \frac{1}{b_{\varrho}M_{\mathrm{Pl}}}k\cdot x\,.
\end{equation}
In Eqs.~\eqref{eq:suppression-momentum-transfer-phi}, \eqref{eq:suppression-momentum-transfer-photon} $x$ is an arbitrary but fixed four-vector in
configuration space.

\subsection{Defect distribution with large separation between individual defects}

\begin{figure}[t]
\centering
\subfigure[\,\,\,dense distribution]{\includegraphics[scale=0.75]{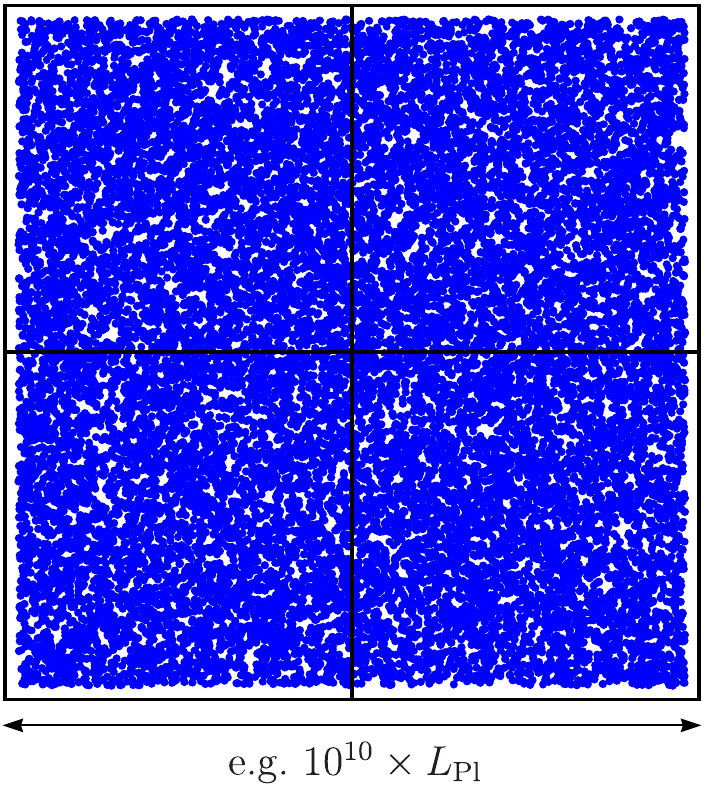}}
\hspace{4cm}\subfigure[\,\,\,distribution not being dense]{\includegraphics[scale=0.75]{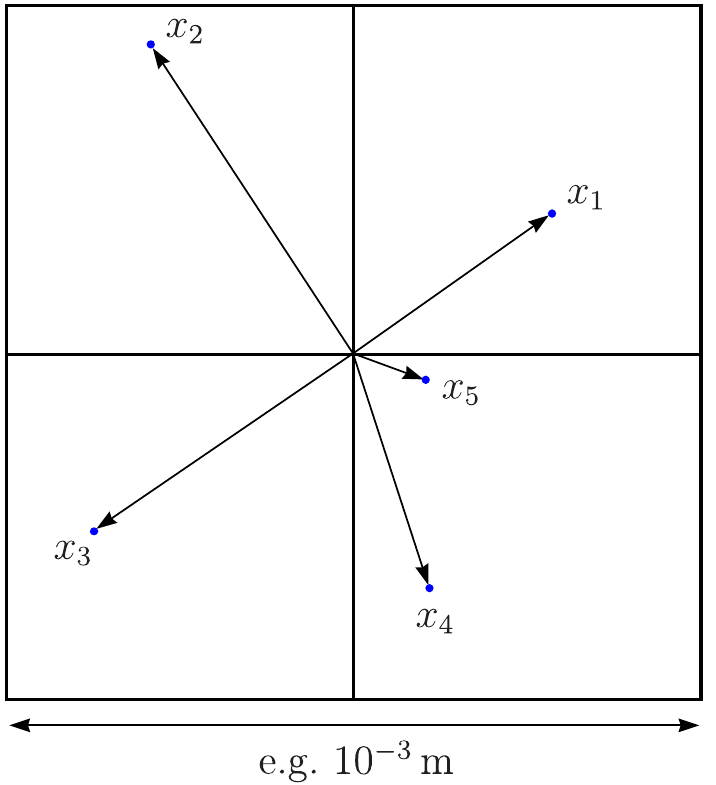}}
\caption{Each panel in the current figure shows a distribution of spacetime defects. The scales of the panels are chosen in order to
make clear that the distances between the defects in the left square correspond to multiples of the Planck scale, whereas in the right
square they are macroscopic. Hence the right panel illustrates a foam in which individual defects are separated by distances that lie many
orders of magnitude above the typical distances in the left panel.}
\label{fig:dense-defect-distribution}
\end{figure}

In Eq.~\eqref{eq:sum-of-random-phases} we assumed the distribution of spacetime defects to be dense (see the left panel of
Fig.~\ref{fig:dense-defect-distribution}). In this case the defect distribution can be approximated by an effective background field
and the sum over all defects results in a spatial integral.
Discarding Ass.~(2c) means that the sum mentioned can no longer be approximated by such an integral and we cannot define a background field, any more.
As a result, the $\delta$-function in Eq.~\eqref{eq:sum-of-random-phases}, which helps to get rid of the second integral in
Eq.~\eqref{eq:photon-field-equation-second-order}, has to be replaced by a sum again. However, due to Ass.~(4) the momentum transfer is suppressed by
the Planck scale and the integration momentum $p$ does not appear
\begin{equation}
\label{eq:delta-function-replacement-rule}
\int \mathrm{d}^4p\,\varrho\delta(k-p)f(p)\mapsto \frac{1}{\mathcal{V}}\sum^{\mathcal{N}}_{i=1}\exp\left(\mathrm{i}\frac{1}{bM_{\mathrm{Pl}}}k\cdot x_i\right)f(k)\,.
\end{equation}
The mass of the $\phi$-field given by Eq.~\eqref{eq:phi-field-modified-mass} then changes as follows
\begin{equation}
\frac{1}{b^2} \mapsto \frac{1}{b^2}+\frac{b^2}{\mathcal{V}}\sum^{\mathcal{N}}_{i=1} \exp\left(\mathrm{i}\frac{1}{bM_{\mathrm{Pl}}}k\cdot x_i\right)\equiv \frac{1}{b(k,x_i)^2}\,.
\end{equation}
It now explicitly depends on the $\phi$-momentum $k$ and the defect positions $x_i$ (see the right panel of Fig.~\ref{fig:dense-defect-distribution}).
In the loop integral of Eq.~\eqref{eq:definition-bnugamma} the shifted mass has to be inserted. Then the integral depends on the defect positions
$x_i$. This can be stated as follows
\begin{subequations}
\begin{align}
B^{\nu}_{\phantom{\nu}\gamma}(k) \mapsto B^{\nu}_{\phantom{\nu}\gamma}(k,x_i)&=\sum^{\mathcal{N}}_{i=1} \int \frac{\mathrm{d}^4q}{(2\pi)^4}\,\exp\left(\mathrm{i}\frac{1}{b(q,x_i)M_{\mathrm{Pl}}}k\cdot x_i\right) \notag \\
&\phantom{{}={}\sum^{\mathcal{N}}_{i=1} \int\frac{\mathrm{d}^4q}{(2\pi)^4}}\,\times \left[\frac{1}{b(q,x_i)^2}\widetilde{H}(q,x_i)\right]^2\frac{1}{(k-q)^2+\mathrm{i}\epsilon}K^{\nu}_{\phantom{\nu}\gamma}\,,
\end{align}
with
\begin{equation}
\frac{1}{b(q,x_i)^2}\widetilde{H}(q,x_i)=\frac{-1}{q^2-1/b^2(q,x_i)+\mathrm{i}\epsilon}\,,
\end{equation}
\end{subequations}
and $K^{\nu}_{\phantom{\nu}\gamma}$ of Eq.~\eqref{eq:definition-knugamma}.

So the result of the loop integral $B^{\nu\gamma}(k)$ will also depend on the positions $x_i$ of the defects. Because of this its tensor structure
may involve terms such as $x_i^{\nu}x_i^{\gamma}$, $x_i^{\nu}k^{\gamma}$ etc. and they will spoil the form of the standard tensor structure
$(k^{\nu}k^{\gamma}-\eta^{\nu\gamma}k^2)$. This leads to a dispersion relation of the photon that differs from $k^2=0$.

\subsection{Defect distribution in a finite spacetime volume}

Now we stick to Ass.~(2c) but drop Ass.~(2d) and consider defects in a spacetime with finite volume $\mathcal{V}$. For computational reasons the
shape of the volume shall be a four-dimensional cube\footnote{with side length $\mathscr{R}$ and centered at the origin} containing $\mathcal{N}$ 
defects. Because of the finiteness of $\mathcal{V}$ the positions $x_i$ are bounded and the argument of the exponential function can be
assumed to be small. So we expand the complex exponential function in Eq.~\eqref{eq:delta-function-replacement-rule} with respect to the (small)
momentum transfer. In the remainder $f(p)$ of the integral over $p$ in Eq.~\eqref{eq:photon-field-equation-second-order} we effectively replace
all $p$ by $k$ and drop the integral over $p$.

This procedure leads to a new replacement rule for the $\delta$-function
\begin{subequations}
\begin{equation}
\int \mathrm{d}^4p\,\varrho\delta(k-p)f(p) \mapsto \frac{1}{\mathcal{V}}\left(\mathcal{N}+\sum_{n=1}^{\infty} S_n\right)f(k)\,,
\end{equation}
\begin{equation}
\label{eq:delta-function-replacement-nth-term}
S_n=\frac{\mathrm{i}^n}{n!}(k-p)_{\mu_{\scaletoheight{0.11cm}{1}}}(k-p)_{\mu_{\scaletoheight{0.11cm}{2}}}\dots (k-p)_{\mu_n}\sum_{i=1}^\mathcal{N} x_i^{\mu_{\scaletoheight{0.11cm}{1}}}x_i^{\mu_{\scaletoheight{0.11cm}{2}}}\dots x_i^{\mu_n}\,.
\end{equation}
\end{subequations}
Since we assume the treatment of the spacetime defect distribution to be a good approximation the sum in
Eq.~\eqref{eq:delta-function-replacement-nth-term} can be replaced by integrals. These are restricted to the finite spacetime volume $\mathcal{V}$.
\begin{subequations}
\begin{equation}
\int \mathrm{d}^4p\,\varrho\delta(k-p)f(p) \mapsto \varrho\left(1+\sum_{n=1}^{\infty} I_{n}\right)f(k)\,,
\end{equation}
\begin{equation}
I_n=\frac{\mathrm{i}^n}{n!\mathcal{N}\mathcal{V}}(k-p)_{\mu_{\scaletoheight{0.11cm}{1}}}(k-p)_{\mu_{\scaletoheight{0.11cm}{2}}}\dots (k-p)_{\mu_n} \int_{\mathcal{V}} \mathrm{d}^4x\,x^{\mu_{\scaletoheight{0.11cm}{1}}}x^{\mu_{\scaletoheight{0.11cm}{2}}}\dots x^{\mu_n}\,.
\end{equation}
\end{subequations}
It is evident that this expansion only makes sense when the integrals run over a finite space, otherwise the integrals would be divergent as the 
integrands are not suppressed for $x\mapsto \infty$.

If we also keep Ass.~(2a) and (2b) for the moment we end up with a ``random distribution'' of spacetime defects. This means that no preferred
direction is defined by the distribution. In this case all integrals that are odd in $x$ vanish, $I_{2k-1}=0$ for $k\in \{1,2,\dots\}$, since there
is no counterpart that could make up the index structure of the result.

The tensor structure of the even integrals can be generated by the metric tensor only. That is why the even integrals do not vanish but depend
on combinations of metric tensors. For example, in four spacetime dimensions the integral with two indices is given by
\begin{equation}
\int \mathrm{d}^4x\,x^{\mu}x^{\nu}=\frac{\eta^{\mu\nu}}{4}\int \mathrm{d}^4x\,x^2=-\frac{\eta^{\mu\nu}}{24}\mathscr{R}^2\mathcal{V}\,.
\end{equation}
Neglecting constant prefactors, the behavior of integrals that are even in $x$ is as follows
\begin{equation}
\label{eq:order-result-integral-2k}
I_{2k}\sim \frac{1}{\mathcal{N}}\left(\frac{\sqrt{\mathcal{V}}}{b^3M_{\mathrm{Pl}}}\right)^k\,, \quad k\in\{1,2,\dots\}\,.
\end{equation}
So further terms in the expansion are suppressed by powers of ratios of the Planck scale and the square root of the spacetime volume. Finally
Eq.~\eqref{eq:phi-field-modified-mass} has to be replaced by
\begin{equation}
\frac{1}{b^2} \mapsto \frac{1}{b^2}+\varrho b^2+\frac{b^2}{\mathcal{V}}\sum_{k=1}^{\infty} C_{2k}\left(\frac{\sqrt{\mathcal{V}}}{b^3M_{\mathrm{Pl}}}\right)^{k}\equiv \frac{1}{b^2}\,,
\end{equation}
where $C_{2k}$ are mere numbers. Since the new mass of $\phi$ neither involves any preferred directions nor any defect positions $x_i$ the
resulting $B^{\nu\gamma}$ of Eq.~\eqref{eq:definition-bnugamma} will still have the gauge-invariant tensor structure
$(k^{\nu}k^{\gamma}-\eta^{\nu\gamma}k^2)$. Although the shape of a finite spacetime volume is not Lorentz-invariant, the photon is not 
affected by the finiteness of the spacetime. The reason is that we do not set any boundary conditions on the photon field in the framework 
of the simple spacetime foam model considered. Thus the dispersion relation of the photon stays $k^2=0$.

\subsection{Anisotropic or inhomogeneous distribution}

\begin{figure}[b]
\centering
\subfigure[\,\,\,one preferred spacetime direction]{\includegraphics[scale=0.75]{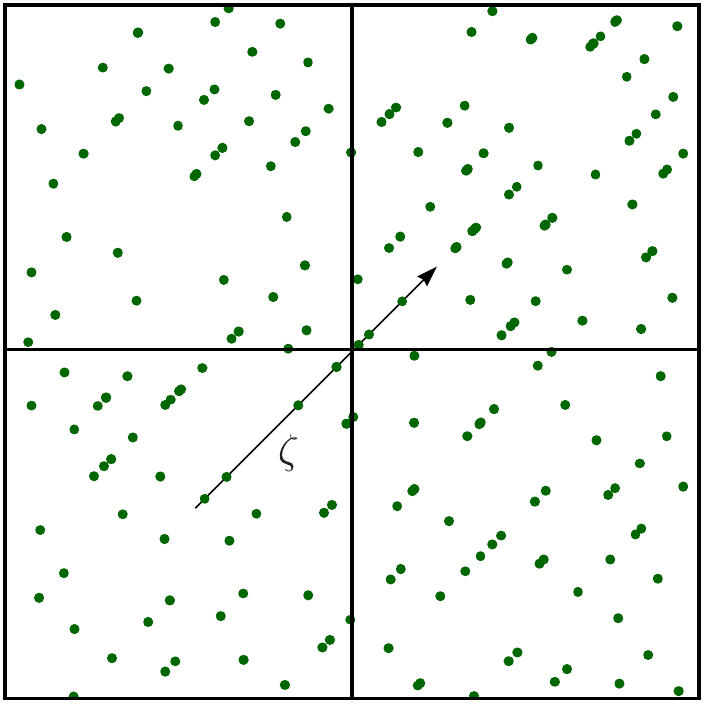}}
\hspace{2cm}\subfigure[\,\,\,two preferred spacetime directions]{\includegraphics[scale=0.75]{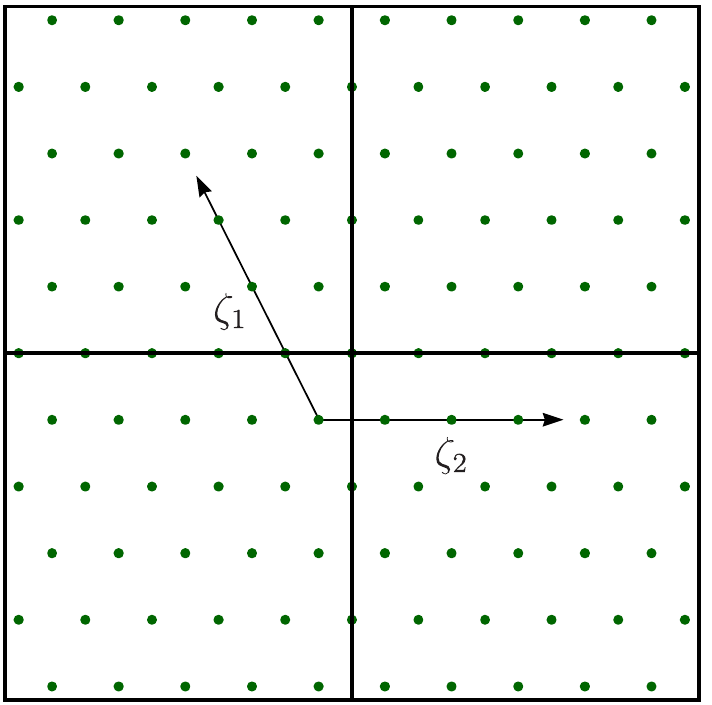}}
\hspace{2cm}\subfigure[\,\,\,inhomogeneous distribution]{\includegraphics[scale=0.75]{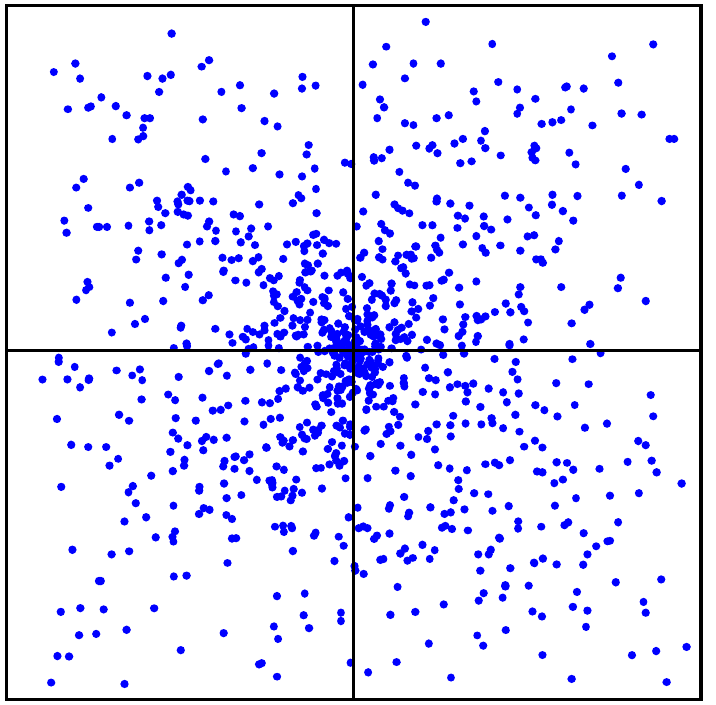}}
\caption{Here, the first panel illustrates an anisotropic distribution defining a single preferred direction $\zeta$, whereas the second panel
contains a regular lattice of defects defining two preferred directions $\zeta_1$ and $\zeta_2$. The third panel depicts a section of an
inhomogeneous defect distribution, where in the center of the region shown the density is higher than near the margin.}
\label{fig:anisotropic-defect-distribution}
\end{figure}

If we additionally drop Ass.~(2a), the spacetime defect distribution may be anisotropic and, therefore, define preferred directions in
spacetime (see the first two panels of Fig.~\ref{fig:anisotropic-defect-distribution}). Let us assume that there is one such direction:
$(\zeta^{\mu})=(\zeta^0,\zeta^1,\zeta^2,\zeta^3)$.
Then the result of all spatial integrals will involve $\zeta$. For example the result for the integral with two indices is then made
up of the metric tensor and the tensor product of preferred directions
\begin{subequations}
\begin{equation}
\int \mathrm{d}^4x\,x^{\mu}x^{\nu}=-\frac{\mathscr{R}^2\mathcal{V}}{36}\left(C_2^{(0)}(\zeta)\eta^{\mu\nu}+C_2^{(2)}(\zeta)\zeta^{\mu}\zeta^{\nu}\right)\,,
\end{equation}
\begin{equation}
C_2^{(0)}(\zeta)=\frac{\zeta^2+2(\zeta^0)^2}{\zeta^2}\,,\quad C_2^{(2)}(\zeta)=\frac{2\big[\zeta^2-4(\zeta^0)^2\big]}{\zeta^4}\,.
\end{equation}
\end{subequations}
Using Ass.~(4), the second integral in the expansion can be written as follows
\begin{align}
I_2&\sim \frac{\sqrt{\mathcal{V}}}{\mathcal{N}}\left(C_2^{(0)}(\zeta)(k-p)^2+C_2^{(2)}(\zeta)[(k-p)\cdot \zeta]^2\right) \notag \\
&\simeq \frac{1}{\mathcal{N}}\frac{\sqrt{\mathcal{V}}}{b^3M_{\mathrm{Pl}}}\left[C_2^{(0)}(\zeta)+C_2^{(2)}(\zeta)(k\cdot \zeta)^2\frac{b}{M_{\mathrm{Pl}}}\right]\,.
\end{align}
It can be shown that despite the existence of a preferred spacetime direction the integrals $I_{2k+1}$ still vanish because they are odd
with respect to the integration variable. Since the tensor structure of each even integral can involve at the most $2k$ preferred directions
$\zeta$, we obtain the following general behavior for the new mass of the $\phi$-field
\begin{align}
\frac{1}{b^2}&\mapsto \frac{1}{b^2}+\varrho b^2 \notag \\
&\phantom{{}\mapsto{}\frac{1}{b^2}}+\frac{b^2}{\mathcal{V}}\sum^{\infty}_{k=1} \left(\frac{\sqrt{\mathcal{V}}}{b^3M_{\mathrm{Pl}}}\right)^{k}\sum_{l=0}^{k} C_{2k}^{(l)}(\zeta)(k\cdot \zeta)^{2l}\left(\frac{b}{M_{\mathrm{Pl}}}\right)^l\equiv \frac{1}{b(k,\zeta)^2}\,,
\end{align}
where $C_{2k}^{(l)}(\zeta)$ are functions with respect to the preferred direction $\zeta$. The loop integral now depends on $\zeta$ and its
tensor structure may involve terms such as $\zeta^{\nu}\zeta^{\gamma}$, $\zeta^{\nu}k^{\gamma}$ etc., which destroy the standard structure
$(k^{\nu}k^{\gamma}-\eta^{\nu\gamma}k^2)$. This case may also produce a deviation from $k^2=0$ in the dispersion relation of the photon. For
example, the appearance of a preferred spacetime direction is the reason for the modification computed in \cite{hep-th/0312032}.

Although the calculation presented here has been performed for a finite volume there is nothing to suggest that the physical effect originates from
this fact. We have just seen that a finite volume does not have any effect on the dispersion relation of the photon so long as
the defect distribution is isotropic and homogeneous.

Keeping Ass.~(2a), but dropping (2b) would result in an inhomogeneous defect distribution (see the third panel of
Fig.~\ref{fig:anisotropic-defect-distribution}). This means that the defect density cannot be considered
as constant and, therefore, it cannot be pulled in front of the integral in Eq.~\eqref{eq:sum-of-random-phases}. There is no physical input for a
varying density function $\varrho=\varrho(x)$, since we are not aware of a mechanism resulting in an inhomogeneous distribution of defects.
The resulting integral in Eq.~\eqref{eq:sum-of-random-phases} will certainly be much more complicated if $\varrho$ is a
function of $x$. If there are regions with a high (constant) density $\varrho_h$ and regions with a low (constant) density $\varrho_l$ such that
$\varrho_h\gg \varrho_l$, the main contributions from the integral will come from the regions with $\varrho(x)=\varrho_h$. Effectively, this means
that the integration volume is reduced. How the photon dispersion relation is affected by such changes, will not be further investigated in this
paper.

\section{\textit{PT}-symmetric extension and the percolation of defects}
\label{sec:pt-symmetric-extension}
\setcounter{equation}{0}

In this section we would like to deliver a brief discussion on an interesting issue. Replacing the real coupling constant $\lambda^{(0)}$ in the
action \eqref{sec:action-effective-theory} by an imaginary one we find that the equation
\begin{equation}
k^2[1+\mathrm{i}^2\mathcal{C}\,\widehat{\Pi}_{\mathrm{ren}}(k^2)]=0
\end{equation}
has a second physically acceptable solution different from the standard dispersion relation:
\begin{equation}
\label{eq:photon-dispersion-relation-pt-symmetric-theory}
k^2=\alpha(\gamma)\frac{1}{b_{\varrho}^2}\,,
\end{equation}
with a function $\alpha(\gamma)$ and $\gamma\propto\lambda^2\varrho$. Equation \eqref{eq:photon-dispersion-relation-pt-symmetric-theory}
means that the photon becomes massive.
\begin{table}[t]
\centering
\begin{tabular}{c|c|c|c||c||c}
\hline
object & \textit{C} & \textit{P} & \textit{T} & \textit{PT} & \textit{CPT} \\
\hline
\hline
vector potential $A^{\mu}$ & $-A^{\mu}$ & $A_{\mu}$ & $A_{\mu}$ & $A^{\mu}$ & $-A^{\mu}$ \\
field strength tensor $F^{\mu\nu}$ & $-F^{\mu\nu}$ & $F_{\mu\nu}$ & $F_{\mu\nu}$ & $F^{\mu\nu}$ & $F^{\mu\nu}$ \\
dual field strength tensor $\widetilde{F}^{\mu\nu}$ & $-\widetilde{F}^{\mu\nu}$ & $-\widetilde{F}_{\mu\nu}$ & $\widetilde{F}_{\mu\nu}$ & $-\widetilde{F}^{\mu\nu}$ & $\widetilde{F}^{\mu\nu}$ \\
four-momentum $k^{\mu}$ & $k^{\mu}$ & $k_{\mu}$ & $k_{\mu}$ & $k^{\mu}$ & $k^{\mu}$ \\
imaginary unit $\mathrm{i}$ & $\mathrm{i}$ & $\mathrm{i}$ & $-\mathrm{i}$ & $-\mathrm{i}$ & $-\mathrm{i}$ \\
\hline
\end{tabular}
\caption{Transformation properties of the vector potential $A^{\mu}$, field strength tensor $F^{\mu\nu}$, dual field strength tensor
$\widetilde{F}^{\mu\nu}$, four-momentum $k^{\mu}$, and the imaginary unit with respect to \textit{P}, \textit{C}, \textit{T}, and their
combinations \textit{PT} and \textit{CPT}.}
\label{tab:tabular-cpt}
\end{table}

This solution will exist if $\gamma$ is larger than a critical value $\gamma_c$. Furthermore $\alpha$ seems to lie in the interval $[0,1]$.
The replacement $\lambda^{(0)}\rightarrow \mathrm{i}\lambda^{(0)}$ makes the interacting part of the Lagrangian non-Hermitian. However
this may not be a problem so long as the Lagrangian is symmetric under a combined parity transformation \textit{P} and time reversal
transformation \textit{T}: $(\mathit{PT})\mathcal{L}=\mathcal{L}$. Non-Hermitian but \textit{PT}-symmetric quantum mechanics has been
thoroughly studied over the past few years \cite{Bender:1998ke}. It was found that it can serve as a description of real physical systems,
see e.g. \cite{Rubinstein:2007,Lin:2011,Bittner:2012}. \textit{PT}-symmetric quantum field theories have not been investigated that
profoundly. Nevertheless there are indications that such extensions of ordinary Hermitian theories are physically meaningful~\cite{Bender:2004sa}.
According to Table~\ref{tab:tabular-cpt} the last term of the action \eqref{eq:action-effective-theory} is indeed \textit{PT}-symmetric for
imaginary coupling constant $\lambda^{(0)}$.

\begin{figure}[t]
\centering
\includegraphics[scale=0.4]{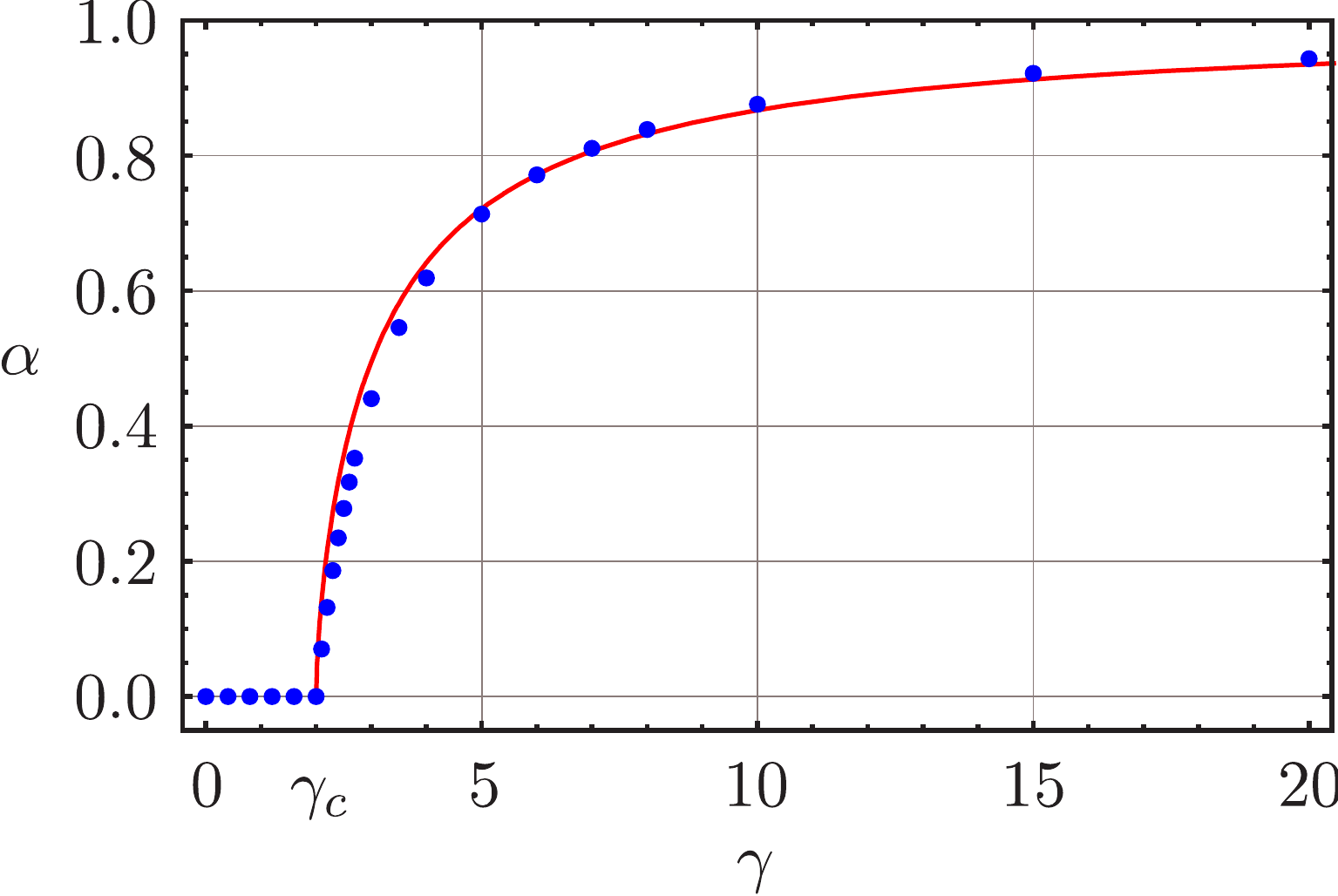}
\caption{Several data points obtained for the coefficient $\alpha(\gamma)$ in Eq.~\eqref{eq:photon-dispersion-relation-pt-symmetric-theory} are
compared with the expected behavior for the order parameter: $\alpha(\gamma)\propto (1-\gamma_c/\gamma)^{\beta}$. The exponent used is $\beta=0.64$
which is also obtained in the context of a four-dimensional lattice percolation. From the plot it becomes evident that $\gamma_c\approx 2$. The
latter point depends on the renormalization scale we choose.}
\label{fig:percolation-phase-transition}
\end{figure}
Observing the relation between $\gamma$ and $\varrho$ we speculate that this solution could be related to the percolation of defects.
Percolation theory studies the formation and properties of clusters of objects randomly distributed over a lattice or a continuous space.
This theory exhibits a phase transition. As the density of the objects increases so does the mean size of the clusters, and it becomes
infinitely large at some critical density $\varrho_c$. For densities smaller than $\varrho_c$ only clusters of finite size exist, whilst
for densities larger than $\varrho_c$ clusters of infinite size also appear. At the critical point the system is scale invariant.

This assumption is supported by the observation that the behavior of $\alpha(\gamma)$ is well described by the critical exponent $\beta$ of
the percolation phase transition in 4 dimensions (cf. Fig.~\ref{fig:percolation-phase-transition}). This issue deserves further study.

Here we see that the general argument about the occurrence of a photon mass given at the end of Sec.~\eqref{sec:general-remarks-about-results}
does not hold. This is because the argument relied on a photon mass continuously depending on the model parameters, which is not the case here
(as can be seen in Fig.~\ref{fig:percolation-phase-transition}).

\section{Summary and conclusions}
\label{sec:summary-and-conclusions}
\setcounter{equation}{0}

To summarize, we have investigated photon propagation through a spacetime foam made up of a Lorentz-invariant distribution of time-dependent,
pointlike defects. From the action of Eq.~\eqref{eq:action-modified-theory-complete}, photons do not interact directly with the defects but
via a scalar mediator field.

If the spacetime volume is assumed to be infinite and the distribution of defects is dense, isotropic, and homogeneous, the only effect on
the propagation of the scalar mediator is that of enlarging its mass. The defects then effectively act as a modified background analogous to
the pictorial representation of a particle getting its mass via the interaction with the Higgs field vacuum. So for the scalar field the
defects have been ``integrated out'' and the photon interacts with a scalar that has a mass larger than when compared to the case without defects.

The surprising result is that the photon dispersion law remains standard: $k^2=k_0^2-|\mathbf{k}|^2=0$. Hence the photon does not ``feel''
the random background field that mimics the defects as an effective theory --- at least at leading order in the interaction between the photon and
the scalar field. The reason for this is that on the one hand no preferred spacetime directions will appear if the defects are distributed in a
Lorentz-invariant manner. On the other hand the photon does not become massive as gauge invariance is not violated. In order to rule out the
appearance of a gauge anomaly the perturbative correction to the photon dispersion law must indeed be computed, which leads to the nontrivial
result obtained.

The outcome of the calculation sheds some new light on the very restrictive bounds on Lorentz-violating parameters of the Standard Model Extension
\cite{Kostelecky:2008ts}. If a nontrivial spacetime structure is assumed to be the underlying cause for a possible Lorentz violation, low-energy
experiments are unlikely to detect Lorentz violation if at energies much smaller than the Planck energy this structure can be described by the
effective theory considered.

However one also has to keep in mind that the photon dispersion relation continues to be the conventional one because of several idealized assumptions
taken to ensure a feasible computation. Using physical principles we have tried to predict the effect on the photon dispersion relation when one of the
assumptions is dropped. Discarding the assumption of \dots

\begin{itemize}

\item \dots the action of
  Eq.~\eqref{eq:action-modified-theory-complete}: The model is no longer an appropriate description as the defect structure itself
  is expected to become important. Furthermore, the theory could be modified such that the scalar field is discarded altogether and the
photon be made to interact directly with the defects. Then it would probe the defects with its wavelength. If we still assume the momentum transfer
$(\Delta k)^2$ to the defects to be suppressed by one power of the Planck scale, the photon energy $k_0$ has to appear as a second dimensionful scale.
Then e.g.
\begin{equation}
(\Delta k)^2\sim \frac{k_0^3}{M_{\mathrm{Pl}}}\,.
\end{equation}
Because of $k_0=\xi\cdot k$ with $\xi=(1,0,0,0)$, a preferred spacetime direction comes into play which may lead to a violation of Lorentz invariance.
\item \dots an infinite spacetime volume: In this case the dispersion relation of the photon stays the same since gauge invariance is not violated.
Note that in the simple model proposed no boundary conditions are set on any field and, therefore, the photon does not feel the finite volume. However,
if the sprinkling procedure changes because of the finite volume the photon dispersion relation might also change. Due to calculational
difficulties we have not been able to examine this case.
\item \dots a dense distribution of defects: If the defects are assumed to be separated by a large distance, their distribution can no longer be
considered to be continuous. The defect positions $x_i$ will appear explicitly in the tensor structure of the photon field equation. This may
lead to a modified dispersion relation of the photon that involves these positions.
\item \dots an isotropic and homogeneous (random) defect distribution: If the distribution is dense but anisotropic it defines a preferred direction
$\zeta$ in spacetime. Then $\zeta$ also shows up in the tensor structure of the photon field equation. This may lead to a modified dispersion relation
of the photon and, therefore, Lorentz violation.
\item \dots the photon momentum squared $\ll$ mass of scalar particle squared: The photon gets a mass and electromagnetic waves are damped. We discard
this case as it does not correspond to physical reality for photons with
an energy much smaller than the Planck energy. Otherwise we would not be
able to observe
light from distant galaxies.

\end{itemize}
It has not been possible to derive the modified photon dispersion relation for the more complicated cases mentioned above. This may be done in a future
research project. Besides that, we could not rigorously demonstrate the influence of a finite spacetime volume as we did not impose any boundary
conditions on the photon field. 

The final results are summarized in Table~\ref{tab:summary-final-results}.
\begin{table}[t]
\centering
\begin{tabular}{cccccccc}
\hline
keep & (1)        & (2a), (2b) & (2c)       & (2d)       & (3)        & (4)        & Lorentz violation \\
\hline
\hline
     & \checkmark & \checkmark & \checkmark & \checkmark & \checkmark & not needed & absent \\
     & \checkmark &            & \checkmark & \checkmark & \checkmark & \checkmark & \checkmark \\
     & \checkmark & \checkmark &            & \checkmark & \checkmark & \checkmark & \checkmark \\
     & \checkmark & \checkmark & \checkmark &            & \checkmark & \checkmark & (?) \\
     & \checkmark & \checkmark & \checkmark & \checkmark &            & not needed & massive photon \\
\hline
\end{tabular}
\caption{The fate of Lorentz invariance where, in short, the assumptions taken are listed as follows. I: effective theory of
Eq.~\eqref{eq:action-modified-theory-complete}, (2a)/(2b): homogeneous/isotropic defect distribution, (2c): dense defect distribution,
(2d): infinite spacetime volume, (3): modified photon momentum square $\ll$ mass of scalar field square, (4): momentum transfer from
particles to defects suppressed by Planck scale.}
\label{tab:summary-final-results}
\end{table}

\section*{Acknowledgments}

It is a pleasure to thank F.~R.~Klinkhamer for helpful discussions. The authors acknowledge support by the German Research Foundation
(\textit{Deutsche Forschungsgemeinschaft, DFG}) within the grant KL 1103/2-1.

\begin{appendix}
\numberwithin{equation}{section}

\section{How good is the assumption of a dense defect distribution?}
\label{sec:table-dense-distribution}

In this section we consider four-dimensional Minkowski space with dimensionless coordinates $t^{\mu}$. The numbers
\begin{equation}
\label{eq:numbers-dense-distribution}
w_1=\sum_{k=1}^{n} \exp(\mathrm{i}t_k^{\mu}1_{\mu})\,,\quad w_2=\varrho \int_{\mathbb{H}} \exp(\mathrm{i}t^{\mu}1_{\mu})\,,
\end{equation}
shall be computed, where $(1^{\mu})=(1,1,1,1)$. The sum in $w_1$ runs over $n$ defects contained in a four-dimensional unit-hypercube $\mathbb{H}$ whose edges 
are supposed to lie parallel to the axes of the coordinate system. The space diagonal of the hypercube runs from the point $(0,0,0,0)$ to $(1,1,1,1)$. The defects 
are assumed to lie at equal distances $\Delta$. Hence, the hypercube contains $(\Delta^{-1}+1)^4$ defects leading to the density $\varrho=(\Delta^{-1}+1)^4$. The 
integral in $w_2$ runs over the same hypercube, as well. We compare the values $w_1$ to $w_2$ for different defect spacings $\Delta$. In principle $w_1$ corresponds
to $w_2$ for infinitesimal defect separation. In Table \ref{tab:separation-between-defects} we see that the integral is already a good approximation for the 
dimensionless distance $\Delta=0.01$. 

We can now replace the dimensionless scalar product $t^{\mu}1_{\mu}$ by the scalar product of two dimensionful quantities: the wave vector $k^{\mu}$ and the spatial 
four-vector $x_{\mu}$. Then $\Delta$ corresponds to a product of a wave vector $K$ and a distance $\Delta X$ in configuration space. Assuming the photon energy 
$E=1\,\mathrm{TeV}$ the dimensionless distance $\Delta$ is in accordance with the following $\Delta X$:\footnote{We assume the standard dispersion relation of the 
photon.}
\begin{equation}
\Delta X=\frac{\Delta}{K}=\frac{1/100}{10^{12}\,\mathrm{eV}\cdot 1.602\cdot 10^{-19}\,\mathrm{J/eV}}\hbar c\approx 2\cdot 10^{-21}\,\mathrm{m}\,.
\end{equation}
Thus the approximation we use is already very good even if the defect separation lies many orders of magnitude above the Planck length.
\begin{table}[t]
\centering
\begin{tabular}{r|c}
\hline
\multicolumn{1}{c}{$\Delta$} & $w_1/w_2$ \\
\hline
\hline
1 & 0.701693 \\
0.1 & 0.966576 \\
0.01 & 0.996615 \\
0.004 & 0.998645 \\
\hline
\end{tabular}
\caption{In the current table different values $\Delta$ of dimensionless defect distances are considered. For specific values the numbers of Eq.~\eqref{eq:numbers-dense-distribution}
are computed and compared to each other.}
\label{tab:separation-between-defects}
\end{table}

\section{Perturbative Feynman rules of the modified theory}
\label{sec:perturbative-feynman-rules}

The following Feynman rules \eqref{eq:Feynman-rule-1} -- \eqref{eq:Feynman-rule-4b} directly follow from the action given by Eq.~\eqref{eq:action-effective-theory}
\begin{subequations}
\begin{equation}
\label{eq:Feynman-rule-1}
\begin{array}{c}
\begin{fmfgraph*}(75,35)
\fmfpen{thin}
\fmfleft{i1}
\fmfright{o1}
\fmf{photon,label=$\overrightarrow{\phantom{x}k\phantom{x}}$}{i1,o1}
\fmfv{label=$\mu$}{i1}
\fmfv{label=$\nu$}{o1}
\fmfdot{i1,o1}
\end{fmfgraph*}
\end{array}\qquad=
-\mathrm{i}\eta^{\mu\nu}\widetilde{\Delta}(k)\,,\quad \widetilde{\Delta}(k)=\frac{1}{k^2+\mathrm{i}\epsilon}\,,
\end{equation}
\begin{equation}
\label{eq:Feynman-rule-2}
\begin{array}{c}
\begin{fmfgraph*}(75,35)
\fmfpen{thin}
\fmfleft{i1}
\fmfright{o1}
\fmf{plain,label=$\overrightarrow{\phantom{x}k\phantom{x}}$}{i1,o1}
\fmfdot{i1}
\fmfv{decoration.shape=cross,decoration.filled=full,decoration.size=15}{o1}
\end{fmfgraph*}
\end{array}\qquad=-\mathrm{i}\widetilde{H}(k)\,,\quad \widetilde{H}(k)=\frac{-(b^{(0)})^2}{k^2-1/(b^{(0)})^2+\mathrm{i}\epsilon}\,,
\end{equation}
\begin{equation}
\label{eq:Feynman-rule-3}
\begin{array}{c}
\begin{fmfgraph*}(75,75)
\fmfpen{thin}
\fmfleft{i1}
\fmfright{o1,o2}
\fmf{plain,label=$\overrightarrow{\phantom{i}k_3\phantom{i}}$}{i1,v1}
\fmf{photon,label=$k_2\nwarrow$,label.side=right}{v1,o1}
\fmf{photon,label=$k_1\swarrow$,label.side=left}{v1,o2}
\fmfforce{0.0w,0.5h}{i1}
\fmfforce{1.0w,0.0h}{o1}
\fmfforce{1.0w,1.0h}{o2}
\fmfforce{0.586w,0.5h}{v1}
\fmfv{label=$\beta$}{o1}
\fmfv{label=$\alpha$}{o2}
\fmfdot{v1}
\end{fmfgraph*}
\end{array}\qquad=\mathrm{i}\lambda^{(0)}\varepsilon^{\mu\alpha\varrho\beta}k_{1,\mu}k_{2,\varrho}=-\mathrm{i}\lambda^{(0)}\varepsilon^{\mu\alpha\varrho\beta}k_{1,\mu}k_{3,\varrho}\,.
\end{equation}
\begin{equation}
\label{eq:Feynman-rule-4a}
\begin{array}{c}
\begin{fmfgraph*}(75,35)
\fmfpen{thin}
\fmfleft{i1}
\fmfright{o1}
\fmf{plain,label=$\overrightarrow{\phantom{i}p\phantom{i}}$}{i1,v1}
\fmf{plain,label=$\overleftarrow{\phantom{i}q\phantom{i}}$}{v1,o1}
\fmfv{decoration.shape=cross,decoration.filled=full,decoration.size=15,label=$\mathscr{R}$,label.angle=90,label.dist=0.25cm}{v1}
\end{fmfgraph*}
\end{array}\qquad=\int\limits^{\Lambda'} \mathrm{d}\Lambda\,\widetilde{G}_{\mathscr{R}}(p)\widetilde{G}_{\mathscr{R}}(q)\,.
\end{equation}
The first Feynman rule gives the photon propagator of the modified theory in Feynman gauge, which corresponds to the photon propagator of
standard QED (in this gauge). The second gives the propagator of the scalar field $\phi$, where this has to be connected to a single defect.
The third describes the interaction between $\phi$ and the photon and the fourth the ``defect vertex'' for a finite number $\mathcal{N}$ of
defects in a box. Performing the limit $\mathcal{N}\mapsto \infty$ (consider Ass.~(2a) -- (2d)) combined with the integration over all
Lorentz transformations to an upper cut-off leads to the fifth Feynman rule that we use to compute the one-loop correction in this limit
\begin{equation}
\label{eq:Feynman-rule-4b}
\begin{array}{c}
\begin{fmfgraph*}(75,35)
\fmfpen{thin}
\fmfleft{i1}
\fmfright{o1}
\fmf{plain,label=$\overrightarrow{\phantom{i}p\phantom{i}}$}{i1,v1}
\fmf{plain,label=$\overleftarrow{\phantom{i}q\phantom{i}}$}{v1,o1}
\fmfv{decoration.shape=cross,decoration.filled=full,decoration.size=15}{v1}
\end{fmfgraph*}
\end{array}\qquad=\Xi\varrho\,\delta^{(4)}(p+q)\,.
\end{equation}
So each $\phi$ line has to begin or end at one defect. If this is not
the case one of the momenta has to be set to zero.

Respecting Ass.~(2a) -- (2d) the following effective vertex can be introduced for the interaction of a photon with the
defect via the $\phi$-field. The latter has been integrated out at one-loop level resulting in
\begin{equation}
\label{eq:Feynman-rule-5}
\begin{array}{c}
\begin{fmfgraph*}(75,25)
\fmfleft{i1}
\fmflabel{$\nu$}{i1}
\fmfright{o1}
\fmflabel{$\gamma$}{o1}
\fmf{photon,label=$\overrightarrow{\phantom{i}k\phantom{i}}$}{i1,v1}
\fmf{photon,label=$\overrightarrow{\phantom{i}k\phantom{i}}$}{v1,o1}
\fmfblob{0.5cm}{v1}
\fmfforce{0.5w,0.5h}{v1}
\fmfforce{0.5w,0.5h}{v2}
\end{fmfgraph*}
\end{array}\quad=-\mathcal{C}\mathrm{i}\Pi^{\nu\gamma}(k)\,,\quad \mathcal{C}=\Xi b_{\varrho}^4\lambda^2\varrho\,,
\end{equation}
\end{subequations}
with $\Pi^{\nu\gamma}(k)$ given by Eq.~\eqref{eq:result-one-loop-eq2}.

\section{Derivation of the Passarino--Veltman reduction}
\label{sec:derivation-passarino-veltman}

In this section the Passarino--Veltman reduction of the tensor integral $I_{\varrho\sigma}$ given by Eq.~\eqref{eq:tensor-one-loop-integral}
in Sec.~\ref{subsec:passarino-veltman-decomposition} will be presented in detail.
\begin{equation}
I_{\varrho\sigma}=\int \mathrm{d}^dq\,\frac{q_{\varrho}q_{\sigma}}{\left[q^2-1/(b^{(0)})^2+\mathrm{i}\epsilon\right]^2\left[(k-q)^2+\mathrm{i}\epsilon\right]}\,.
\end{equation}
Performing the reduction we can neglect all $\mathrm{i}\epsilon$. First, we contract $I_{\varrho\sigma}$ with $-2k^{\varrho}$ and this leads to
\begin{align}
-2k^{\varrho}I_{\varrho\sigma}&=\int \mathrm{d}^dq\,\frac{-(2k\cdot q)q_{\sigma}}{\left[q^2-1/(b^{(0)})^2\right]^2(k-q)^2} \notag \\
&=\underbrace{\int \mathrm{d}^dq\,\frac{q_{\sigma}}{\left[q^2-1/(b^{(0)})^2\right]^2}}_{=0}-\int \mathrm{d}^dq\,\frac{(k^2+q^2)q_{\sigma}}{\left[q^2-1/(b^{(0)})^2\right]^2(k-q)^2}\,.
\end{align}
Second, by contracting the latter result again with $-2k^{\sigma}$ we obtain the quantity $4K_1$ defined in Sec.~\ref{subsec:passarino-veltman-decomposition}:
\begin{align}
4K_1&=4k^{\varrho}k^{\sigma}I_{\varrho\sigma}=-\int \mathrm{d}^dq\,\frac{(k^2+q^2)(-2k\cdot q)}{\left[q^2-1/(b^{(0)})^2\right]^2(k-q)^2} \notag \\
&=-\underbrace{\int \mathrm{d}^dq\,\frac{k^2+q^2}{\left[q^2-1/(b^{(0)})^2\right]^2}}_{\equiv \widehat{I}_3}+\underbrace{\int \mathrm{d}^dq\,\frac{(k^2+q^2)^2}{\left[q^2-1/(b^{(0)})^2\right]^2(k-q)^2}}_{\equiv \widehat{I}_{4}}\,.
\end{align}
Now we still have to decompose the integrals $\widehat{I}_3$ and $\widehat{I}_4$ into the master integrals of Eq.~\eqref{eq:master-integrals}
\begin{align}
\widehat{I}_{3}&=\int \mathrm{d}^dq\,\frac{k^2+q^2}{\left[q^2-1/(b^{(0)})^2\right]^2}=k^2\int \mathrm{d}^dq\,\frac{1}{\left[q^2-1/(b^{(0)})^2\right]^2}+\int \mathrm{d}^dq\,\frac{q^2}{\left[q^2-1/(b^{(0)})^2\right]^2} \notag \\
&=\left(k^2+\frac{1}{(b^{(0)})^2}\right)\int \mathrm{d}^dq\,\frac{1}{\left[q^2-1/(b^{(0)})^2\right]^2}+\int \mathrm{d}^dq\,\frac{1}{q^2-1/(b^{(0)})^2}\,.
\end{align}
\begin{align}
\widehat{I}_{4}&=\int \mathrm{d}^dq\,\frac{k^4+2k^2q^2+q^4}{\left[q^2-1/(b^{(0)})^2\right]^2(k-q)^2} \notag \\
&=k^4\int \mathrm{d}^dq\,\frac{1}{\left[q^2-1/(b^{(0)})^2\right]^2(k-q)^2}+2k^2\underbrace{\int \mathrm{d}^dq\,\frac{q^2}{\left[q^2-1/(b^{(0)})^2\right]^2(k-q)^2}}_{\equiv \widehat{I}_5} \notag \\
&\phantom{{}={}}+\underbrace{\int \mathrm{d}^dq\,\frac{q^4}{\left[q^2-1/(b^{(0)})^2\right]^2(k-q)^2}}_{\equiv \widehat{I}_6}\,.
\end{align}
\begin{equation}
\widehat{I}_5=\int \mathrm{d}^dq\,\frac{1}{\left[q^2-1/(b^{(0)})^2\right](k-q)^2}+\frac{1}{(b^{(0)})^2}\int \mathrm{d}^dq\,\frac{1}{\left[q^2-1/(b^{(0)})^2\right]^2(k-q)^2}\,.
\end{equation}
\begin{align}
\widehat{I}_6&=\underbrace{\int \mathrm{d}^dq\,\frac{1}{(k-q)^2}}_{=0}+\frac{2}{(b^{(0)})^2}\int \mathrm{d}^dq\,\frac{q^2}{\left[q^2-1/(b^{(0)})^2\right]^2(k-q)^2} \notag \\
&\phantom{{}={}\int \mathrm{d}^dq\frac{1}{(k-q)^2}}-\frac{1}{(b^{(0)})^4}\int \mathrm{d}^dq\,\frac{1}{\left[q^2-1/(b^{(0)})^2\right]^2(k-q)^2} \notag \\
&=\frac{2}{(b^{(0)})^2}\int \mathrm{d}^dq\,\frac{1}{\left[q^2-1/(b^{(0)})^2\right](k-q)^2}+\frac{1}{(b^{(0)})^4}\int \mathrm{d}^dq\,\frac{1}{\left[q^2-1/(b^{(0)})^2\right]^2(k-q)^2}\,.
\end{align}
The contraction of $I_{\varrho\sigma}$ with the metric tensor leads to $K_2$ which was also defined in Sec.~\ref{subsec:passarino-veltman-decomposition}
\begin{align}
\eta^{\varrho\sigma}I_{\varrho\sigma}&=\int \mathrm{d}^dq\,\frac{q^2}{\left[q^2-1/(b^{(0)})^2\right]^2(k-q)^2} \notag \\
&=\int \mathrm{d}^dq\,\frac{1}{\left[q^2-1/(b^{(0)})^2\right](k-q)^2}+\frac{1}{(b^{(0)})^2} \int \mathrm{d}^dq\,\frac{1}{\left[q^2-1/(b^{(0)})^2\right]^2(k-q)^2}\,.
\end{align}

\section{Computation of scalar integrals}
\label{sec:computation-scalar-integrals}

Finally we evaluate the scalar integrals of Eq.~\eqref{eq:master-integrals} in Sec.~\ref{subsec:passarino-veltman-decomposition}.
We use dimensional regularization with $d=4-2\widehat{\varepsilon}$ and later on we employ the reasonable redefinition
\begin{equation}
\frac{1}{\varepsilon}\equiv \frac{1}{\widehat{\varepsilon}}-\gamma_{\scriptscriptstyle{E}}+\ln(4\pi)\,.
\end{equation}
The simplest integral without any external momenta can be computed as follows
\begin{align}
A_0\left(\frac{1}{(b^{(0)})^2}\right)&=-\mathrm{i}(2\pi\mu)^{4-d}\int \mathrm{d}\Omega_d \int\limits_0^{\infty} \mathrm{d}q\,\frac{q^{d-1}}{q^2+1/(b^{(0)})^2} \notag \\
&=-\mathrm{i}(2\mu)^{4-d}\frac{2\pi^{4-d/2}}{\Gamma(d/2)}(b^{(0)})^{2-d}\frac{\Gamma(1-d/2)\Gamma(d/2)}{2} \notag \\
&=-\mathrm{i}(2\mu)^{4-d}\pi^{4-d/2}(b^{(0)})^{2-d}\Gamma\left(1-\frac{d}{2}\right) \notag \\
&=-\mathrm{i}(2\mu)^{2\widehat{\varepsilon}}\pi^{2+\widehat{\varepsilon}}(b^{(0)})^{2(\widehat{\varepsilon}-1)}\Gamma(\widehat{\varepsilon}-1)
\notag \\
&=\mathrm{i}\pi^2\frac{1}{(b^{(0)})^2}\left[\frac{1}{\varepsilon}-\ln\left(\frac{1}{(b^{(0)})^2\mu^2}\right)+1\right]+\mathcal{O}(\varepsilon)\,.
\end{align}
The remaining integrals can be evaluated in the same way
\begin{align}
B_0\left(0,\frac{1}{(b^{(0)})^2},\frac{1}{(b^{(0)})^2}\right)&=\mathrm{i}(2\pi\mu)^{4-d}\int \mathrm{d}\Omega_d\int\limits_0^{\infty} \mathrm{d}q\,\frac{q^{d-1}}{\left(q^2+1/(b^{(0)})^2\right)^2} \notag\displaybreak[0] \\
&=\mathrm{i}(2\pi\mu)^{4-d}\frac{2\pi^{d/2}}{\Gamma(d/2)}(b^{(0)})^{4-d}\frac{\Gamma(2-d/2)\Gamma(d/2)}{2\Gamma(2)} \notag\displaybreak[0] \\
&=\mathrm{i}(2\mu)^{4-d}\pi^{4-d/2}(b^{(0)})^{4-d}\Gamma\left(2-\frac{d}{2}\right) \notag \\
&=\mathrm{i}(2\mu)^{2\widehat{\varepsilon}}\pi^{2+\widehat{\varepsilon}}(b^{(0)})^{2\widehat{\varepsilon}}\Gamma(\widehat{\varepsilon})
\notag\displaybreak[0] \\
&=\mathrm{i}\pi^2\left[\frac{1}{\varepsilon}-\ln\left(\frac{1}{(b^{(0)})^2\mu^2}\right)\right]+\mathcal{O}(\varepsilon)\,.
\end{align}
\begin{align}
\hspace{-0.5cm}B_0\left(-k,\frac{1}{(b^{(0)})^2},0\right)&=(2\pi\mu)^{4-d}\int \mathrm{d}^dq\,\frac{1}{\left(q^2-1/(b^{(0)})^2+\mathrm{i}\epsilon\right)\left[(k-q)^2+\mathrm{i}\epsilon\right]} \notag \\
&=(2\pi\mu)^{4-d}\int_0^1 \mathrm{d}x \int \mathrm{d}^dq\,\left[q^2-\frac{1}{(b^{(0)})^2}+\left(k^2-2k\cdot q+\frac{1}{(b^{(0)})^2}\right)x+\mathrm{i}\epsilon\right]^{-2} \notag \\
&=(2\pi\mu)^{4-d}\int_0^1 \mathrm{d}x \int \mathrm{d}^dq\,\left[(q-kx)^2-M^2\right]^{-2}\,,
\end{align}
where
\begin{equation}
\label{eq:definition-m2}
-M^2\equiv -k^2x^2+\left(k^2+\frac{1}{(b^{(0)})^2}\right)x-\frac{1}{(b^{(0)})^2}+\mathrm{i}\epsilon\,.
\end{equation}
This then leads to
\begin{align}
\hspace{-0.5cm}B_0\left(-k,\frac{1}{(b^{(0)})^2},0\right)&=\mathrm{i}\pi^2\left\{\frac{1}{\varepsilon}-\int_0^1 \mathrm{d}x\,\ln\left[\frac{k^2x^2-\left(k^2+1/(b^{(0)})^2\right)x+1/(b^{(0)})^2-\mathrm{i}\epsilon}{\mu^2}\right]\right\}+\mathcal{O}(\varepsilon) \notag \\
&\equiv \mathrm{i}\pi^2\left(\frac{1}{\varepsilon}-I_{B_{0,2}}\right)+\mathcal{O}(\varepsilon)\,.
\end{align}
As already mentioned, the $C_0$-integral is both infrared and ultraviolet convergent. Hence we can set $d=4$ at the beginning and we obtain
\begin{align}
C_0\left(-k,0,\frac{1}{(b^{(0)})^2},0,\frac{1}{(b^{(0)})^2}\right)&=\int \mathrm{d}^4q\,\frac{1}{\left(q^2-1/(b^{(0)})^2+\mathrm{i}\epsilon\right)\left[(k-q)^2+\mathrm{i}\epsilon\right]} \notag \\
&=\frac{\Gamma(2+1)}{\Gamma(2)\Gamma(1)}\int_0^1 \mathrm{d}x_1 \int_0^1 \mathrm{d}x_2\,\delta(1-x_1-x_2) \notag \\
&\phantom{{}={}}\times x_1\left\{\left(q^2-\frac{1}{(b^{(0)})^2}+\mathrm{i}\epsilon\right)x_1+\left[(k-q)^2+\mathrm{i}\epsilon\right]x_2\right\}^{-3} \notag \\
&=2\int\limits_0^1 \mathrm{d}x_2\,\frac{1-x_2}{[(q-kx_2)^2-M^2]^3}\,,
\end{align}
with $M^2$ given by Eq.~\eqref{eq:definition-m2}. Finally this results in
\begin{align}
C_0\left(-k,0,\frac{1}{(b^{(0)})^2},0,\frac{1}{(b^{(0)})^2}\right)&=-\mathrm{i}\pi^2\int^1_0 \mathrm{d}x\,\frac{1-x}{k^2x^2-\left(k^2+1/(b^{(0)})^2\right)x+1/(b^{(0)})^2} \notag \\
&\equiv -\mathrm{i}\pi^2 I_{C_0}\,.
\end{align}
Now we want to compute the remaining one-dimensional integrals
\begin{align}
I_{B_{0,2}}&=\int_0^1 \mathrm{d}x\,\ln\left[\frac{k^2x^2-\left(k^2+1/(b^{(0)})^2\right)+1/(b^{(0)})^2-\mathrm{i}\epsilon}{\mu^2}\right]\overset{x\mapsto 1-x}{=} \notag \\
&=\int\limits_0^1 \mathrm{d}x\,\ln\left[\frac{k^2x^2-\left(k^2-1/(b^{(0)})^2\right)x-\mathrm{i}\epsilon}{\mu^2}\right] \notag \\
&=-\ln(\mu^2)+\int_0^1 \mathrm{d}x\,\ln(x)+\int_0^1 \mathrm{d}x\,\ln\left[k^2x-\left(k^2-1/(b^{(0)})^2\right)-\mathrm{i}\epsilon\right] \notag \\
&=-\ln(\mu^2)-1+\frac{1}{k^2}\left[y\ln(y)-y\right]^{1/(b^{(0)})^2-\mathrm{i}\epsilon}_{-k^2+1/(b^{(0)})^2-\mathrm{i}\epsilon} \notag \displaybreak[0] \\
&=-\ln(\mu^2)-1 \notag \\
&\phantom{{}={}}+\frac{1}{k^2}\left[\frac{1}{(b^{(0)})^2}\ln\left(\frac{1}{(b^{(0)})^2}-\mathrm{i}\epsilon\right)-\frac{1}{(b^{(0)})^2}\right. \notag \displaybreak[0] \\
&\phantom{{}={}}+\left.\left(k^2-\frac{1}{(b^{(0)})^2}\right)\ln\left(-k^2+\frac{1}{(b^{(0)})^2}-\mathrm{i}\epsilon\right)-k^2+\frac{1}{(b^{(0)})^2}\right] \notag \displaybreak[0] \\
&=\ln\left(\frac{1}{(b^{(0)})^2\mu^2}\right)-2+\frac{k^2-1/(b^{(0)})^2}{k^2}\ln\left(\frac{-k^2+1/(b^{(0)})^2}{1/(b^{(0)})^2}-\mathrm{i}\epsilon\right)\,.
\end{align}
\begin{align}
k^2I_{C_0}&=\int^1_0 \mathrm{d}x\,\frac{1-x}{k^2x^2-\left(k^2+1/(b^{(0)})^2\right)x+1/(b^{(0)})^2-\mathrm{i}\epsilon} \overset{x\mapsto 1-x}{=} \notag \\
&=\int^1_0 \mathrm{d}x\,\frac{1}{x-1+1/[(b^{(0)})k]^2-\mathrm{i}\epsilon}=\left[\ln\left(x-1+\frac{1}{[(b^{(0)})k]^2}-\mathrm{i}\epsilon\right)\right]^1_0 \notag \\
&=\ln\left(\frac{1}{[(b^{(0)})k]^2}-\mathrm{i}\epsilon\right)-\ln\left(\frac{1}{[(b^{(0)})k]^2}-1-\mathrm{i}\epsilon\right)=-\ln\Big[1-(b^{(0)})^2k^2-\mathrm{i}\epsilon\Big]\,.
\end{align}

\end{appendix}

\end{fmffile}

\newpage



\begin{thebibliography}{99}

\bibitem{Wheeler:1957mu}
J.~A.~Wheeler,
``On the nature of quantum geometrodynamics,''
Annals Phys.\ {\bf 2}, 604 (1957).

\bibitem{Wheeler:1968}
J.~A.~Wheeler,
\emph{Relativity, Gravitation and Topology},
in \emph{Battelle Rencontres 1967}, Editors C.M. DeWitt and J.~A.~ Wheeler (Benjamin, New York, 1968),
pp. 242-307.

\bibitem{Hawking:1979zw}
S.~W.~Hawking,
``Space-time foam,''
Nucl.\ Phys.\ B {\bf 144}, 349 (1978).

\bibitem{Hawking:1979pi}
S.~W.~Hawking, D.~N.~Page, and C.~N.~Pope,
 ``Quantum gravitational bubbles,''
Nucl.\ Phys.\ B {\bf 170}, 283 (1980).

\bibitem{Friedman:1990xc}
J.~Friedman, M.~S.~Morris, I.~D.~Novikov, F.~Echeverria, G.~Klinkhammer, K.~S.~Thorne, and U.~Yurtsever,
``Cauchy problem in space-times with closed timelike curves,''
Phys.\ Rev.\ D {\bf 42}, 1915 (1990).

\bibitem{Visser:1996}
M.~Visser,
\emph{Lorentzian Wormholes: From Einstein to Hawking}
(Springer, New York, 1996).

\bibitem{Schwarz:2010}
M.~Schwarz,
\emph{Nontrivial Spacetime Topology, Modified Dispersion Relations, and an SO(3)-Skyrme Model} (Ph.D. thesis)
(Dr. Hut, Munich, 2010).

\bibitem{BernadotteKlinkhamer2007}
S.~Bernadotte and F.~R.~Klinkhamer,
``Bounds on length scales of classical spacetime foam models,''
Phys.\ Rev.\ D {\bf 75}, 024028 (2007),
arXiv:hep-ph/0610216.

\bibitem{hep-th/0312032}
F.~R.~Klinkhamer and C.~Rupp,
``Space-time foam, CPT anomaly, and photon propagation,''
Phys.\ Rev.\ D\ {\bf 70}, 045020 (2004),
hep-th/0312032.

\bibitem{Klinkhamer:2012pr}
F.~R.~Klinkhamer,
private communication.

\bibitem{Kibble:1976sj}
T.~W.~B.~Kibble,
``Topology of cosmic domains and strings,''
J.\ Phys.\ A {\bf 9}, 1387 (1976).

\bibitem{Dowker:2003hb}
F.~Dowker, J.~Henson, and R.~D.~Sorkin,
``Quantum gravity phenomenology, Lorentz invariance and discreteness,''
Mod.\ Phys.\ Lett.\ A {\bf 19}, 1829 (2004),
gr-qc/0311055.

\bibitem{Bombelli:2006nm}
L.~Bombelli, J.~Henson, and R.~D.~Sorkin,
``Discreteness without symmetry breaking: A Theorem,''
Mod.\ Phys.\ Lett.\ A {\bf 24}, 2579 (2009),
gr-qc/0605006.

\bibitem{Komatsu:2008hk}
E.~Komatsu {\it et al.} [WMAP Collaboration],
``Five-year Wilkinson Microwave Anisotropy Probe (WMAP) observations: Cosmological interpretation,''
Astrophys.\ J.\ Suppl.\ {\bf 180}, 330 (2009),
arXiv:0803.0547 [astro-ph].

\bibitem{Kostelecky:2008ts}
V.~A.~Kosteleck\'{y} and N.~Russell,
``Data Tables for Lorentz and CPT Violation,''
Rev.\ Mod.\ Phys.\ {\bf 83}, 11 (2011),
arXiv:0801.0287 [hep-ph].

\bibitem{Steinhauser:2003}
M.~Steinhauser,
``\"{U}bungen zu Strahlungskorrekturen in Eichtheorien,''
in German,
available at \url{http://maria-laach.physik.uni-siegen.de/2003/programm.html}.

\bibitem{PeskinSchroeder1995}
M.~E.~Peskin and D.~V.~Schroeder, 
\emph{An Introduction to Quantum Field Theory}
(Addison--Wesley, Reading, USA, 1995).

\bibitem{ChadhaNielsen1983}
S.~Chadha and H.B.~Nielsen,
``Lorentz invariance as a low-energy phenomenon,''
Nucl. Phys. B {\bf 217}, 125 (1983).

\bibitem{ColladayKostelecky1998}
D.~Colladay and V.~A.~Kosteleck\'{y},
``Lorentz-violating extension of the standard model,''
Phys. Rev. D~{\bf 58}, 116002 (1998),
arXiv:hep-ph/9809521.

\bibitem{KosteleckyMewes2002}
V.~A.~Kosteleck\'{y} and M.~Mewes,
``Signals for Lorentz violation in electrodynamics,''
Phys. Rev. D {\bf 66}, 056005 (2002),
arXiv:hep-ph/0205211.

\bibitem{BaileyKostelecky2004}
Q.~G.~Bailey and V.~A.~Kosteleck\'{y},
``Lorentz-violating electrostatics and magnetostatics,''
Phys. Rev. D {\bf 70}, 076006 (2004),
arXiv:hep-ph/0407252.

\bibitem{Carroll-etal1990}
S.~M.~Carroll, G.~B.~Field, and R.~Jackiw,
``Limits on a {L}orentz- and parity-violating modification of electrodynamics,''
Phys. Rev. D {\bf 41}, 1231 (1990).


\bibitem{Bender:1998ke}
C.~M.~Bender and S.~Boettcher,
``Real spectra in non-Hermitian Hamiltonians having \textit{PT} symmetry,''
Phys.\ Rev.\ Lett.\ {\bf 80}, 5243 (1998),
physics/9712001.

\bibitem{Bittner:2012}
S.~Bittner, B.~Dietz, U.~G\"unther, H.~L.~Harney, M.~Miski-Oglu, A.~Richter, and F.~Sch\"afer,
``\textit{PT} symmetry and spontaneous symmetry breaking in a microwave billiard,''
Phys.~Rev.~Lett.~{\bf 108}, 024101 (2012).

\bibitem{Lin:2011}
Z.~Lin, H.~Ramezani, T.~Eichelkraut, T.~Kottos, H.~Cao, and D.~N.~Christodoulides,
``Unidirectional invisibility induced by \textit{PT}-symmetric periodic structures,''
Phys.~Rev.~Lett.~{\bf 106}, 213901 (2011).

\bibitem{Rubinstein:2007}
J.~Rubinstein, P.~Sternberg, and Q.~Ma,
``Bifurcation diagram and pattern formation of phase slip centers in superconducting wires driven with electric currents,''
Phys.~Rev.~Lett.~{\bf 99}, 167003 (2007).

\bibitem{Bender:2004sa}
C.~M.~Bender, D.~C.~Brody, and H.~F.~Jones,
``Extension of \textit{PT} symmetric quantum mechanics to quantum field theory with cubic interaction,''
Phys.\ Rev.\ D {\bf 70}, 025001 (2004),
Erratum-ibid.\ D {\bf 71}, 049901 (2005),
hep-th/0402183.

\end{thebibliography}
\end{document}